\documentclass{aa}

\usepackage{graphicx}
\usepackage[fleqn]{amsmath}
\def\oc{$\omega$~Cen}
\def\d{\mathrm{d}}
\def\masyr{mas\,yr$^{-1}$}
\def\kms{km\,s$^{-1}$}
\def\sbrunits{mas\,yr$^{-1}$\,arcmin$^{-1}$}
\def\dgr{$^\circ$}

\def\Msun{M$_\odot$}
\def\Lsun{L$_\odot$}
\def\Lsunpc2{L$_\odot$\,pc$^{-2}$}
\def\MLsun{M$_\odot$/L$_\odot$}
\def\farcd{\hbox{$.\!\!^\circ$}}

\title{The dynamical distance and intrinsic structure of the globular
  cluster $\omega$ Centauri}
\titlerunning{The dynamical distance and intrinsic structure of \oc}
\authorrunning{G. van de Ven et al.}
\author{G. van de Ven, R.C.E. van den Bosch, E.K. Verolme, P.T. de Zeeuw}
\offprints{G. van de Ven}
\institute{Sterrewacht Leiden, Postbus 9513, 2300 RA Leiden, The Netherlands\\
\email{glenn@strw.leidenuniv.nl}
}
\date{Received 0000 0000, Accepted 0000 0000}

\begin{document}



\abstract{We determine the dynamical distance $D$, inclination $i$,
  mass-to-light ratio $M/L$ and the intrinsic orbital structure of the
  globular cluster \oc, by fitting axisymmetric dynamical models to
  the ground-based proper motions of van Leeuwen et al.\ and
  line-of-sight velocities from four independent data-sets. We bring
  the kinematic measurements onto a common coordinate system, and
  select on cluster membership and on measurement error. This provides
  a homogeneous data-set of 2295 stars with proper motions accurate to
  0.20 \masyr\ and 2163 stars with line-of-sight velocities accurate to 2
  \kms, covering a radial range out to about half the tidal radius.

  We correct the observed velocities for perspective rotation caused
  by the space motion of the cluster, and show that the residual
  solid-body rotation component in the proper motions (caused by
  relative rotation of the photographic plates from which they were
  derived) can be taken out without any modelling other than assuming
  axisymmetry. This also provides a tight constraint on $D\tan i$. The
  corrected mean velocity fields are consistent with regular rotation,
  and the velocity dispersion fields display significant deviations
  from isotropy.

  We model \oc\ with an axisymmetric implementation of Schwarzschild's
  orbit superposition method, which accurately fits the surface
  brightness distribution, makes no assumptions about the degree of
  velocity anisotropy in the cluster, and allows for radial variations
  in $M/L$. We bin the individual measurements on the plane of the sky
  to search efficiently through the parameter space of the models.
  Tests on an analytic model demonstrate that this approach is capable
  of measuring the cluster distance to an accuracy of about 6 per
  cent. Application to \oc\ reveals no dynamical evidence for a
  significant radial dependence of $M/L$, in harmony with the
  relatively long relaxation time of the cluster. The best-fit
  dynamical model has a stellar $V$-band mass-to-light ratio
  $M/L_V=2.5\pm0.1$ \MLsun\ and an inclination $i=50^\circ\pm4^\circ$,
  which corresponds to an average intrinsic axial ratio of $0.78\pm0.03$.
  The best-fit dynamical distance $D=4.8\pm0.3$ kpc (distance modulus
  $13.75\pm0.13$ mag) is significantly larger than obtained by means
  of simple spherical or constant-anisotropy axisymmetric dynamical
  models, and is consistent with the canonical value $5.0\pm0.2$ kpc
  obtained by photometric methods. The total mass of the cluster is
  $(2.5\pm0.3)\times10^6$ \Msun.

  The best-fit model is close to isotropic inside a radius of about 10
  arcmin and becomes increasingly tangentially anisotropic in the
  outer region, which displays significant mean rotation. This
  phase-space structure may well be caused by the effects of the tidal
  field of the Milky Way. The cluster contains a separate disk-like
  component in the radial range between 1 and 3 arcmin, contributing
  about 4\% to the total mass.

} 

\maketitle

\keywords{Galaxy: globular clusters: individual: NGC~5139, galaxy:
  kinematics and dynamics}


\section{Introduction}
\label{sec:intro}

The globular cluster \oc\ (NGC~5139) is a unique window into
astrophysics (van Leeuwen, Hughes \& Piotto
2002\nocite{2002ocuw.conf.....V}).  It is the most massive globular
cluster of our Galaxy, with an estimated mass between $2.4\times10^6
M_\odot$ (Mandushev et al.\ 1991\nocite{1991A&A...252...94M}) and $5.1
\times 10^6 M_\odot$ (Meylan et al.\
1995\nocite{1995A&A...303..761M}). It is also one of the most
flattened globular clusters in the Galaxy (e.g. Geyer, Nelles \& Hopp
1983\nocite{1983A&A...125..359G}) and it shows clear differential
rotation in the line-of-sight (Merritt, Meylan \& Mayor
1997\nocite{1997AJ....114.1074M}).  Furthermore, multiple stellar
populations can be identified (e.g.  Freeman \& Rodgers
1975\nocite{1975ApJ...201L..71F}; Lee et al.\
1999\nocite{1999Natur.402...55L}; Pancino et al.\
2000\nocite{2000ApJ...534L..83P}; Bedin et al.\
2004\nocite{2004ApJ...605L.125B}).  Since this is unusual for a
globular cluster, a whole range of different formation scenarios of
\oc\ have been suggested, from self-enrichment in an
isolated cluster or in the nucleus of a tidally stripped dwarf galaxy,
to a merger between two or more globular clusters (e.g. Icke \&
Alcaino 1988\nocite{1988A&A...204..115I}; Freeman
1993\nocite{1993ASPC...48..608F}; Lee et al.\
2002\nocite{2002ASPC..265..305L}; Tsuchiya, Korchagin \& Dinescu et
al.\ 2004\nocite{2004MNRAS.350.1141T}).

\oc\ has a core radius of $r_c=2.6$ arcmin, a half-light (or
effective) radius of $r_h=4.8$ arcmin and a tidal radius of $r_t=45$
arcmin (e.g.  Trager, King \& Djorgovski
1995\nocite{1995AJ....109..218T}). The resulting concentration index
$\log(r_t/r_c)\sim1.24$ implies that \oc\ is relatively loosely bound.
In combination with its relatively small heliocentric distance of
$5.0\pm0.2$ kpc (Harris et al.
1996\nocite{1996AJ....112.1487H})\footnote{Throughout the paper we use
  this distance of $5.0\pm0.2$ kpc, obtained with photometric methods,
  as the canonical distance.}. This makes it is possible to observe
individual stars over almost the entire extent of the cluster,
including the central parts. Indeed, line-of-sight velocity
measurements\footnote{Instead of the often-used term \textit{radial}
  velocities, we adopt the term \textit{line-of-sight} velocities, to
  avoid confusion with the decomposition of the proper motions in the
  plane of the sky into a radial and tangential component.}  have been
obtained for many thousands of stars in the field of \oc\
(Suntzeff \& Kraft 1996, hereafter
SK96\nocite{1996AJ....111.1913S}; Mayor et al.\ 1997, hereafter
M97\nocite{1997AJ....114.1087M}; Reijns et al.\
2005\nocite{2005A&A...paperII}, hereafter Paper~II; Xie, Gebhardt
et al.\ in preparation, hereafter XGEA). Recently, also
high-quality measurements of proper motions of many thousands of
stars in \oc\ have become available, based on ground-based
photographic plate observations (van Leeuwen et al.\ 2000,
hereafter Paper~I\nocite{2000A&A...360..472V}) and Hubble Space
Telescope (HST) imaging (King \& Anderson
2002\nocite{2002ocuw.conf...21K}).

The combination of proper motions with line-of-sight velocity
measurements allows us to obtain a dynamical estimate of the distance
to \oc\ and study its internal dynamical structure. While
line-of-sight velocity observations are in units of \kms, proper
motions are angular velocities and have units of
(milli)arcsec\,yr$^{-1}$. A value for the distance is required to
convert these angular velocities to \kms. Once this is done, the
proper motion and line-of-sight velocity measurements can be combined
into a three-dimensional space velocity, which can be compared to
kinematic observables that are predicted by dynamical models. By
varying the input parameters of these models, the set of model
parameters (including the distance) that provides the best-fit to the
observations can be obtained. Similar studies for other globular
clusters, based on comparing modest numbers of line-of-sight velocity
and proper motion measurements with simple spherical dynamical models,
were published for M3 (Cudworth 1979\nocite{1979AJ.....84.1312C}), M22
(Peterson \& Cudworth 1994\nocite{1994ApJ...420..612P}), M4 (Peterson,
Rees \& Cudworth 1995\nocite{1995ApJ...443..124P}; see also Rees
1997\nocite{1997ASPC..127..109R}), and M15 (McNamara, Harrison \&
Baumgardt 2004\nocite{2004ApJ...602..264M}).

A number of dynamical models which reproduce the line-of-sight
velocity measurements for \oc\ have been published. As no proper
motion information was included in these models, the distance could
not be fitted and had to be assumed. Furthermore, all these models
were limited by the flexibility of the adopted techniques and assumed
either spherical geometry (Meylan 1987\nocite{1987A&A...184..144M},
Meylan et al.\ 1995\nocite{1995A&A...303..761M}) or an isotropic
velocity distribution (Merritt et al.\ 
1997\nocite{1997AJ....114.1074M}).  Neither of these assumptions is
true for \oc\ (Geyer, Nelles \& Hopp 1983\nocite{1983A&A...125..359G};
Merrifield \& Kent 1990\nocite{1990AJ.....99.1548M}). Recent work,
using an axisymmetric implementation of Schwarzschild's
(1979\nocite{1979ApJ...232..236S}) orbit superposition method, shows
that it is possible to fit anisotropic dynamical models to
(line-of-sight) kinematic observations of non-spherical galaxies (van
der Marel et al.\ 1998\nocite{1998ApJ...493..613V}; Cretton et al.\ 
1999\nocite{1999ApJS..124..383C}; Cappellari et al.\ 
2002\nocite{2002ApJ...578..787C}; Verolme et al.\ 
2002\nocite{2002MNRAS.335..517V}; Gebhardt et al.\ 
2003\nocite{2003ApJ...583...92G}; Krajnovi{\'c} et al.\ 
2005\nocite{2005MNRAS...krajnovic}). In this paper, we extend
Schwarzschild's method in such a way that it can deal with a
combination of proper motion and line-of-sight velocity measurements
of individual stars. This allows us to derive an accurate dynamical
distance and improve our understanding of the internal structure of
\oc.

It is possible to incorporate the discrete kinematic measurements of
\oc\ directly in dynamical models by using maximum likelihood
techniques (Merritt \& Saha 1993\nocite{1993ApJ...409...75M}; Merritt
1993\nocite{1993ApJ...413...79M}; Merritt
1997\nocite{1997AJ....114..228M}; Romanowsky \& Kochanek
2001\nocite{2001ApJ...553..722R}; Kleyna et al.\
2002\nocite{2002MNRAS.330..792K}), but these methods are non-linear,
are not guaranteed to find the global best-fitting model, and are very
CPU-intensive for data-sets consisting of several thousands of
measurements. We therefore decided to bin the measurements instead and
obtain the velocity moments in a set of apertures on the plane of the
sky. While this method is (in principle) slightly less accurate, as
some information in the data may be lost during the binning process,
it is much faster, which allows us to make a thorough investigation of
the parameter space of \oc\ in a relatively short time. It should also
give a good starting point for a subsequent maximum likelihood model
using the individual measurements.

This paper is organised as follows. In \S~\ref{sec:observations},
we describe the proper motion and line-of-sight velocity
measurements and transform them to a common coordinate system. The
selection of the kinematic measurements on cluster membership and
measurement error is outlined in \S~\ref{sec:selection}. In
\S~\ref{sec:kinematics}, we correct the kinematic measurements for
perspective rotation and show that a residual solid-body rotation
component in the proper motions can be taken out without any
modelling other than assuming axisymmetry. This also provides a
tight constraint on the inclination of the cluster.
In \S~\ref{sec:schwarzschild}, we describe our axisymmetric dynamical
modelling method, and test it in \S~\ref{sec:tests} on an analytical
model. In \S~\ref{sec:dynmodels}, we construct the mass model for \oc,
bin the individual kinematic measurements on the plane of the sky and
describe the construction of dynamical models that we fit to these
observations. The resulting best-fit parameters for \oc\ are presented
in \S~\ref{sec:bestfitpar}. We discuss the intrinsic structure of the
best-fit model in \S~\ref{sec:intstructure}, and draw conclusions in
\S~\ref{sec:conclusions}.\looseness=-1


\section{Observations}
\label{sec:observations}

We briefly describe the stellar proper motion and line-of-sight
velocity observations of \oc\ that we use to constrain our dynamical
models (see Table~\ref{tab:datasets}). We then align and transform
them to a common coordinate system.

\begin{table}
\caption{
  Overview of the proper motions and line-of-sight
  velocity data-sets for \oc. The last
  row describes the four different line-of-sight
  velocity data-sets merged together, using the
  stars in common. The precision
  is estimated as the median of the (asymmetric)
  velocity error distribution. If a selection on the
  velocity errors is applied (\S~\ref{sec:selection}), the upper limit is
  given. For the proper motions, we assume a canonical
  distance of 5 kpc to convert from \masyr\ to \kms.
  }
\label{tab:datasets}
\begin{center}
\begin{tabular}{llrrr}
\hline
Source & Extent   & Observed  & Selected  & Precision\\
       & (arcmin) & (\#stars) & (\#stars) & (\kms)   \\
\hline
\multicolumn{5}{c}{proper motions}\\
Paper~I   & 0--30      & 9847   & 2295  & $<4.7$\\
\hline
\multicolumn{5}{c}{line-of-sight velocities}\\
SK96     & 3--23     &  360 &  345 & 2.2\\
M97      & 0--22     &  471 &  471 & 0.6\\
Paper~II & 0--38     & 1966 & 1588 & 2.0\\
XGEA     & 0--3      & 4916 & 1352 & 1.1\\
Merged   & 0--30     &      & 2163 & $<2.0$\\
\hline
\end{tabular}
\end{center}
\end{table}

\subsection{Proper motions}
\label{sec:propermotions}

The proper motion study in Paper~I is based on 100 photographic
plates of \oc, obtained with the Yale-Columbia 66 cm refractor
telescope. The first-epoch observations were taken between 1931
and 1935, for a variable star survey of \oc\ (Martin 1938\nocite{1938AnLei..17b...1M}).
Second-epoch plates, specifically meant for the proper motion
study, were taken between 1978 and 1983. The plates from both
periods were compared and proper motions were measured for 9847
stars. The observations cover a radial range of about 30 arcmin
from the cluster centre.

\subsection{Line-of-sight velocities}
\label{sec:losvelocities}

We use line-of-sight velocity observations from four different
data-sets: the first two, by SK96 and M97, from the literature,
the third is described in the companion Paper~II and the fourth
data-set (XGEA) was kindly provided by Karl Gebhardt in advance of
publication.

SK96 used the \texttt{ARGUS} multi-object spectrograph on the
CTIO 4 m Blanco telescope to measure, from the Ca II triplet range
of the spectrum, the line-of-sight velocities of bright giant and
subgiant stars in the field of \oc. They found respectively 144 and
199 line-of-sight velocity members, and extended the bright sample to
161 with measurements by Patrick Seitzer. The bright giants cover a
radial range from 3 to 22 arcmin, whereas the subgiants vary in
distance between 8 and 23 arcmin. From the total data-set of 360
stars, we remove the 6 stars without (positive) velocity error
measurement together with the 9 stars for which we do not have a
position (see \S~\ref{sec:poscoordproj}), leaving a total of 345
stars.

M97 published 471 high-quality line-of-sight velocity measurements
of giants in \oc, taken with the photoelectric spectrometer
\texttt{CORAVEL}, mounted on the 1.5 m Danish telescope at Cerro
La Silla. The stars in their sample are located between 10 arcsec
and 22 arcmin from the cluster centre.

In Paper~II, we describe the line-of-sight velocity measurements
of 1966 individual stars in the field of \oc, going out in radius
to about 38 arcmin.
Like SK96, we also observed with
\texttt{ARGUS}, but used the Mg$b$ wavelength range. We use the
1589 cluster members, but exclude the single star for which no
positive velocity error measurement is available.

Finally, the data-set of XGEA contains the line-of-sight
velocities of 4916 stars in the central 3 arcmin of \oc. These
measurements were obtained in three epochs over a time span of
four years, using the Rutgers Imaging Fabry-Perot
Spectrophotometer on the CTIO 1.5 m telescope. During the
reduction process, some slightly smeared out single stars were
accidentally identified as two fainter stars. Also, contaminating
light from surrounding stars can lead to offsets in the
line-of-sight velocity measurements. To exclude (most of) these
misidentifications (Gebhardt, priv. comm.), we select the 1352
stars with a measured (approximately $R$-band) magnitude brighter
than 14.5.

\subsection{Coordinate system: positions}
\label{sec:coordsyspositions}

We constrain our dynamical models by merging all the above data-sets.
We convert all stellar positions to the same projected Cartesian
coordinates and align the different data-sets with respect to each
other by matching the stars in common between the different data-sets.
Next, we rotate the coordinates over the observed position angle of
\oc\ to align with its major and minor axis, and give the relation
with the intrinsic axisymmetric coordinate system we assume for our
models.

\subsubsection{Projected Cartesian coordinates ($x'',y''$)}
\label{sec:poscoordproj}

The stellar positions in Paper~I are given in equatorial coordinates
$\alpha$ and $\delta$ (in units of degrees for J2000), with the cluster centre
at $\alpha_0=201\farcd69065$ and $\delta_0=-47\,\farcd47855$. For
objects with small apparent sizes, these equatorial coordinates can be
converted to Cartesian coordinates by setting $x''=
-\Delta\alpha\cos\delta$ and $y''=\Delta\delta$, with $x''$ in the
direction of West and $y''$ in the direction of North, and
$\Delta\alpha\equiv\alpha-\alpha_0$ and
$\Delta\delta\equiv\delta-\delta_0$. However, this transformation
results in severe projection effects for objects that have a large
angular diameter or are located at a large distance from the
equatorial plane. Since both conditions are true for \oc, we must
project the coordinates of each star on the plane of the sky along the
line-of-sight vector through the cluster centre
\begin{eqnarray}
\label{eq:radec2cart}
x'' & = & -r_0\cos\delta\sin\Delta\alpha, \nonumber
\\*[-7.5pt]\\*[-7.5pt]
y'' & = & r_0\left( \sin\delta\cos\delta_0 -
\cos\delta\sin\delta_0\cos\Delta\alpha \right) , \nonumber
\end{eqnarray}
with scaling factor $r_0\equiv10800/\pi$ to have $x''$ and $y''$ in
units of arcmin. The cluster centre is at $(x'',y'')=(0,0)$.

The stellar observations by SK96 are tabulated as a function of
the projected radius to the centre only. However, for each star
for which its ROA number (Woolley
1966\nocite{1966ROAn....2....1W}) appears in the Tables of Paper~I
or M97, we can reconstruct the positions from these data-sets. In
this way, only nine stars are left without a position. The
positions of the stars in the M97 data-set are given in terms of
the projected polar radius $R''$ in arcsec from the cluster centre
and the projected polar angle $\theta''$ in radians from North to
East, and can be straightforwardly converted into Cartesian
coordinates $x''$ and $y''$. For Paper~II, we use the Leiden
Identification (LID) number of each star, to obtain the stellar
positions from Paper~I. The stellar positions in the XGEA data-set
are already in the required Cartesian coordinates $x''$ and $y''$.

\subsubsection{Alignment between data-sets}
\label{sec:poscoordalign}

Although for all data-sets the stellar positions are now in terms
of the projected Cartesian coordinates $(x'',y'')$, (small)
misalignments between the different data-sets are still present.
These misalignments can be eliminated using the stars in common
between the different data-sets. As the data-set of Paper~I covers
\oc\ fairly uniformly over much of its extent, we take their
stellar positions as a reference frame.

All the positions for the Paper~II data-set and most of the
positions for the SK96 data-set come directly from Paper~I, and
hence are already aligned. For the M97 and XGEA data-set, we use
the \texttt{DAOMASTER} program (Stetson 1992), to obtain the
transformation (horizontal and vertical shift plus rotation) that
minimises the positional difference between the stars that are in
common with those in Paper~I: 451 for the M97 data-set and 1667
for the XGEA data-set.

\subsubsection{Major-minor axis coordinates ($x',y'$)}
\label{sec:poscoordmajmin}

With all the data-sets aligned, we finally convert the stellar
positions into the Cartesian coordinates $(x',y')$, with the $x'$-axis
and $y'$-axis aligned with respectively the observed major and minor
axis of \oc. Therefore we have to rotate $(x'',y'')$ over the position
angle of the cluster. This angle is defined in the usual way as the
angle between the observed major axis and North (measured
counterclockwise through East).

To determine the position angle, we fit elliptic isophotes to the
smoothed Digital Sky Survey (\texttt{DSS}) image of \oc, while
keeping the centre fixed. In this way, we find a nearly constant
position angle of 100\dgr\ between 5 and 15 arcmin from the centre
of the cluster. This is consistent with an estimate by Seitzer
(priv. comm.) from a $U$-band image, close to the value of 96\dgr\
found by White \& Shawl (1987\nocite{1987ApJ...317..246W}), but
significantly larger than the position angle of 91.3\dgr\ measured
in Paper~I from star counts.

\subsubsection{Intrinsic axisymmetric coordinates $(x,y,z)$}
\label{sec:poscoordintr}

Now that we have aligned the coordinates in the plane of the sky
$(x',y')$ with the observed major and the major axis, the definition
of the intrinsic coordinate system of our models and the relation
between both becomes straightforward. We assume the cluster to be
axisymmetric and express the intrinsic properties of the model in
terms of Cartesian coordinates $(x,y,z)$, with the $z$-axis the
symmetry axis. The relation between the intrinsic and projected
coordinates is then given by
\begin{eqnarray}
\label{eq:poscartint2obs}
x' & = &  y, \nonumber\\
y' & = & -x\cos i + z\sin i, \\
z' & = & -x\sin i - z\cos i. \nonumber
\end{eqnarray}
The $z'$-axis is along the line-of-sight in the direction away from
us\footnote{In the common (mathematical) definition of a Cartesian
  coordinate system the $z'$-axis would point towards us, but here we
adopt the astronomical convention to have positive line-of-sight
away from us.}, and $i$ is the inclination along which the object
is observed, from $i=0^\circ$ face-on to $i=90^\circ$ edge-on.

\subsection{Coordinate system: velocities}
\label{sec:coordsysvelocities}

After the stellar positions have been transformed to a common
coordinate system, we also convert the proper motions and
line-of-sight velocities to the same (three-dimensional) Cartesian
coordinate system. We centre it around zero (mean) velocity by
subtracting the systemic velocity in all three directions, and
relate it to the intrinsic axisymmetric coordinate system.

\subsubsection{Proper motions}
\label{sec:velcoordpmxy}

The proper motions (in \masyr) of Paper~I are given in
the directions East and North, i.e. in the direction of $-x''$ and
$y''$ respectively. After rotation over the position angle of
100\dgr, we obtain the proper motion components $\mu_{x'}$ and
$\mu_{y'}$, aligned with the observed major and minor axis of \oc,
and similarly, for the proper motion errors.

\subsubsection{Multiple line-of-sight velocity measurements}
\label{sec:velcoordvlos}

In Paper~II, the measured line-of-sight velocities are compared
with those of SK96 and M97 for the stars in common. A systematic
offset in velocity between the different data-sets is clearly
visible in Figure~1 of that paper. We measure this offset with
respect to the M97 data-set, since it has the highest velocity
precision and more than a hundred stars in common with the other
three data-sets: 129 with SK96, 312 with Paper~II\footnote{In
Paper~II, we report only 267 stars in common with the data-set of
M97. The reason is that there the comparison is based on matching
ROA numbers, and since not all stars from Paper~II have a ROA
number, we find here more stars in common by matching in
position.} and 116 with XGEA. As in Paper~II, we apply four-sigma
clipping, i.e., we exclude all stars for which the measured
velocities differ by more than four times the combined velocity
error. This leaves respectively 117, 284 and 109 stars in common
between M97 and the three data-sets of SK96, Paper~II and XGEA.
The (weighted\footnote{To calculate the mean and dispersion of a
sample, we use the weighted estimators and corresponding
uncertainties as described in Appendix A of Paper~II.})  mean
velocity offsets of the data-set of M97 minus the three data-sets
of SK96, Paper~II and XGEA, are respectively $-0.41 \pm 0.08$
\kms, $1.45 \pm 0.07$ \kms\ and $0.00 \pm 0.12$ \kms. For each of
the latter three data-sets, we add these offsets to all observed
line-of-sight velocities.

Next, for each star that is present in more than one data-set, we
combine the multiple line-of-sight velocity measurements. Due to
non-overlapping radial coverage of the data-set of SK96 and XGEA,
there are no stars in common between these two data-sets, and
hence no stars that appear in all four data-sets. There are 138
stars with position in common between three data-sets and 386
stars in common between two data-sets.

For the 138 stars in common between three data-sets, we check if
the three pairwise velocity differences satisfy the four-sigma
clipping criterion. For 6 stars, we find that two of the three
pairs satisfy the criterion, and we select the two velocities that
are closest to each other. For 7 stars, we only find a single pair
that satisfies the criterion, and we select the corresponding two
velocities. Similarly, we find for the 386 stars in common between
two data-sets, 13 stars for which the velocity difference does not
satisfy the criterion, and we choose the velocity measurement with
the smallest error. This means from the 524 stars with multiple
velocity measurements, for 26 stars (5\%) one of the velocity
measurements is removed as an outlier. This can be due to a chance
combination of large errors, a misidentification or a binary;
Mayor et al.\ (1996\nocite{1996oedb.conf..190M}) estimated the
global frequency of short-period binary systems in \oc\ to be
3--4\%.

As pointed out in \S~2.6 of Paper~II, we can use for the stars in
common between (at least) three data-sets, the dispersion of the
pairwise differences to calculate the external (instrumental)
dispersion for each of the data-sets. In this way, we found in
Paper~II that the errors tabulated in SK96 are under-estimated by
about 40\% and hence increased them by this amount, whereas those
in M97 are well-calibrated. Unfortunately, there are too few stars
in common with the XGEA data-set for a similar (statistically
reliable) external error estimate.

In the final sample, we have 125 stars with the weighted mean of
three velocity measurements and 373 stars with the weighted mean
of two velocity measurements. Together with the 2596 single
velocity measurements, this gives a total of 3094 cluster stars
with line-of-sight velocities.

\subsubsection{Systemic velocities}
\label{sec:velcoordvsys}

To centre the Cartesian velocity system around zero mean velocity,
we subtract from both the proper motion data-sets and the merged
line-of-sight data-set the (remaining) systemic velocities. In
combination with the cluster proper motion values from Table 4 of
Paper~I, we find the following systemic velocities
\begin{eqnarray}
  \label{eq:sysvel}
  \mu_{x'}^\mathrm{sys} & = &   3.88 \pm 0.41
  \quad \mathrm{mas\,yr}^{-1}, \nonumber\\
  \mu_{y'}^\mathrm{sys} & = &  -4.44 \pm 0.41
  \quad \mathrm{mas\,yr}^{-1}, \\
  v_{z'}^\mathrm{sys}  & = &  232.02 \pm 0.03
  \quad \mathrm{km\,s}^{-1}. \nonumber
\end{eqnarray}

\subsubsection{Intrinsic axisymmetric coordinate system}
\label{sec:velcoordintr}

In our models, we calculate the velocities in units of \kms. If we
assume a distance $D$ (in units of kpc), the conversion of the proper
motions in units of \masyr\ into units of \kms\ is given by
\begin{equation}
\label{eq:masyr2kms} v_{x'} = 4.74\,D\,\mu_{x'} \quad \mathrm{and}
\quad v_{y'} = 4.74\,D\,\mu_{y'}.
\end{equation}
The relation between observed $(v_{x'},v_{y'},v_{x'})$ and
intrinsic $(v_x,v_y,v_z)$ velocities is the same as in equation~
(\ref{eq:poscartint2obs}), with the coordinates replaced by the
corresponding velocities.

In addition to Cartesian coordinates, we also describe the intrinsic
properties of our axisymmetric models in terms of the usual
cylindrical coordinates $(R,\phi,z)$, with $x=R\cos\phi$ and
$y=R\sin\phi$. In these coordinates the relation between the observed
and intrinsic velocities is
\begin{eqnarray}
\label{eq:velintpol2obscart}
v_{x'} &=& v_R\sin\phi + v_\phi\cos\phi, \nonumber\\
v_{y'} &=& (-v_R\cos\phi + v_\phi\sin\phi)\cos i + v_z\sin i, \\
v_{z'} &=& (-v_R\cos\phi + v_\phi\sin\phi)\sin i + v_z\cos i. \nonumber
\end{eqnarray}
%


\section{Selection}
\label{sec:selection}

We discuss the selection of the cluster members from the different
data-sets, as well as some further removal of stars that cause
systematic deviations in the kinematics.

\subsection{Proper motions}
\label{sec:selgbpm}

In Paper~I, a membership probability was assigned to each star. We
use the stars for which we also have line-of-sight velocity
measurements to investigate the membership determination.
Furthermore, in Paper~I the image of each star was inspected and
classified according to its separation from other stars. We study
the effect of the disturbance by a neighbouring star on the
kinematics. Finally, after selection of the undisturbed cluster
members, we exclude the stars with relatively large uncertainties
in their proper motion measurements, which cause a systematic
overestimation of the mean proper motion dispersion.

\begin{figure}
\includegraphics{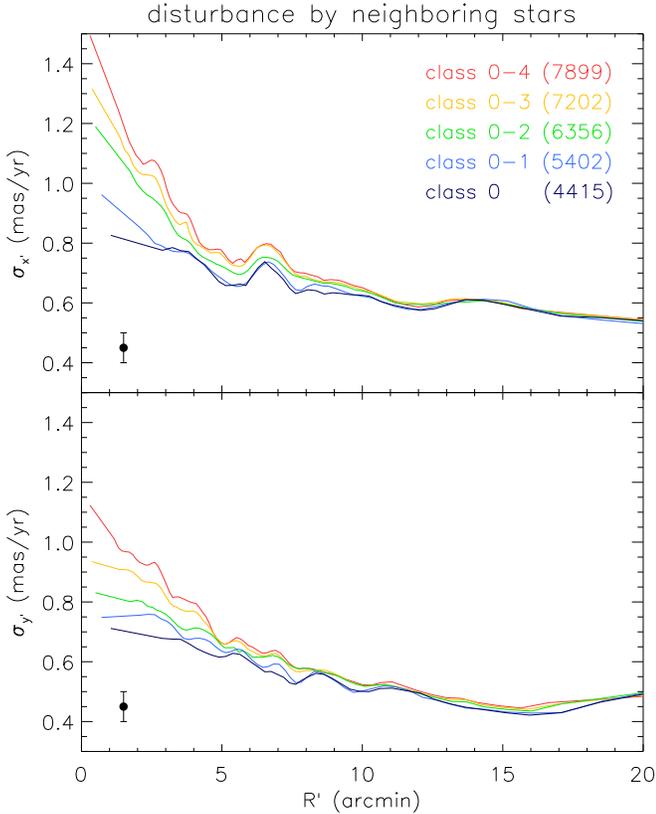}
\caption{
  Velocity dispersion profiles, calculated along concentric rings,
  from the proper motions of Paper~I. The dispersion profiles from the
  proper motions in the $x'$-direction ($y'$-direction) are shown in
  the top (bottom). The error bar at the bottom-left indicates
  the typical uncertainty
  The red curves are the dispersion profiles for all 7899 cluster
  stars with proper motion measurements. The other coloured curves
  show how the dispersion decreases significantly, especially in the
  crowded centre of \oc, when sequentially stars of class 4 (severely
  disturbed) to class 1 (slightly disturbed) are removed.  We select
  the 4415 undisturbed stars of class 0.}
\label{fig:rad_pmclass}
\end{figure}

\begin{figure}
\includegraphics{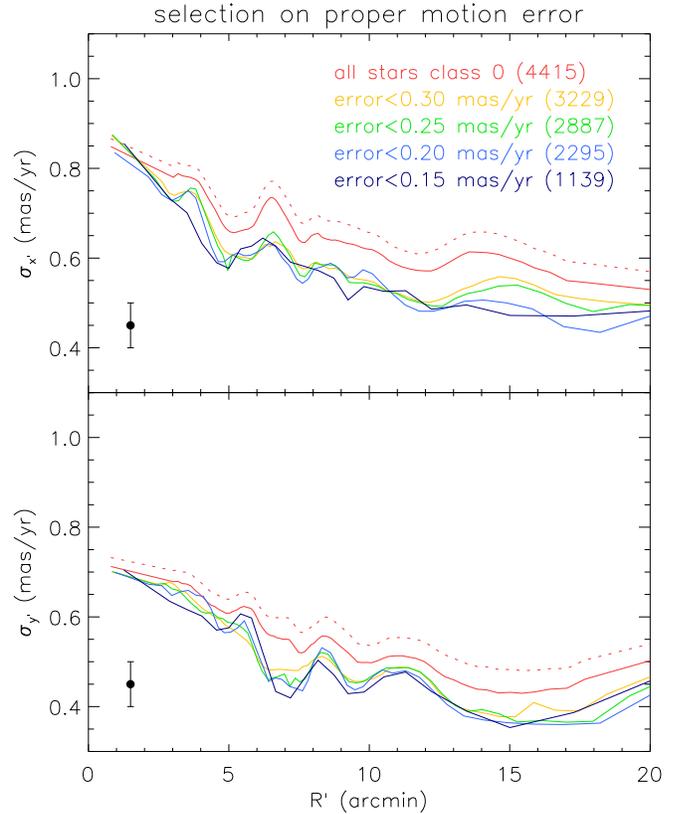}
\caption{
  Proper motion dispersion profiles as in
  Figure~\ref{fig:rad_pmclass}. Starting with all undisturbed (class
  0) cluster stars (red solid curve), sequentially a smaller number of
  stars is selected by setting a tighter limit on the allowed error in
  their proper motion measurements. The dispersion decreases if the
  stars with uncertain proper motion measurements are excluded. This effect is
  significant and larger than the dispersion broadening due to the
  individual velocity errors, indicated by the red dotted curve. We
  select the 2295 stars with proper motion error smaller than 0.20
  \masyr, since below this limit the kinematics stay similar.}
\label{fig:rad_pmvelerr}
\end{figure}

\subsubsection{Membership determination}
\label{sec:selgbpmmpperc}

The membership probability in Paper~I was assigned to each star in
the field by assuming that the distribution of stellar velocities
is Gaussian. In most studies, this is done by adopting one common
distribution for the entire cluster. However, this does not take
into account that the internal dispersion, as well as the relative
number of cluster stars, decreases with radius. To better
incorporate these two effects, the membership probability in
Paper~I was calculated along concentric rings.

By matching the identification numbers and the positions of stars,
we find that there are 3762 stars for which both proper motions
and line-of-sight velocities are measured. This allows us to
investigate the quality of the membership probability assigned in
Paper~I, as the separation of cluster and field stars is very
clean in line-of-sight velocities (see e.g. Paper~II, Figure~4).

From the line-of-sight velocities, we find that of the 3762 matched
stars, 3385 are cluster members. Indeed, most of these cluster stars,
3204 (95\%), have a membership probability based on their proper
motions of at least 68 per cent.
Based on the latter criterion, the remaining 181
(5\%) cluster stars are wrongly classified as field stars in Paper~I.
From the 3762 matched stars, 377 stars are field stars from the
line-of-sight velocity data-set of Paper~II. Based on a membership
probability of 68 per cent, 54 (14\%) of these field stars are wrongly
classified as cluster members in Paper~I. This fraction of field stars
misclassified as cluster stars is an upper limit, since the obvious
field stars are already removed from the proper motion data-set of
Paper~I.

Wrongly classifying cluster stars as field stars is relatively
harmless for our purpose, since it only reduces the total cluster
data-set. However, classifying field stars as members of the cluster
introduces stars from a different population with different
(kinematical) properties. With a membership probability of 99.7 per
cent 
the fraction of field stars misclassified as
cluster stars reduces to 5\%. However, at the same time we expect to
miss almost 30\% of the cluster stars as they are wrongly classified
as field stars. Taking also into account that the additional
selections on disturbance by neighbouring stars and velocity error
below remove (part of) the field stars misclassified as cluster stars,
we consider stars with a membership probability of at least 68 per
cent as cluster members.

While for the 3762 matched stars, the line-of-sight velocities
confirm 3385 stars as cluster members, from the remaining 6084
(unmatched) stars of Paper~I, 4597 stars have a proper motion
membership probability of at least 68 per cent. From the resulting
proper motion distribution, we remove 83 outliers with proper
motions five times the standard deviation away from the mean,
leaving a total of 7899 cluster stars.

\subsubsection{Disturbance by neighbouring stars}
\label{sec:selgbpm_class}

In Paper~I, each star was classified according to its separation
from other stars on a scale from 0 to 4, from completely free to
badly disturbed by a neighbouring star. In
Figure~\ref{fig:rad_pmclass}, we show the effect of the
disturbance on the proper motion dispersion. The (smoothed)
profiles are constructed by calculating the mean proper motion
dispersion of the stars binned in concentric rings, taking the
individual measurement errors into account
(Appendix~\ref{sec:mlvelocitymoments}).  The proper motions in the
$x'$-direction give rise to the velocity dispersion profiles
$\sigma_{x'}$ in the upper panel. The proper motions in the
$y'$-direction yield the velocity dispersion profiles
$\sigma_{y'}$ in the bottom panel. The red curves are the velocity
dispersion profiles for all 7899 cluster stars with proper motion
measurements. The other coloured curves show how, especially in
the crowded centre of \oc, the dispersion decreases significantly
when sequentially stars of class 4 (severely disturbed) to class 1
(slightly disturbed) are removed. We select the 4415 undisturbed
stars of class 0.

The membership determination is cleaner for undisturbed stars, so that
above fraction of 5\% of the cluster stars misclassified as field
stars becomes smaller than 3\% if only stars of class 0 are selected.
The velocity dispersion profiles $\sigma_{x'}$ and $\sigma_{y'}$ in
Figure~\ref{fig:rad_pmclass} are systematically offset with respect to
each other, demonstrating that the velocity distribution in \oc\ is
anisotropic. We discuss this further in \S~\ref{sec:veldispprofiles}
and \S~\ref{sec:anisotropy}.

\subsubsection{Selection on proper motion error}
\label{sec:selgbpmvelerr}

After selection of the cluster members that are not disturbed by
neighbouring stars, it is likely that the sample of 4415 stars still
includes (remaining) interlopers and stars with uncertain proper
motion measurements, which can lead to systematic deviations in the
kinematics. Figure~\ref{fig:rad_pmvelerr} shows that the proper motion
dispersion profiles decrease if we sequentially select a smaller
number of stars by setting a tighter limit on the allowed error in
their proper motion measurements.

Since the proper motion errors are larger for the fainter stars
(see also Figure~11 of Paper~I), a similar effect happens if we
select on magnitude instead. The decrease in dispersion is most
prominent at larger radii as the above selection on disturbance by
a neighbouring star already removed the uncertain proper motion
measurements in the crowded centre of \oc. All dispersion profiles
in the above are corrected for the broadening due to the
individual proper motion errors (cf.
Appendix~\ref{sec:mlvelocitymoments}). The effect of this
broadening, indicated by the dotted curve, is less than the
decrease in the dispersion profiles due to the selection on proper
motion error.

Since the kinematics do not change anymore significantly for a
limit on the proper motion errors lower than 0.20 \masyr, we
select the 2295 stars with proper motion errors below this limit.
The preliminary HST proper motions of King \& Anderson
(2002\nocite{2002ocuw.conf...21K}) in the centre of \oc\
($R'\sim1$ arcmin) give rise to mean proper motion dispersion
$\sigma_{x'}=0.81\pm0.08$ \masyr\ and $\sigma_{y'}=0.77\pm0.08$
\masyr, depending on the magnitude cut-off. In their outer
calibration field ($R'\sim14$ arcmin), the average dispersion is
about $0.41\pm0.03$ \masyr. These values are consistent with the
mean proper motion dispersion of the 2295 selected stars at those
radii (light blue curves in Figure~\ref{fig:rad_pmvelerr}). We are
therefore confident that the proper motion kinematics have
converged.

The spatial distribution of the selected stars is shown in the
top panel of Figure \ref{fig:positionstars}. In the two upper
panels of Figure~\ref{fig:velhist}, the distributions of the two
proper motion components (left panels) and the corresponding
errors (right panels) of the $N_\mathrm{sel}=2295$ selected stars
are shown as shaded histograms, on top of the histograms of the
$N_\mathrm{mem}=7899$ cluster members. The selection removes the
extended tails, making the distribution narrower with an
approximately Gaussian shape.

\subsection{Line-of-sight velocities}
\label{sec:selvlos}

\begin{figure}
\includegraphics{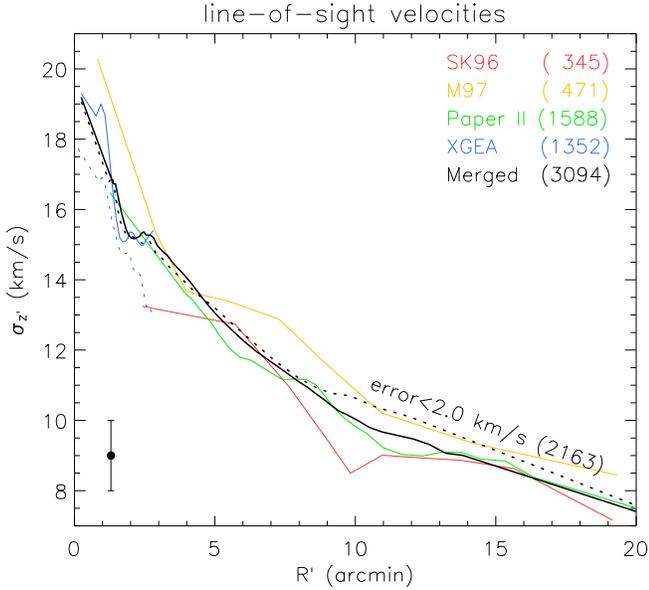}
\caption{
  Velocity dispersion profiles, calculated along concentric rings, for
  the four different line-of-sight velocity data-sets separately and
  after they have been merged. The blue dotted curve shows the
  under-estimated dispersion for the XGEA data-set if also the faint
  stars are included. From the merged data-set of 3094 stars we select
  the 2163 stars with line-of-sight velocity errors smaller than 2.0
  \kms, resulting in a dispersion profile (black dotted curve) that is not
  under-estimated due to
  uncertain line-of-sight velocity measurements.}
\label{fig:rad_vlosmag}
\end{figure}

For each of the four different line-of-sight velocity data-sets
separately, the velocity dispersion profiles of the selected (cluster)
stars (\S~\ref{sec:losvelocities} and Table~\ref{tab:datasets}) are
shown as solid coloured curves in Figure~\ref{fig:rad_vlosmag}.  The
dotted blue curve is the dispersion profile of all the 4916 stars
observed by XGEA, whereas the solid blue curve is based on the 1352
selected stars with a measured magnitude brighter than 14.5, showing
that fainter misidentified stars lead to an under-estimation of the
line-of-sight velocity dispersion. Although the dispersion profile of
the M97 data-set (yellow curve) seems to be systematically higher than
those of the other data-sets, it is based on a relatively small number
of stars, similar to the SK96 data-set, and the differences are still
within the expected uncertainties indicated by the error bar.

The solid black curve is the dispersion profile of the 3094 stars
after merging the four line-of-sight velocity data-sets
(\S~\ref{sec:velcoordvlos}). Due to uncertainties in the line-of-sight
velocity measurements of especially the fainter stars, the latter
dispersion profile is (slightly) under-estimated in the outer parts.
By sequentially lowering the limit on the line-of-sight velocity
errors, we find that below 2.0 \kms\ the velocity dispersion (dotted
black curve) converges. Hence, we select the 2163 stars with
line-of-sight velocity errors smaller than 2.0 \kms.

\begin{figure}
\includegraphics{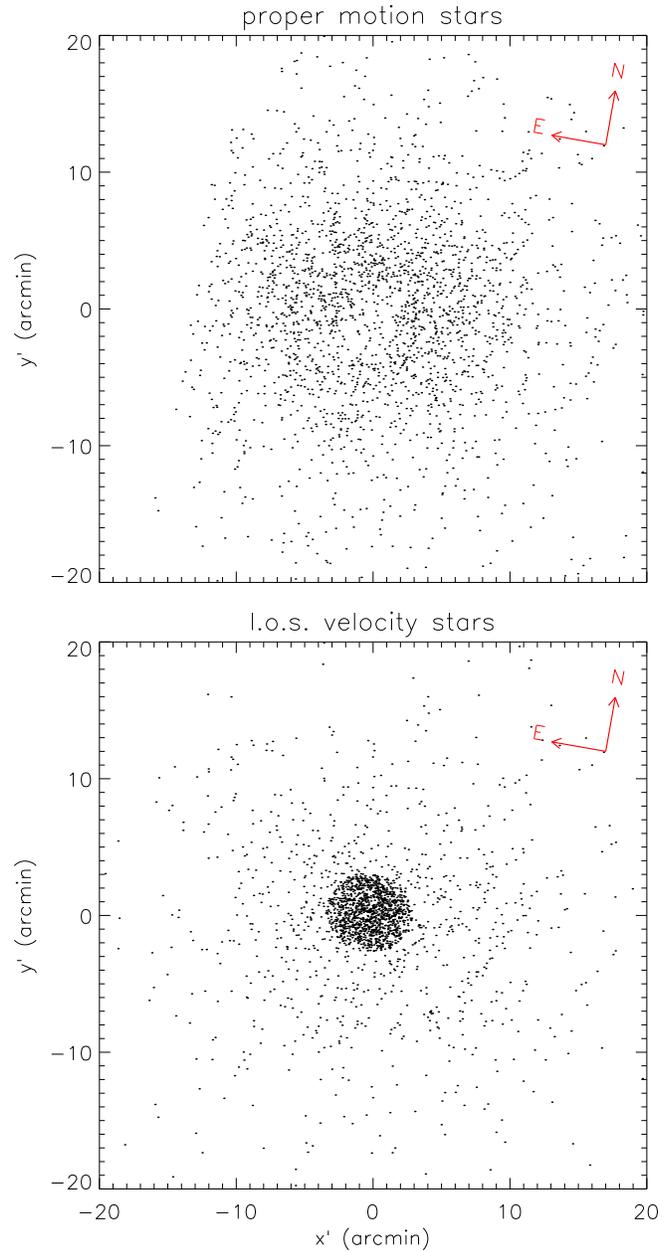}
\caption{
  The stars in \oc\ with proper motion measurements (top) and
  line-of-sight velocity measurements (bottom), that are used in our
  analysis. The stellar positions are plotted as a function of the
  projected Cartesian coordinates $x'$ and $y'$, with the $x'$-axis
  aligned with the observed major axis and the $y'$-axis aligned with
  the observed minor axis of \oc. The excess of stars with
  line-of-sight velocities inside the central 3 arcmin in the bottom
  panel is due to the XGEA data-set.}
\label{fig:positionstars}
\end{figure}

\begin{figure}
\includegraphics{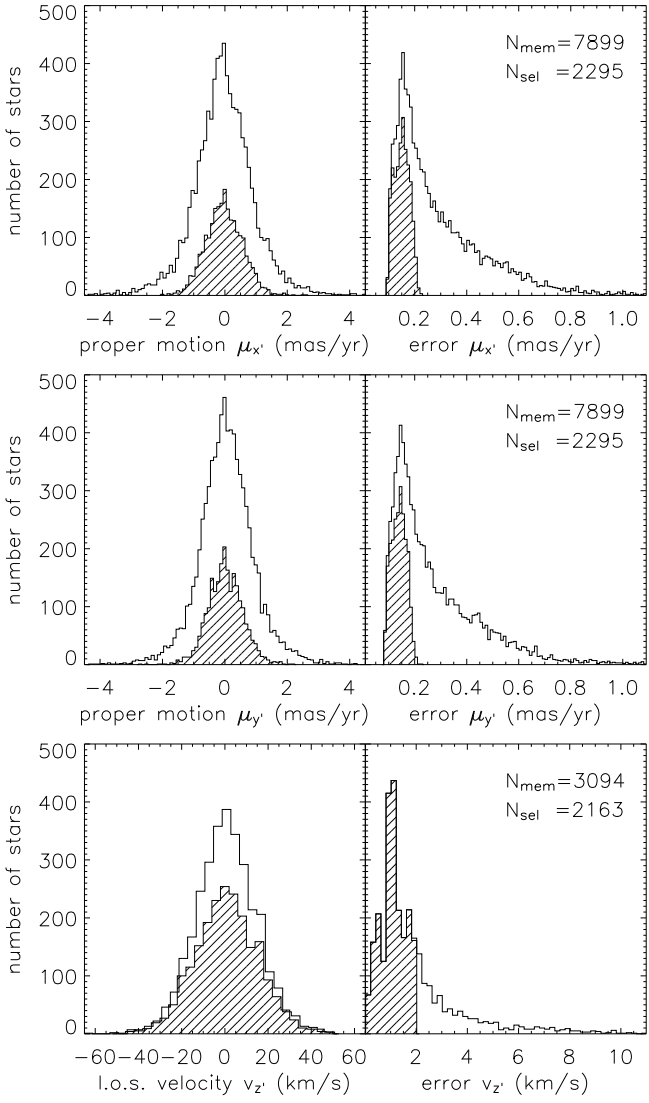}
\caption{
  Histograms of measured velocities (left panels) and corresponding
  velocity errors (right panels). The proper motion components
  $\mu_{x'}$ (upper panels) and $\mu_{y'}$ (middle panels), in the
  direction of the observed major and minor axis of \oc\ respectively,
  come from the photographic plate observations in Paper~I. The
  line-of-sight velocities (lower panels) are taken from four
  different data-sets (\S~\ref{sec:losvelocities}). The shaded
  histograms for the $N_\mathrm{sel}$ selected stars
  (\S~\ref{sec:selection}) are overlayed on the histograms of the
  $N_\mathrm{mem}$ cluster member stars.  } \label{fig:velhist}
\end{figure}

The spatial distribution of the selected stars is shown in the
bottom panel of Figure \ref{fig:positionstars}.
In the bottom panels of Figure~\ref{fig:velhist}, the
distribution of the line-of-sight velocities (left) and corresponding
errors (right) of the $N_\mathrm{sel}=2163$ selected stars are shown
as filled histograms, on top of the histograms of the
$N_\mathrm{mem}=3094$ cluster members in the merged data-set.


\section{Kinematics}
\label{sec:kinematics}

We compute the mean velocity fields for the selected stars and correct
the kinematic data for perspective rotation and for residual
solid-body rotation in the proper motions. At the same time, we place
a tight constraint on the inclination. Finally, we calculate the mean
velocity dispersion profiles from the corrected kinematic data.

\subsection{Smoothed mean velocity fields}
\label{sec:smoothmeanvel}

\begin{figure*}
  \includegraphics{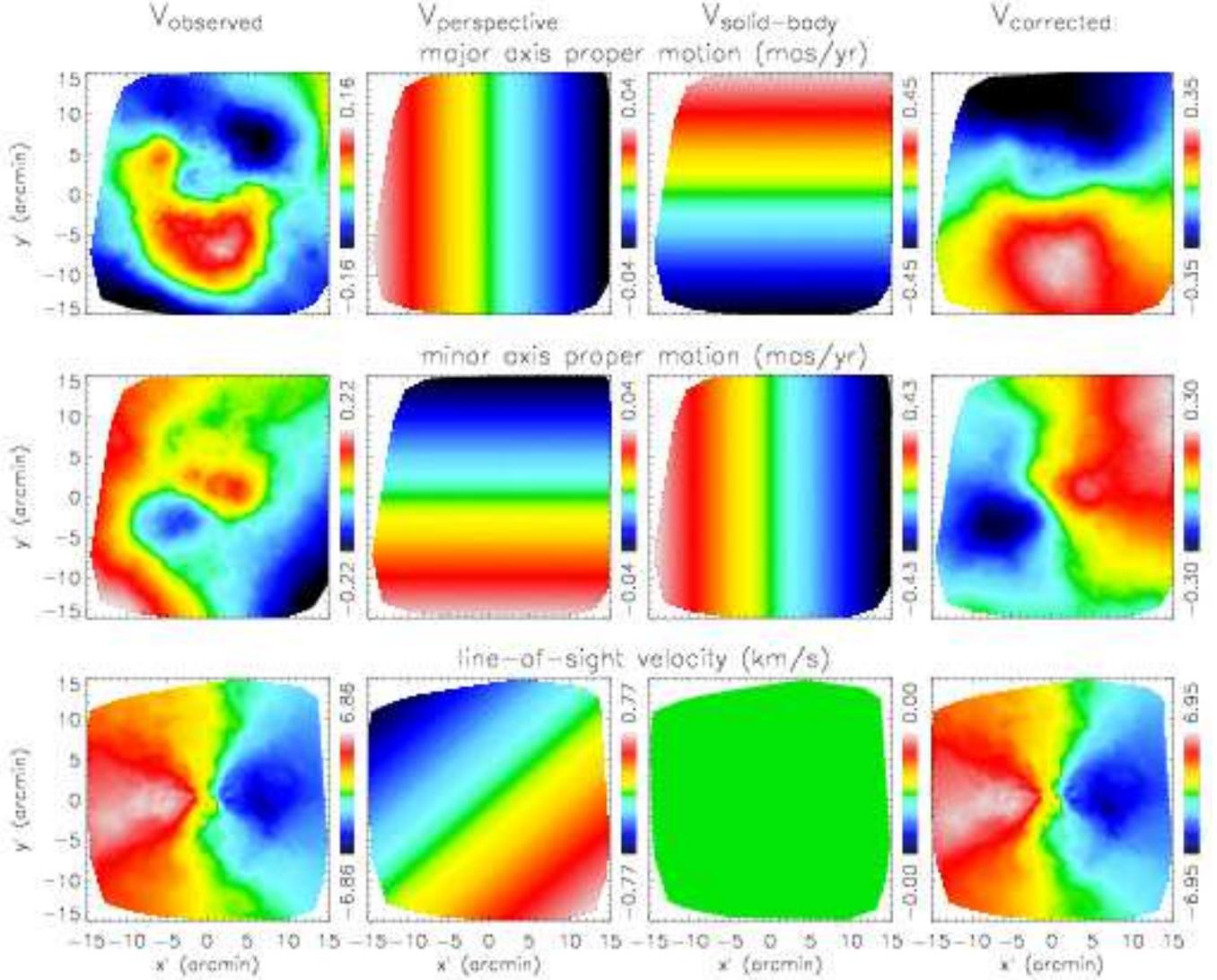}
\caption{
  The mean velocity fields of \oc\ corrected for perspective and
  solid-body rotation. The individual measurements are smoothed using
  adaptive kernel smoothening. From top to bottom: The mean
  ground-based proper motion in the major axis $x'$-direction and in the
  minor axis $y'$-direction, and the mean line-of-sight velocity. From left to
  right: Observed velocity fields of \oc, contribution from
  perspective rotation, contribution from solid-body rotation and the
  velocity fields after correcting for both.  The perspective rotation
  is caused by the space motion of \oc. The solid-body rotation in the
  proper motions is due to relative rotation of the first and second
  epoch photographic plates by an amount of $0.029$ \sbrunits\
  (\S~\ref{sec:amountsbr}).}
\label{fig:smoothV}
\end{figure*}

The left-most panels of Figure~\ref{fig:smoothV} show the smoothed
mean velocity fields for the 2295 selected stars with proper motion
measurements and the 2163 selected stars with line-of-sight velocity
measurements. This adaptive kernel smoothening is done by selecting
for each star its 200 nearest neighbours on the plane of the sky, and
then calculating the mean velocity (and higher order velocity moments)
from the individual velocity measurements
(Appendix~\ref{sec:mlvelocitymoments}). The contribution of each
neighbour is weighted with its distance to the star, using a Gaussian
distribution with zero mean and the mean distance of the 200 nearest
neighbours as the dispersion.

The upper-left panel shows the mean proper motion (in \masyr) in the
major axis $x'$-direction, i.e., the horizontal component of the
streaming motion on the plane of the sky. The colour coding is such
that red (blue) means that the stars are moving on average to the
right (left) and green shows the region where the horizontal component
of the mean proper motion vanishes. Similarly, the middle-left panel
shows the mean proper motion in the minor axis $y'$-direction, i.e.
the vertical component of the streaming motion on the plane of the
sky, with red (blue) indicating average proper motion upwards
(downwards). Finally, the lower-left panel shows the mean velocity (in
\kms) along the line-of-sight $z'$-axis, where red (blue) means that
the stars are on average receding (approaching) and green indicates
the zero-velocity curve, which is the rotation axis of \oc.

Apart from a twist in the (green) zero-velocity curve, the latter
line-of-sight velocity field is as expected for a (nearly)
axisymmetric stellar system. However, both proper motion fields show a
complex structure, with an apparently dynamically decoupled inner
part, far from axisymmetric. We now show that it is, in fact, possible
to bring these different observations into concordance.

\subsection{Perspective rotation}
\label{sec:perspectiverotation}

The non-axisymmetric features in the observed smoothed mean velocity
fields in the left-most panels of Figure~\ref{fig:smoothV}, might be
(partly) caused by perspective rotation. Because \oc\ has a large
extent on the plane of the sky (with a diameter about twice that of
the full moon), its substantial systemic (or space) motion
(eq.~\ref{eq:sysvel}) produces a non-negligible amount of apparent
rotation: the projection of the space motion onto the principal axis
$(x',y',z')$ is different at different positions on the plane of the
sky (Feast et al. 1961\nocite{1961MNRAS.122..433F}).  We expand this
perspective rotation in terms of the reciprocal of the distance $D$.
Ignoring the negligible terms of order $1/D^2$ or smaller, we find the
following additional velocities
\begin{eqnarray}
  \label{eq:velpr}
  \mu_{x'}^\mathrm{pr} & = & -6.1363\times\!10^{-5}\, x' v_{z'}^\mathrm{sys}/D
  \quad \mathrm{mas\,yr}^{-1}, \nonumber\\
  \mu_{y'}^\mathrm{pr} & = & -6.1363\times\!10^{-5}\, y' v_{z'}^\mathrm{sys}/D
  \quad \mathrm{mas\,yr}^{-1}, \\
  v_{z'}^\mathrm{pr}   & = & 1.3790\times\!10^{-3}
  \left( x' \mu_{x'}^\mathrm{sys} + y' \mu_{y'}^\mathrm{sys} \right) D
  \quad \mathrm{km\,s}^{-1}, \nonumber
\end{eqnarray}
with $x'$ and $y'$ in units of arcmin and $D$ in kpc. For the
canonical distance of 5 kpc, the systemic motion for \oc\ as given
in eq.~\eqref{eq:sysvel} and the data typically extending to 20
arcmin from the cluster centre, we find that the maximum amplitude
of the perspective rotation for the proper motions is about 0.06
\masyr\ and for the line-of-sight velocity about 0.8 \kms. These
values are a significant fraction of the observed mean velocities
(left panels of Figure~\ref{fig:smoothV}) and of the same order as
the uncertainties in the extracted kinematics (see
Appendix~\ref{sec:mlvelocitymoments}). Therefore, the perspective
rotation as shown in the second column panels of
Figure~\ref{fig:smoothV}, cannot be ignored and we correct the
observed stellar velocities by subtracting it. Since we use the
more recent and improved values for the systemic proper motion
from Paper~I, our correction for perspective rotation is different
from that of Merritt et al.\ (1997\nocite{1997AJ....114.1074M}).
The amplitude of the correction is, however, too small to explain
all of the complex structure in the proper motion fields and we
have to look for an additional cause of non-axisymmetry.

\subsection{Residual solid-body rotation}
\label{sec:residualsbr}

Van Leeuwen \& Le Poole (2002\nocite{2002ocuw.conf...41V}) already
showed that a possible residual solid-body rotation component in
the ground-based proper motions of Paper~I can have an important
effect on the kinematics. The astrometric reduction process to
measure proper motions removes the ability to observe an overall
rotation on the plane of the sky (e.g. Vasilevskis et al.\
1979\nocite{1979A&AS...37..333V}).  This solid-body rotation
results in a transverse proper motion $v_t=\Omega\,R'$, with
$\Omega$ the amount of solid-body rotation (in units of \sbrunits)
and $R'$ the distance from the cluster centre in the plane of the
sky (in units of arcmin). Decomposition of $v_t$ along the
observed major and minor axis yields
\begin{eqnarray}
  \label{eq:velsbr}
  \mu_{x'}^\mathrm{sbr} & = & +\Omega\,y'
  \quad \mathrm{mas\,yr}^{-1}, \nonumber
  \\*[-7.5pt]\\*[-7.5pt]
  \mu_{y'}^\mathrm{sbr} & = & -\Omega\,x'
  \quad \mathrm{mas\,yr}^{-1}. \nonumber
\end{eqnarray}
Any other reference point than the cluster centre results in a
constant offset in the proper motions, and is removed by setting
the systemic proper motions to zero. Also an overall expansion (or
contraction) cannot be determined from the measured proper
motions, and results in a radial proper motion in the plane of the
sky. Although both the amount of overall rotation and expansion
are in principle free parameters, they can be constrained from the
link between the measured (differential) proper motions to an
absolute proper motion system, such as defined by the Hipparcos
and Tycho-2 catalogues (Perryman et al.\
1997\nocite{1997hity.book.....P}; H\o g et al.
2000\nocite{2000yCat.1259....0H}). In Paper~I, using the 56 stars
in common with these two catalogues, the allowed amount of
residual solid-body rotation was determined to be no more than
$\Omega=0.02\pm0.02$ \sbrunits\ and no significant expansion was
found.

As the amplitude of the allowed residual solid-body rotation is of
the order of the uncertainties in the mean proper motions already
close to the centre, and can increase beyond the maximum amplitude
of the mean proper motions in the outer parts, correcting for it
has a very important effect on the proper motions. We use a
general relation for axisymmetric objects to constrain $\Omega$,
and at the same find a constraint on the inclination.

\begin{figure*}[t]
\includegraphics{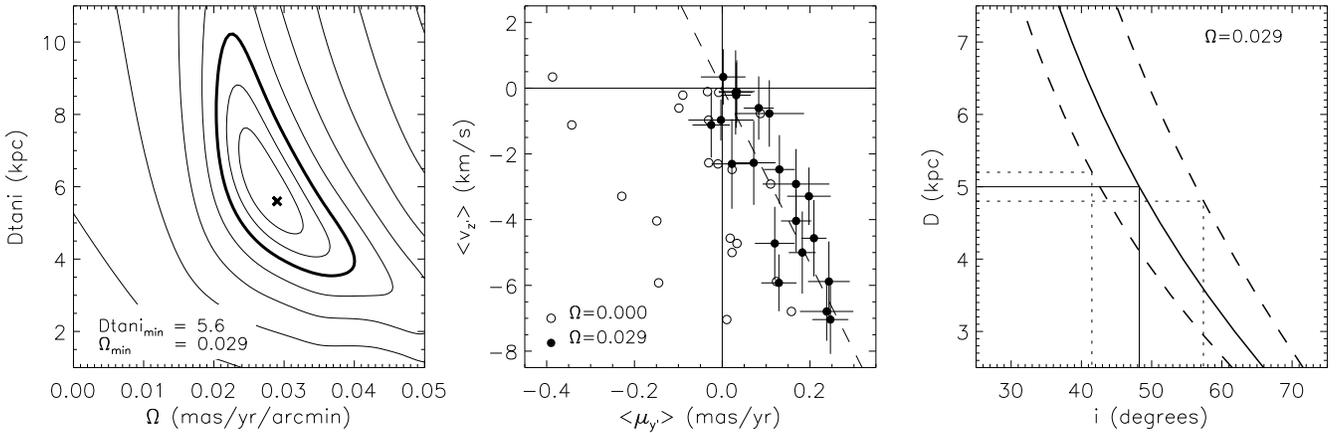}
\caption{
  Constraints on the amount of residual solid-body body rotation
  $\Omega$ and via $D\tan i$, on the distance $D$ (in kpc) and
  inclination $i$, using the general relation (\ref{eq:Dtani}) for
  axisymmetric objects. The left panel shows the contour map of the
  goodness-of-fit parameter $\Delta\chi^2$. The inner three contours
  are drawn at the 68.3\%, 95.4\% and 99.7\% (thick contour) levels of a
  $\Delta\chi^2$-distribution with two degrees of freedom. Subsequent
  contours correspond to a factor of two increase in $\Delta\chi^2$.
  The overall minimum is indicated by the cross. The middle panel
  shows the mean line-of-sight velocity $\langle v_{z'} \rangle$ and
  mean short-axis proper motion $\langle \mu_{y'} \rangle$ within the
  same polar apertures, before (open circles) and after (filled
  circles) correction for residual solid-body rotation with the
  best-fit value of $\Omega=0.029\pm0.004$ \sbrunits. The best-fit
  value for $D\tan i=5.6$ (+1.9/-1.0) kpc gives rise to the relation
  in the right panel (sold line), bracketed (at the 68.3\%-level) by
  the dashed lines.  Given the canonical distance of $D=5.0\pm0.2$
  kpc, the dotted lines indicate the constraint on inclination of
  $i=48$ (+9/-7) degrees.}
  \label{fig:sbr}
\end{figure*}

\subsection{The amount of residual solid-body rotation directly from
  the mean velocities}
\label{sec:amountsbr}

For any axisymmetric system, there is, at each position $(x',y')$ on
the plane of the sky, a simple relation between the mean proper motion
in the $y'$-direction $\langle \mu_{y'} \rangle$ and the mean
line-of-sight velocity $\langle v_{z'} \rangle$ (see e.g. Appendix~A
of Evans \& de Zeeuw 1994\nocite{1994MNRAS.271..202E}, hereafter
EZ94).
Using relation~(\ref{eq:velintpol2obscart}), with for an axisymmetric
system $\langle v_R \rangle = \langle v_z \rangle = 0$, we see that,
while the mean velocity component in the $x'$-direction includes the
spatial term $\cos\phi$, those in the $y'$-direction and line-of-sight
$z'$-direction both contain $\sin\phi$. The latter implies that, by
integrating along the line-of-sight to obtain the observed mean
velocities, the expressions for $\langle v_{y'} \rangle$ and $\langle
v_{z'} \rangle$ only differ by the $\cos i$ and $\sin i$ terms. Going
from $\langle v_{y'} \rangle$ to $\langle \mu_{y'} \rangle$ via
equation~(\ref{eq:masyr2kms}), we thus find the following general
relation for axisymmetric objects
\begin{equation}
  \label{eq:Dtani}
  \langle v_{z'} \rangle(x',y') = 4.74\; D \tan i \;
  \langle \mu_{y'} \rangle(x',y'),
\end{equation}
with distance $D$ (in kpc) and inclination $i$.

This relation implies that, at each position on the plane of the sky,
the only difference between the mean short-axis proper motion field
and the mean line-of-sight velocity field should be a constant scaling
factor equal to $4.74~D\tan i$. Comparing the left-most
middle and bottom panel in Figure~\ref{fig:smoothV}
($V_\mathrm{observed}$), this is far from what we see, except perhaps
for the inner part. We ascribe this discrepancy to the residual
solid-body rotation, which causes a perturbation of $\langle \mu_{y'}
\rangle$ increasing with $x'$ as given in eq.~\eqref{eq:velsbr}.  In
this way, we can objectively quantify the amount of solid body
rotation $\Omega$ needed to satisfy the above relation
(\ref{eq:Dtani}), and at the same time find the best-fit value for
$D\tan i$.

To compute uncorrelated values (and corresponding errors) for the
mean short-axis proper motion $\langle \mu_{y'} \rangle$ and mean
line-of-sight velocity $\langle v_{z'} \rangle$ at the same positions
on the plane of the sky, we bin the stars with proper motion and
line-of-sight velocity measurements in the same polar grid of
apertures (see also Appendix~\ref{sec:mlvelocitymoments}).  We plot
the resulting values for $\langle v_{z'} \rangle$ against $\langle
\mu_{y'} \rangle$ and fit a line (through the origin) by minimising
the $\chi^2$, taking into account the errors in both directions
(\S~15.3 of Press et al.\ 1992\nocite{Press92..numrecipies}).

By varying the amount of solid-body rotation $\Omega$ and the slope of
the line, which proportional to $D\tan i$ (eq.~\ref{eq:Dtani}), we
obtain the $\Delta\chi^2 = \chi^2 - \chi_\mathrm{min}^2$ contours in
the left panel of Figure~\ref{fig:sbr}. The inner three
  contours are drawn at the levels containing 68.3\%, 95.4\% and
  99.7\% (thick contour) of a $\Delta\chi^2$-distribution with two
  degrees of freedom.\footnote{For a Gaussian distribution with
    dispersion $\sigma$, these percentages correspond to the
    $1\sigma$, $2\sigma$ and $3\sigma$ confidence intervals
    respectively. For the (asymmetric) $\chi^2$-distribution there is
    in general no simple relation between dispersion and confidence
    intervals. Nevertheless, the 68.3\%, 95.4\% and 99.7\% levels of
    the $\chi^2$-distribution are often referred to as the $1\sigma$,
    $2\sigma$ and $3\sigma$ levels.} Subsequent contours correspond to
    a factor of two increase in $\Delta\chi^2$. The overall minimum
  $\chi_\mathrm{min}^2$, indicated by a cross, implies (at the
  68.3\%-level) a best-fit value of $\Omega=0.029\pm0.004$ \sbrunits.
  This is fully consistent with the upper limit of
  $\Omega=0.02\pm0.02$ \sbrunits\ from Paper~I.

The middle panel of Figure~\ref{fig:sbr} shows that without any
correction for residual solid-body rotation, the values for
$\langle v_{z'} \rangle$ and $\langle \mu_{y'} \rangle$ are
scattered (open circles), while they are nicely correlated after
correction with $\Omega=0.029$ \sbrunits\ (filled circles). The
resulting solid-body rotation, shown in the third column of
Figure~\ref{fig:smoothV}, removes the cylindrical rotation that is
visible in the outer parts of the observed proper motion fields
(first column). After subtracting this residual solid-body
rotation, together with the perspective rotation (second column),
the complex structures disappear, resulting in (nearly)
axisymmetric mean velocity fields in the last column. Although the
remaining non-axisymmetric features, such as the twist of the
(green) zero-velocity curve, might indicate deviations from true
axisymmetry, they can also be (partly) artifacts of the
smoothening, which, especially in the less dense outer parts, is
sensitive to the distribution of stars on the plane of the sky.

This shows that the application of eq.~\eqref{eq:Dtani} to the
combination of proper motion and line-of-sight measurements provides a
powerful new tool to determine the amount of solid body rotation. At
the same time, it also allows us to place a constraint on the
inclination.

\subsection{Constraint on the inclination}
\label{sec:constraintincl}

From the left panel of Figure~\ref{fig:sbr} we obtain (at the
68.3\%-level) a best-fit value for $D\tan i$ of 5.6 (+1.9/-1.0)
kpc. The right panel shows the resulting relation (solid line)
between the distance $D$ and the inclination $i$, where the dashed
lines bracket the 68.3\%-level uncertainty. If we assume the
canonical value $D=5.0\pm0.2$ kpc, then the inclination is
constrained to $i=48$ (+9/-7) degrees.

Although we apply the same polar grid to the proper motions and
line-of-sight velocities, the apertures contain different (numbers
of) stars. To test that this does not significantly influence the
computed average kinematics and hence the above results, we
repeated the analysis but now only include the 718 stars for which
both the proper motions and line-of-sight velocity are measured.
The results are equivalent, but less tightly constrained due to
the smaller number of apertures.

Van Leeuwen \& Le Poole (2002\nocite{2002ocuw.conf...41V}) compared,
for different values for the amount of residual solid-body rotation
$\Omega$, the shape of the radial profile of the mean transverse
component of proper motions from Paper~I, with that of the mean
line-of-sight velocities calculated by Merritt et al.\
(1997\nocite{1997AJ....114.1074M}) from the line-of-sight velocity
data-set of M97. They found that $\Omega\sim0.032$ \sbrunits\ provides
a plausible agreement. Next, assuming a distribution for the density
and the rotational velocities in the cluster, they computed projected
transverse proper motion and line-of-sight velocity profiles, and by
comparing them to the observed profiles, they derived a range for the
inclination $i$ from 40 to 60 degrees. Their results are consistent
with our best-fit values $\Omega=0.029\pm0.004$ \sbrunits\ and $i=48$
(+9/-7) degrees. Our method is based on a general relation for
axisymmetric objects, without any further assumptions about the
underlying density and velocity distribution. Moreover, instead of
comparing shapes of mean velocity profiles, we actually fit the
(two-dimensional) mean velocity fields.

In the above analysis, we assume that all of the solid-body rotation
in the proper motion is the result of a (non-physical) residual from
the photographic plate reduction in Paper~I. This raises the question
what happens if a (physical) solid-body rotation component is present
in \oc. Such a solid-body rotation component is expected to be aligned
with the intrinsic rotation axis, inclined at about 48\dgr, and
therefore also present in the line-of-sight velocities. Except for the
perspective rotation correction, we leave the mean line-of-sight
velocities in the above analysis unchanged, so that any such
solid-body rotation component should also remain in the proper motion.

Still, since we are fitting the residual solid-body rotation $\Omega$
and the slope $D\tan i$ simultaneously, we show next that these
parameters can become (partly) degenerate. Combining
eq.~\eqref{eq:velsbr} with \eqref{eq:degenerateOmegaDtani}, we obtain
the best-fit values for $D\tan i$ and $\Omega$ by minimising
\begin{equation}
  \label{eq:Dtani_chi2}
  \chi^2 = \sum_j^n \frac{\left[ \langle v_{z'}^\mathrm{obs} \rangle_j
    - 4.74\,D \tan i\,\left( \langle \mu_{y'}^\mathrm{obs} \rangle_j +
      \Omega\,x'_j \right) \right]^2}{
  \left[\Delta\langle v_{z'}^\mathrm{obs} \rangle_j\right]^2 +
  \left[4.74\,D \tan i\,
    \Delta\langle \mu_{y'}^\mathrm{obs} \rangle_j\right]^2},
\end{equation}
where $\langle v_{z'}^\mathrm{obs} \rangle_j$ and $\langle
\mu_{y'}^\mathrm{obs} \rangle_j$ are respectively the observed mean
line-of-sight velocity and the observed mean proper motion in the
$y'$-direction, measured in aperture $j$ of a total of $n$ apertures,
with their centres at $x'_j$. $\Delta\langle v_{z'}^\mathrm{obs}
\rangle_j$ and $\Delta\langle \mu_{y'}^\mathrm{obs} \rangle_j$ are the
corresponding measurement errors. Suppose now the extreme case that
all of the observed mean motion is due to solid-body rotation: an
amount of $\Omega_0$ residual solid-body rotation in the plane of the
sky, and an amount of $\omega_0$ intrinsic solid-body rotation, around
the intrinsic $z$-axis in \oc, which we assume to be inclined at $i_0$
degrees.  At a distance $D_0$, the combination yields per aperture
$\langle v_{z'}^\mathrm{obs} \rangle_j = 4.74 D_0 \omega_0 \sin i_0
x'_j$ and $\langle \mu_{y'}^\mathrm{obs} \rangle_j = (\omega_0\cos i_0
- \Omega_0) x'_j$. Substitution of these quantities in the above
eq.~\eqref{eq:Dtani_chi2}, and ignoring the (small) variations in the
measurements errors, yields that $\chi^2=0$ if
\begin{equation}
  \label{eq:degenerateOmegaDtani}
  D\tan i = D_0\tan i_0 \left[ 1 +
    \frac{\Omega-\Omega_0}{\omega_0\cos i_0} \right]^{-1}.
\end{equation}
This implies a degeneracy between $D\tan i$ and $\Omega$, which left
panel of Figure~\ref{fig:sbr}, would result in the same minimum all
along a curve. However, in the case the motion in \oc\ consists of
more than only solid-body rotation, this degeneracy breaks down and we
expect to find a unique minimum. The latter seems to be the case here,
and we conclude that the degeneracy and hence the intrinsic solid-body
rotation are not dominant, if present at all.

\begin{figure}[t]
\includegraphics{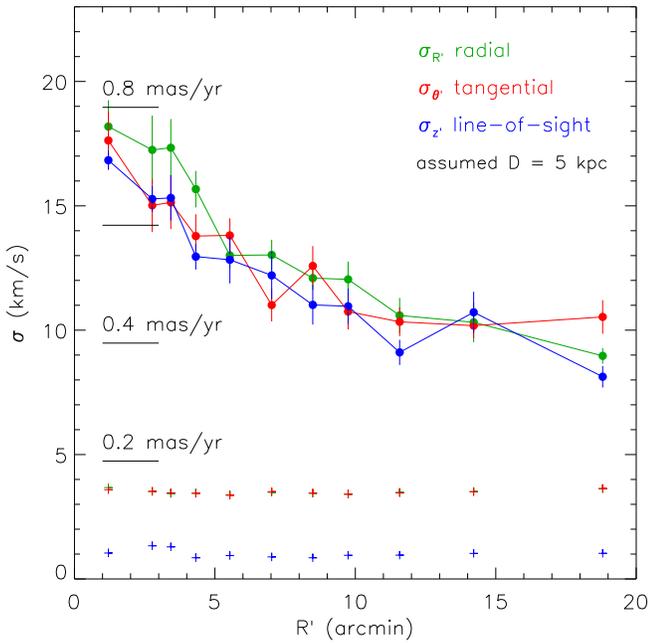}
\caption{
  Mean velocity dispersion profiles calculated along concentric rings.
  Assuming the canonical distance of 5 kpc, the profiles of the radial
  $\sigma_{R'}$ (green) and tangential $\sigma_{\theta'}$ (red)
  components of the proper motion dispersion are converted into the
  same units of \kms\ as the profile of the line-of-sight velocity
  dispersion $\sigma_{z'}$ (blue). The black horizontal lines indicate
  the corresponding scale in \masyr. The mean velocity error per ring
  is indicated below the profiles by the crosses. The green and red
  crosses mostly overlap, as the errors of the radial and tangential
  components are nearly similar.
 }
  \label{fig:dispprofile}
\end{figure}

\subsection{Mean velocity dispersion profiles}
\label{sec:veldispprofiles}

In Figure~\ref{fig:dispprofile}, we show the mean velocity dispersion
profiles of the radial $\sigma_{R'}$ (green) and tangential
$\sigma_{\theta'}$ (red) components of the proper motions, together
with the line-of-sight velocity dispersion $\sigma_{z'}$ (blue).  The
dispersions are calculated along concentric rings from the selected
sample of 2295 stars with proper motions corrected for perspective and
residual solid-body rotation and 2163 stars with line-of-sight
velocities corrected for perspective rotation.
We obtain similar mean velocity dispersion profiles if we only use the
718 stars for which both proper motions and line-of-sight velocity are
measured.  We assume the canonical distance of 5 kpc to convert the
proper motion dispersion into units of \kms, while the black
horizontal lines indicate the corresponding scale in units of
\masyr. Below the profiles, the crosses represent the corresponding
mean velocity error per ring, showing that the accuracy of the
line-of-sight velocity measurements (blue crosses) is about four times
better than the proper motion measurements (green and red
crosses, which mostly overlap since the errors for the two components
are similar).

In \S~\ref{sec:selgbpm}, we already noticed that since the (smoothed)
profile of the major-axis proper motion dispersion $\sigma_{x'}$ lies
on average above that of the minor-axis proper motion dispersion
$\sigma_{y'}$
(Figure~\ref{fig:rad_pmclass}~and~~\ref{fig:rad_pmvelerr}), the
velocity distribution of \oc\ cannot be fully isotropic. By comparing
in Figure~\ref{fig:dispprofile} the radial (green) with the tangential
(red) component of the proper motion dispersion, \oc\ seems to be
radial anisotropic towards the centre, and there is an indication of
tangential anisotropy in the outer parts. Moreover, if \oc\ would be
isotropic, the line-of-sight velocity dispersion profile (blue) would
have to fall on top of the proper motion dispersion profiles if scaled
with the correct distance. A scaling with a distance lower than the
canonical 5 kpc is needed for the line-of-sight dispersion profile to
be on average the same as those of both proper motion components.

Hence, it is not surprising that we find a distance as low as
$D=4.6\pm0.2$ kpc from the ratio of the average line-of-sight
velocity dispersion and the average proper motion dispersion
(Appendix~\ref{sec:simpledistest}). This often used simple
distance estimate is only valid for spherical symmetric objects.
Whereas the averaged observed flattening for \oc\ is already as
low as $q'=0.879\pm0.007$ (Geyer et al.\
1983\nocite{1983A&A...125..359G}), an inclination of around
48\dgr\ (\S~\ref{sec:constraintincl}), implies an intrinsic
axisymmetric flattening $q<0.8$.

A model with a constant oblate velocity ellipsoid as in
Appendix~\ref{sec:simpledistest}, allows for offsets between the mean
velocity dispersion profiles. However, the model is not suitable to
explain the observed variation in anisotropy with radius. Therefore,
we use in what follows Schwarzschild's method to build general
axisymmetric anisotropic models.


\section{Schwarzschild's method}
\label{sec:schwarzschild}

We construct axisymmetric dynamical models using Schwarzschild's
(1979\nocite{1979ApJ...232..236S}) orbit superposition method.
This approach is flexible and efficient and does not require any
assumptions about the degree of velocity anisotropy. The only
crucial approximations are that the object is collisionless and
stationary. While these assumptions are always valid for a galaxy,
they may not apply to a globular cluster.
The central relaxation time of \oc\ is a few times $10^9$ years and
the half-mass relaxation time a few times $10^{10}$ years (see also
Figure~\ref{fig:timescales} below). The collisionless approximation
should therefore be fairly accurate.

The implementation that we use here is an extension of the method
presented in Verolme et al.
  (2002\nocite{2002MNRAS.335..517V}). In the next subsections, we
outline the method and describe the extensions.

\subsection{Mass model}
\label{sec:massmodel}

Schwarzschild's method requires a mass parameterisation, which we
obtain by using the Multi-Gaussian Expansion method (MGE; Monnet,
Bacon \& Emsellem 1992\nocite{1992A&A...253..366M}; Emsellem et al.\
1994\nocite{1994A&A...285..723E}a,b\nocite{1994A&A...285..739E};
Cappellari 2002\nocite{2002MNRAS.333..400C}). The MGE-method tries to
find the collection of two-dimensional Gaussians that best reproduces
a given surface brightness profile or a (set) of images. Typically, of
the order of ten Gaussians are needed, each with three free
parameters: the central surface brightness $\Sigma_{0,j}$, the
dispersion along the observed major axis $\sigma'_j$ and the observed
flattening $q'_j$. Even though Gaussians do not form a complete set of
functions, in general the surface brightness is well fitted (see also
Fig.~\ref{fig:mge}). Moreover, the MGE-parameterisation has the
advantage that the deprojection can be performed analytically once the
viewing angles (in this case the inclination) are given. Also many
intrinsic quantities such as the potential and accelerations can be
calculated by means of simple one-dimensional integrals.

\subsection{Gravitational potential}
\label{sec:potcalc}

We deproject the set of best-fitting Gaussians by assuming that
the cluster is axisymmetric and by choosing a value of the
inclination $i$. The choice of a distance $D$ to the object then
allows us to convert angular distances to physical units, and
luminosities are transformed to masses by multiplying with the
mass-to-light ratio $M/L$.

The latter quantity is often assumed to be independent of radius. In
the inner regions of most galaxies, where two-body relaxation does not
play an important role, this often is a valid assumption. Generally,
globular clusters have much shorter relaxation times and may therefore
show significant $M/L$-variations. This has been confirmed for post
core-collapse clusters such as M15 (Dull et al.\
1997\nocite{1997ApJ...481..267D}). However, \oc\ has a relatively long
relaxation time of $>10^9$ years, implying that little mass
segregation has occurred and the mass-to-light ratio should be
nearly constant with radius. In our analysis we assume a constant
$M/L$, but we also investigate possible $M/L$-variations.

The stellar potential is then calculated by applying Poisson's
equation to the intrinsic density. The contribution of a dark object
such as a collection of stellar remnants or a central black hole may
be added at this stage. On the basis of the relation between the black
hole mass and the central dispersion (e.g.  Tremaine et al.
2002\nocite{2002ApJ...574..740T}), globular clusters might be expected
to harbour central black holes with intermediate mass of the order
$10^3$--$10^4$ \Msun\ (e.g. van der Marel
2004\nocite{2004cbhg.symp...37V}). With a central dispersion of nearly
20 \kms, the expected black hole mass for \oc\ would be $\sim10^4$
\Msun. The spatial resolution that is required to observe the
kinematical signature of such a black hole is of the order of its
radius of influence, which is around 5 arcsec (at the canonical
distance of 5 kpc). This is approximately an order of magnitude
smaller than the resolution of the ground-based observations we use in
our analysis, so that our measurements are insensitive to such a small
mass. Hence, we do not include a black hole in our dynamical models of
\oc.

\subsection{Initial conditions and orbit integration}
\label{sec:inicondandorbitintg}

After deriving the potential and accelerations, the next step is to
find the initial conditions for a representative orbit library.  This
orbit library must include all types of orbits that the potential can
support, to avoid any bias. This is done by choosing orbits through
their integrals of motion, which, in this case, are the orbital energy
$E$, the vertical component of the angular momentum $L_z$ and the
effective third integral $I_3$. 

For each energy $E$, there is one circular orbit in the equatorial
plane, with radius $R_c$ that follows from $E=\Phi+\frac12
R_c\partial\Phi/\partial R_c$ for $z=0$, and with $\Phi(R,z)$ the
underlying (axisymmetric) potential. We sample the energy by choosing
the corresponding circular radius $R_c$ from a logarithmic grid.  The
minimum radius of this grid is determined by the resolution of the
data, while the maximum radius is set by the constraint that $\geq
99.9$ per cent of the model mass should be included in the grid. $L_z$
is sampled from a linear grid in $\eta = L_z/L_\mathrm{max}$, with
$L_\mathrm{max}$ the angular momentum of the circular orbit. $I_3$ is
parameterised by the starting angle of the orbit and is sampled
linearly between zero and the initial position of the so-called thin
tube orbit (see Figure~3 of Cretton et al.\ 
1999\nocite{1999ApJS..124..383C}).

The orbits in the library are integrated numerically for 200
times the period of a circular orbit with energy $E$. In order to
allow for comparison with the data, the intrinsic density, surface
brightness and the three components of the projected velocity are
stored on grids. During grid storage, we exploit the symmetries of the
projected velocities by folding around the projected axes and store
the observables only in the positive quadrant ($x'\geq 0, y' \geq 0$).
Since the sizes of the polar apertures on which the average kinematic
data is computed (Figure~\ref{fig:apgrids}) are much larger than the
typical seeing FWHM (1--2 arcsec), we do not have to store the
orbital properties on an intermediate grid and after orbit integration
convolve with the point spread function (PSF). Instead, the orbital
observables are stored directly onto the polar apertures.

\subsection{Fitting to the observations}
\label{sec:fit2obs}

After orbit integration, the orbital predictions are matched to
the observational data. We determine the superposition of orbits
whose properties best reproduce the observations. If $O_{ij}$ is
the contribution of the $j$th orbit to the $i$th constraint point,
this problem reduces to solving for the orbital weights $\gamma_j$
in
\begin{equation}
\label{eq:schwarzschildproblem}
\sum_j^{N_O}\gamma_j\,O_{ij} = C_i, \qquad i=1,\ldots,N_C,
\end{equation}
where $N_O$ is the number of orbits in the library, $N_C$ is the
number of constraints that has to be reproduced and $C_i$ is the
$i$th constraint. Since $\gamma_j$ determines the mass of each
individual orbit in this superposition, it is subject to the
additional condition $\gamma_j\geq 0$.

Equation~(\ref{eq:schwarzschildproblem}) can be solved by using
linear or quadratic programming (e.g. Schwarzschild
1979\nocite{1979ApJ...232..236S},
1982\nocite{1982ApJ...263..599S},
1993\nocite{1993ApJ...409..563S}; Vandervoort
1984\nocite{1984ApJ...287..475V}; Dejonghe
1989\nocite{1989ApJ...343..113D}), maximum entropy methods (e.g.
Richstone \& Tremaine 1988\nocite{1988ApJ...327...82R}; Gebhardt
et al.\ 2003\nocite{2003ApJ...583...92G}) or with a linear
least-squares solver [e.g. Non-Negative Least-Squares (NNLS),
Lawson \& Hanson 1974\nocite{1974slsp.book.....L}], which was used
in many of the spherical and axisymmetric implementations (e.g.
Rix et al. 1997\nocite{1997ApJ...488..702R}; van der Marel et al.\
1998\nocite{1998ApJ...493..613V}; Cretton et al.\
1999\nocite{1999ApJS..124..383C}; Cappellari et al.\
2002\nocite{2002ApJ...578..787C}; Verolme et al.\
2002\nocite{2002MNRAS.335..517V}; Krajnovi{\'c} et al.\
2005\nocite{2005MNRAS...krajnovic}), and is also used here. NNLS
has the advantage that it is guaranteed to find the global
best-fitting model and that it converges relatively quickly.

Due to measurement errors, incorrect choices of the model
parameters and numerical errors, the agreement between model and
data is never perfect. We therefore express the quality of the
solution in terms of $\chi^2$, which is defined as
\begin{equation}
\label{eq:chi2} \chi ^2 = \sum_{i=1}^{N_c} \left( \frac{C^\star_i
- C_i}{\Delta C_i} \right)^2.
\end{equation}
Here, $C_i^\star$ is the model prediction of the constraint $C_i$ and
$\Delta C_i$ is the associated error. The value of $\chi^2$ for a
single model is of limited value, since the true number of degrees of
freedom is generally not known. On the other hand, the difference in
$\chi^2$ between a model and the overall minimum value, $\Delta \chi^2
= \chi^2 - \chi^2_{\rm min}$, is statistically meaningful (see Press
et al.\ 1992\nocite{Press92..numrecipies}, \S~15.6), and we can assign
the usual confidence levels to the $\Delta \chi^2$ distribution. The
probability that a given set of model parameters occurs can be
measured by calculating $\Delta\chi^2$ for models with different
values of these model parameters. We determine the overall
best-fitting model by searching through parameter space.

The orbit distribution for the best-fitting model may vary rapidly for
adjacent orbits, which corresponds to a distribution function that is
probably not realistic. This can be prevented by imposing additional
regularisation constraints on the orbital weight distribution. This is
usually done by minimising the $n$th-order partial derivatives of the
orbital weights $\gamma_j(E,L_z,I_3)$, with respect to the integrals
of motion $E$, $L_z$ and $I_3$. The degree of smoothing is determined
by the order $n$ and by the maximum value $\Delta$ that the
derivatives are allowed to have, usually referred to as the
regularisation error. Since the distribution function is well
recovered by minimising the second order derivatives ($n=2$) and
smoothening with $\Delta=4$ (e.g. Verolme \& de Zeeuw
2002\nocite{2002MNRAS.331..959V}; Krajnovi{\'c} et al.\
2005\nocite{2005MNRAS...krajnovic}), we adopt these values.


\section{Tests}
\label{sec:tests}

Before applying our method to observational data, we test it on a
theoretical model, the axisymmetric power-law model (EZ94).

\begin{figure*}
\includegraphics[width=\textwidth]{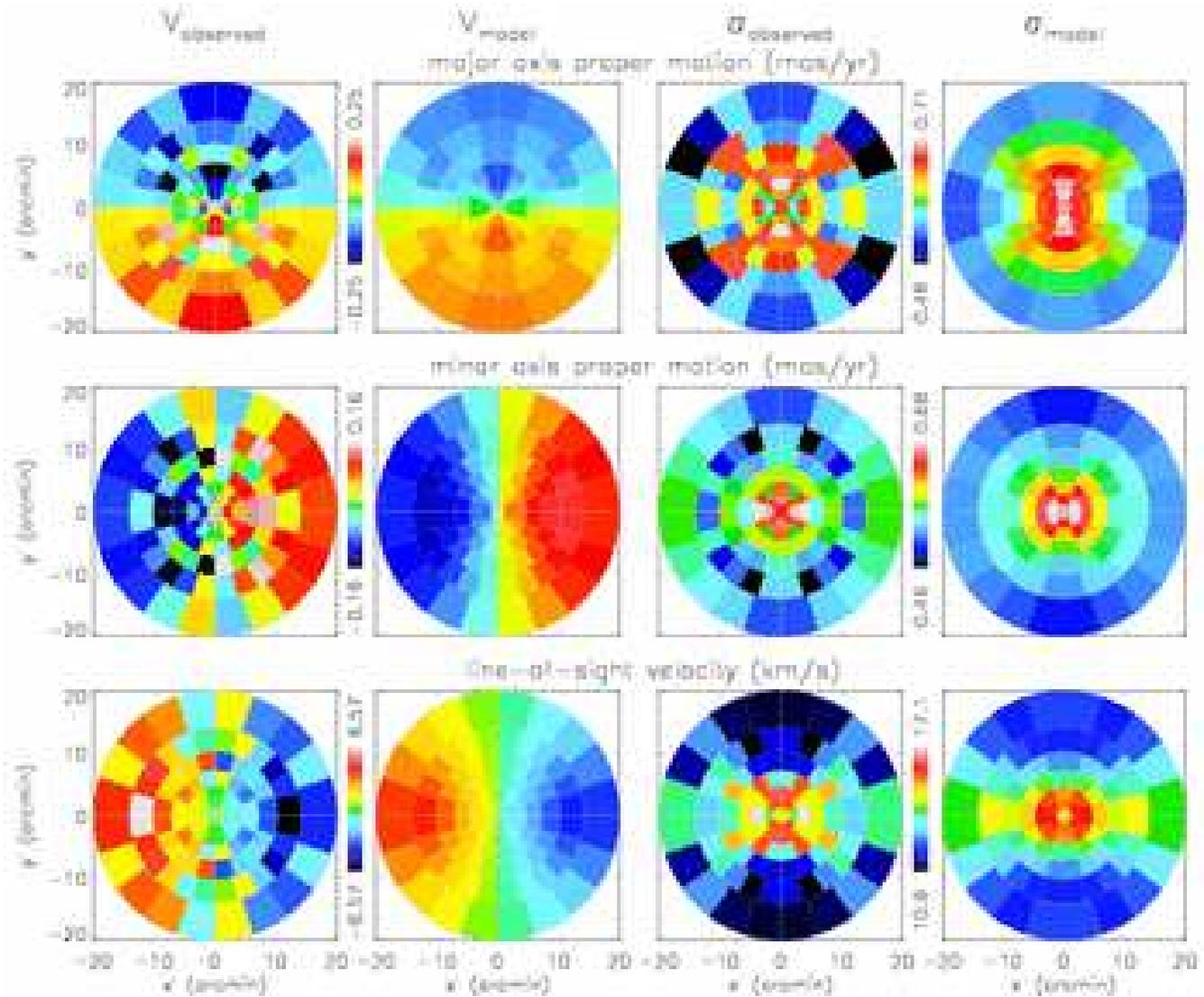}
\caption{
  Mean velocity and velocity dispersion calculated from a power-law
  model (first and third column) and from the best-fit dynamical
  Schwarzschild model with $D=4.9$ kpc, $i=45$\dgr\ and $M/L=2.5$
  \MLsun\ (second and fourth column). The parameters of the
    power-law model are chosen such that its observables resemble
    those of \oc, including the level of noise, which is obtained by
    randomising the observables according to the uncertainties in the
    measurements of \oc\ (see \S~\ref{sec:PLobservables} and
    Appendix~\ref{sec:apgrid} for details). The average proper motion
  kinematics in the $x'$-direction (top row) and $y'$-direction
  (middle row), and the average mean line-of-sight kinematics (bottom
  row), calculated in polar apertures in the first quadrant, are
  unfolded to the other three quadrants to facilitate the
  visualisation.}
\label{fig:bin_pl}
\end{figure*}

\subsection{The power-law model}
\label{sec:PLmodel}

The potential $\Phi$ of the power-law model is given by
\begin{equation}
  \label{eq:PLpot}
  \Phi(R,z)=\frac{\Phi_0\,R_c^\beta}
  {\left(R_c^2+R^2+z^2q_\Phi^{-2}\right)^{\beta/2}},
\end{equation}
in which $(R,z)$ are cylindrical coordinates, $\Phi_0$ is the
central potential, $R_c$ is the core radius and $q_\Phi$ is the
axial ratio of the spheroidal equipotentials. The parameter
$\beta$ controls the logarithmic gradient of the rotation curve at
large radii.

The mass density that follows from applying Poisson's equation to
eq.~\eqref{eq:PLpot} can be expanded as a finite sum of powers of the
cylindrical radius $R$ and the potential $\Phi$. Such a power-law
density implies that the \textit{even} part of the distribution
function is a power-law of the two integrals energy $E$ and angular
momentum $L_z$. For the \textit{odd} part of the distribution
function, which defines the rotational properties, a prescription for
the stellar streaming is needed. We adopt the prescription given in
eq.~(2.11) of EZ94, with a free parameter $k$ controlling the strength
of the stellar streaming, so that the odd part of the distribution
function is also a simple power-law of $E$ and $L_z$. Due to the
simple form of the distribution function, the calculation of the
power-law observables is straightforward.  Analytical expressions for
the surface brightness, the three components of the mean velocity and
velocity dispersion are given in eqs.~(3.1)--(3.8) of EZ94.

\begin{figure*}
\includegraphics{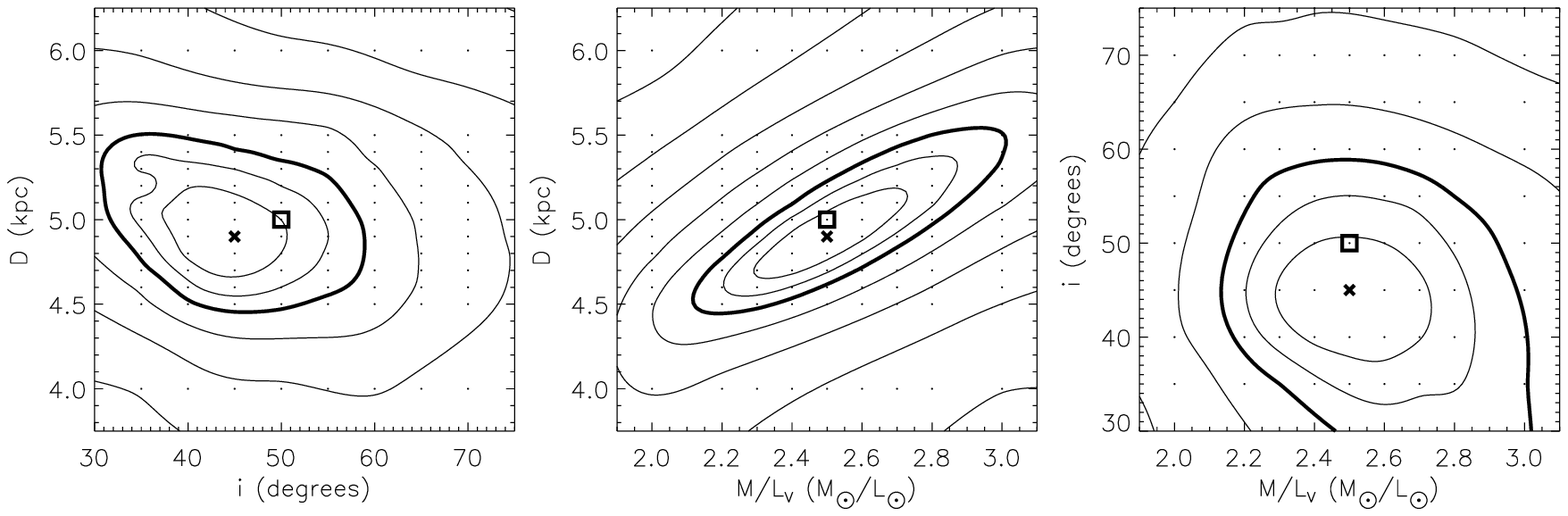}
\caption{
  The (marginalised) goodness-of-fit parameter $\Delta\chi^2$ as a
  function of distance $D$, inclination $i$ and mass-to-light ratio
  $M/L_V$, for different Schwarzschild model fits (indicated by the
  dots) to an axisymmetric power-law model with observables resembling
  those of \oc\ (see text for details). The $\chi^2$-values are offset
  such that the overall minimum, indicated by the cross, is zero. The
  contours are drawn at the confidence levels for a
  $\Delta\chi^2$-distribution with three degrees of freedom, with
  inner three contours corresponding to the 68.3\%, 95.4\% and 99.7\%
  (thick contour) confidence levels.  Subsequent contours correspond
  to a factor of two increase in $\Delta\chi^2$. The input parameters
  $D=5.0$ kpc, $i=50$\dgr\ and $M/L=2.5$ \MLsun, indicated by the open
  square, are recovered within the 68.3\% confidence levels.}
\label{fig:chi2mod_pl}
\end{figure*}

\subsection{Observables}
\label{sec:PLobservables}

We choose the parameters of the power-law model such that its
observable properties resemble those of \oc. We use $\Phi_0$=2500
km$^2$\,s$^{-2}$, which sets the unit of velocity of our models, and a
core radius of $R_c=2.5$ arcmin, which sets the unit of length.  For
the flattening of the potential we take $q_\Phi=0.95$ and we set
$\beta=0.5$, so that the even part of the distribution function is
positive (Fig.~1 of EZ94). The requirement that the total distribution
function (even plus odd part) should be non-negative places an upper
limit on the (positive) parameter $k$.  This upper limit
$k_\mathrm{max}$ is given by eq.~(2.22) of EZ94\footnote{ The
  definition of $\chi$ has a typographical error and should be
  replaced by $\chi=(1-\beta/2)/|\beta|$.}. Their eq.~(2.24) gives the
value $k_\mathrm{iso}$ for which the power-law model has a nearly
isotropic velocity distribution. In our case $k_\mathrm{max}=1.38$ and
$k_\mathrm{iso}=1.44$. We choose $k=1$, i.e., a power-law model that
has a (tangential) anisotropic velocity distribution.

Furthermore, we use an inclination of $i=50$\dgr, a mass-to-light
ratio of $M/L=2.5$ \MLsun\ and a distance of $D=5$ kpc.  At this
inclination the projected flattening of the potential is
$q_\Phi'=0.97$. The isocontours of the projected surface density
are more flattened. Using eq.~(2.9) of Evans
(1994\nocite{1994MNRAS.267..333E}), the central and asymptotic
axis ratios of the isophotes are respectively $q'_0=0.96$ and
$q'_\infty=0.86$, i.e., bracketing the average observed flattening
of \oc\ of $q'=0.88$ (Geyer et al.\
1983\nocite{1983A&A...125..359G}).

Given the above power-law parameters, we calculate the three
components of the mean velocity $V$ and velocity dispersion $\sigma$
on a polar grid of 28 apertures, spanning a radial range of 20 arcmin.
Because of axisymmetry we only need to calculate the observables in
one quadrant on the plane of the sky, after which we reflect the
results to the other quadrants. Next, we use the relative precisions
$\Delta V/\sigma\sim0.11$ and $\Delta\sigma/\sigma\sim0.08$ as
calculated for \oc\ (Appendix~\ref{sec:apgrid}), multiplied with the
calculated $\sigma$ for the power-law model, to attach an error to the
power-law observables in each aperture. Finally, we perturb the
power-law observables by adding random Gaussian deviates with the
corresponding errors as standard deviations.

Without the latter randomisation, the power-law observables are as
smooth as those predicted by the dynamical Schwarzschild models, so
that the goodness-of-fit parameter $\chi^2$ in eq.~\eqref{eq:chi2},
approaches zero. Such a perfect agreement never happens for real data.
Including the level of noise representative for \oc, allows us to use
$\chi^2$ to not only investigate the recovery of the power-law
parameters, but, at the same time, also asses the accuracy with
which we expect to measure the corresponding parameters for \oc\ 
itself.

The resulting mean velocity $V_\mathrm{observed}$ and velocity
dispersion $\sigma_\mathrm{observed}$ fields for the power-law model
are shown in respectively the first and third column of
Figure~\ref{fig:bin_pl}.  They are unfolded to the other three
quadrants to facilitate the visualisation.

\subsection{Schwarzschild models}
\label{sec:PLschwarzschild}

We construct axisymmetric Schwarzschild models based on the power-law
potential \eqref{eq:PLpot}. We calculate a library of 2058 orbits by
sampling 21 energies $E$, 14 angular momenta $L_z$ and 7 third
integrals $I_3$. In this way, the number and variety of the library
of orbits is large enough to be representative for a broad range of
stellar systems, and the set of eqs.~\eqref{eq:schwarzschildproblem}
is still solvable on a machine with 512 Mb memory (including
regularisation constraints).

The resulting three-integral Schwarzschild models include the special
case of dependence on only $E$ and $L_z$ like for the power-law
models.  Schwarzschild's method requires that the orbits in the
library are sampled over a range that includes most of the total mass,
whereas all power-law models have infinite mass. To solve this problem
at least partially, we ensure that there are enough orbits to
constrain the observables at all apertures. We distribute the orbits
logarithmically over a radial range from 0.01 to 100 arcmin (five
times the outermost aperture radius) and fit the intrinsic density out
to a radius of $10^5$ arcmin. The orbital velocities are binned in
histograms with 150 bins, at a velocity resolution of 2 \kms.

To test whether and with what precision we can recover the input
distance of $D=5$ kpc, the inclination of $i=50$\dgr\ and the
mass-to-light ratio $M/L=2.5$ \MLsun, we calculate models for values
of $D$ between 3.5 and 6.5 kpc, $i$ between 35\dgr\ (the asymptotic
isophotal axis ratio $q'_\infty=0.86$ implies that $i>30$\dgr) and
70\dgr, and $M/L$ between 1.5 and 3.5 \MLsun. Additionally, to test
how strongly the best-fitting parameters depend on the underlying mass
model, we also vary the flattening of the power-law potential $q_\Phi$
between 0.90 and 1.00.

We then fit each of the dynamical models simultaneously to the
calculated observables of the power-law model (with
$q_\Phi=0.95$). Comparing these calculated observables with those
predicted by the Schwarzschild models, results for each fitted
Schwarzschild model in a goodness-of-fit parameter $\chi^2$. We
use this value to find the best-fit Schwarzschild model and to
determine the accuracy of the corresponding best-fit parameters.

Calculating the observables for all orbits in the library requires
about an hour on a 1 GHz machine with 512 MB memory and the
NNLS-fit takes about half an hour. No distinct models need to be
calculated for different values of $M/L$, as a simple velocity
scaling prior to the NNLS-fit is sufficient. Making use of (a
cluster of) about 30 computers, the calculations for the full
four-parameter grid of Schwarzschild models takes a few days.

\begin{figure*}
\includegraphics{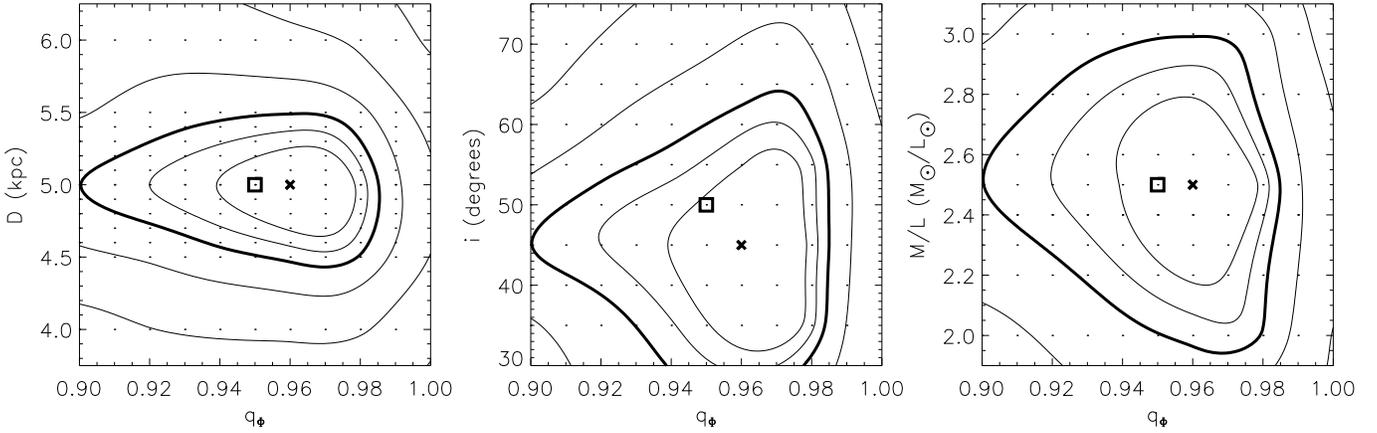}
\caption{
  The (marginalised) goodness-of-fit parameter
  $\Delta\chi^2$ as a function of distance $D$, inclination $i$ and
  mass-to-light ratio $M/L$ against the flattening $q_\Phi$ of the
  underlying potential, for different Schwarzschild model
  fits (indicated by the dots) to the observables of an
  axisymmetric power-law model resembling those of \oc. The contours
  are as in Figure~\ref{fig:chi2mod_pl}, but for a
  $\Delta\chi^2$-distribution with four degrees of freedom.  The cross
  indicates the overall best-fit model ($\Delta\chi^2=0$). The input
  parameters of the power-law model, $q_\Phi=0.95$, $D=5.0$ kpc,
  $i=50$\dgr\ and $M/L=2.5$ \MLsun, are indicated by the open square.
  The input parameters are recovered within the 68.3\% confidence
  levels, even for mass models that assume a (slightly) incorrect
  value for the flattening. However, spherical models ($q_\Phi=1.0$)
  are strongly ruled out.}
\label{fig:chi2q_pl}
\end{figure*}

\subsection{Distance, inclination and mass-to-light ratio}
\label{sec:PLdistinclML}

The Schwarzschild model that best fits the calculated power-law
observables is the one with the (overall) lowest $\chi^2$-value.
After subtraction of this minimum value, we obtain $\Delta\chi^2$ as
function of the three parameters $D$, $i$ and $M/L$ (with
$q_\Phi=0.95$ fixed). To visualise this three-dimensional function, we
calculate for a pair of parameters, say $D$ and $i$, the minimum in
$\Delta\chi^2$ as function of the remaining parameter, $M/L$ in this
case. The contour plot of the resulting marginalised $\Delta\chi^2$ is
shown in the left panel of Figure~\ref{fig:chi2mod_pl}. The dots
indicate the values at which Schwarzschild models have been
constructed and fitted to the power-law observables. The contours are
drawn at the confidence levels for a $\Delta\chi^2$-distribution with
three degrees of freedom, with inner three contours corresponding to
the 68.3\%, 95.4\% and 99.7\% (thick contour) confidence levels.
Subsequent contours correspond to a factor of two increase in
$\Delta\chi^2$.  The minimum ($\Delta\chi^2=0$) is indicated by the
cross. Similarly, we show in the middle and left panel the contour
plots of $\Delta\chi^2$ marginalised for respectively the pair $D$ and
$M/L$ and the pair $i$ and $M/L$.

The input parameters $D=5.0$ kpc, $i=50$\dgr\ and $M/L=2.5$ \MLsun,
indicated by the open square, are well recovered. The mean velocity
$V_\mathrm{model}$ and velocity dispersion $\sigma_\mathrm{model}$
predicted by the best-fit Schwarzschild model are shown in the second
and fourth column of Figure~\ref{fig:bin_pl}. The corresponding
power-law observables are well reproduced within the error bars.

Since the parameters of the power-law model are chosen such that its
observables and corresponding errors resemble those of \oc, the
contours in Figure~\ref{fig:chi2mod_pl} provide an estimate of the
precision with which we expect to measure the best-fitting parameters
for \oc.  At the 68.3\%-level (99.7\%-level) the distance $D$,
inclination $i$ and mass-to-light ratio $M/L$ are retrieved with an
accuracy of respectively 6 (11), 9 (18), 13 (28) per cent. Due the
additional complication of infinite mass in the case of the power-law
models, these estimates most likely are upper limits to the precision
we expect to achieve for \oc. This holds especially for the
inclination and the mass-to-light ratio as they are sensitive to how
well the mass model is fitted. The distance is mainly constrained by
the kinematics, so that the corresponding accuracy is probably an
accurate estimate of the precision with which we expect to measure the
distance to \oc.

\subsection{Flattening}
\label{sec:flattening}

The above investigation of the recovery of the global parameters $D$,
$i$ and $M/L$ is for a known mass model, given by the power-law
potential \eqref{eq:PLpot}. In general, we obtain the mass model from
a MGE-parameterisation of the observed surface brightness
(\S~\ref{sec:massmodel}). There is no guarantee that the resulting MGE
model provides an accurate description of the true mass distribution.
Therefore, we tested the effect of an incorrect mass model on the
best-fit parameters by varying the flattening $q_\Phi$ of the
power-law potential while keeping the calculated observables (for the
power-law model with $q_\Phi=0.95$) fixed.

Since we use these same values for the other power-law parameters
($\Phi_0$=2500 km$^2$\,s$^{-2}$, $R_c=2.5$ arcmin, $\beta=0.5$ and
$k=1$), we have to be careful that by varying $q_\Phi$ the resulting
model is still physical, i.e., that the underlying distribution
function is non-negative. For these parameters and $q_\Phi$ between
0.9 and 1.0 this is the case (EZ94).

As before, for all Schwarzschild models we calculate $\Delta\chi^2$,
which is now a function of the four parameters $D$, $i$, $M/L$ and
$q_\Phi$. In the three panels of Figure~\ref{fig:chi2q_pl}, we show
$\Delta\chi^2$ marginalised for respectively $D$, $i$ and $M/L$
against $q_\Phi$.  The symbols and contours are as in
Figure~\ref{fig:chi2mod_pl}, but now for a $\Delta\chi^2$-distribution
with four degrees of freedom.  The input parameters of the power-law
model (indicated by an open square) are $q_\Phi=0.95$, $D=5.0$ kpc,
$i=50$\dgr\ and $M/L=2.5$ \MLsun.

The distance $D$ is well constrained around the correct input value,
even at values for the flattening of the potential $q_\Phi$ that are
different from the true value of 0.95. This implies that the
best-fitting distance is accurate even for mass models that assume a
(slightly) incorrect value for the flattening. Whereas a potential
with a flattening as low as 0.90 still (just) falls within the contour
at the 99.7\%-level, we conclude, as in \S~\ref{sec:veldispprofiles},
that spherical models ($q_\Phi=1$) are strongly ruled out. The middle
and right panel of Figure~\ref{fig:chi2q_pl} show that the results for
respectively the mass-to-light $M/L$ and inclination $i$ are similar,
although, as before, they are less well constrained due to the
infinite mass of the power-law models.



\section{Dynamical models for \oc}
\label{sec:dynmodels}

We use our method as described in \S~\ref{sec:schwarzschild}, to
construct dynamical models for \oc. We obtain the mass model from
a MGE-parametrisation of the observed surface brightness. We
compute the mean velocity and velocity dispersion of both proper
motion components and along the line-of-sight in polar apertures
on the plane of the sky. For a range of distances, inclinations
and (constant) mass-to-light ratios, we then simultaneously fit
axisymmetric Schwarzschild models to these observations.
Additionally, we also allow for radial variation in the
mass-to-light ratio.

\begin{figure}
\includegraphics{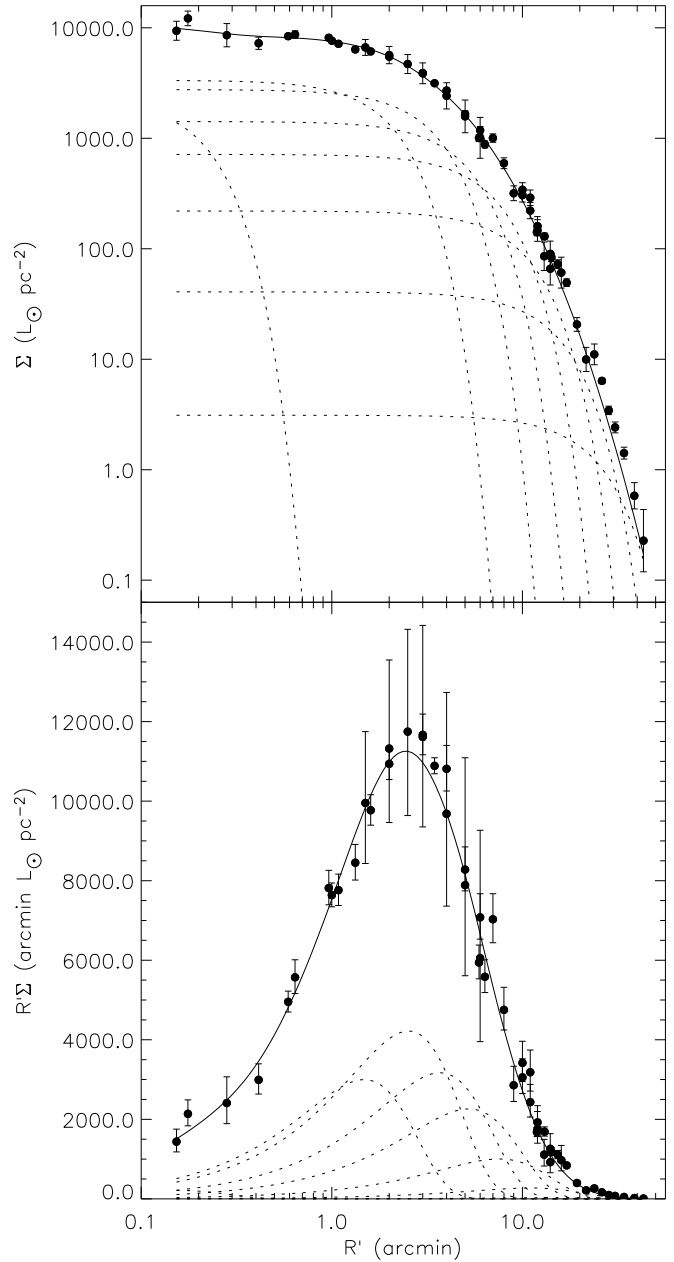}
\caption{
  The Multi-Gaussian Expansion (MGE) of the $V$-band surface
  brightness profile of \oc. The filled circles represent the
  measurements by Meylan (1987), the dotted curves correspond to the
  eight Gaussians in the expansion and the solid curve represents
  their sum. The top panel shows the surface brightness $\Sigma$ as a
  function of projected radius $R'$ (in arcmin). Kalnajs (1999) has
  shown that the quantity $R'\Sigma$ in the bottom panel is a good
  diagnostic of the mass that is enclosed at each radius. }
\label{fig:mge}
\end{figure}

\begin{table}
\caption{The parameters of the 8 Gaussians from the MGE-fit to the
  $V$-band surface brightness profile of \oc\ as derived by Meylan
  (1987). The second
  column gives the central surface brightness (in \Lsunpc2) of each Gaussian
  component, the third column the dispersion (in arcmin) along the major
  axis and the fourth column the observed flattening.}
\label{tab:mge}
\begin{center}
\begin{tabular}{cccc}
\hline
$j$ & $\Sigma_{0V}$ & $\sigma'$ & $q'$\\
    & (\Lsunpc2)     & (arcmin)  &     \\
\hline
1  & 2284.7077 & 0.15311 & 1.000000\\
2  & 3583.7901 & 1.47715 & 0.934102\\
3  & 3143.2029 & 2.52542 & 0.876713\\
4  & 1670.8477 & 3.69059 & 0.848062\\
5  & 840.86244 & 5.21905 & 0.849760\\
6  & 262.69433 & 7.53405 & 0.835647\\
7  & 46.995893 & 11.0685 & 0.866259\\
8  & 3.3583961 & 17.5470 & 0.926328\\
\hline
\end{tabular}
\end{center}
\end{table}

\begin{table*}
\caption{The mean velocity and velocity dispersion calculated in polar
  apertures on the plane of sky from the proper motion observations. Per row the
  information per aperture is given. The first column labels the
  aperture and the second column gives the number of stars $n_\star$ that fall in
  the aperture. Columns 3--6 list the polar coordinates $r$ (in
  arcmin) and the angle $\theta$ (in degrees) of the centroid of the
  aperture and the corresponding
  widths $\Delta r$ (in arcmin) and $\Delta\theta$ (in degrees). The
  remaining columns present the average proper motion kinematics in units of
  \masyr. The
  mean velocity $V$ with error $\Delta V$ and velocity dispersion
  $\sigma$ with error $\Delta\sigma$ are given in columns 7--10 for
  the proper motion component in the $x'$-direction and in columns
  11--14 for the proper motion component in the $y'$-direction.
}
\label{tab:apVsig_pmxy}
\begin{center}
\begin{tabular}{*{14}{r}}
\hline
& $n_\star$ & $r_0$ & $\theta_0$ & $\Delta r$ & $\Delta\theta$ &
$V_{x'}$ & $\Delta V_{x'}$ & $\sigma_{x'}$ & $\Delta\sigma_{x'}$ &
$V_{y'}$ & $\Delta V_{y'}$ & $\sigma_{y'}$ & $\Delta\sigma_{y'}$ \\
\hline
 1 &  80 &  1.14 & 45.0 & 2.28 & 90.0 & -0.15 & 0.09 & 0.80 & 0.07 & -0.01 & 0.09 & 0.70 & 0.05 \\
 2 &  99 &  3.04 & 15.0 & 1.53 & 30.0 & -0.16 & 0.07 & 0.66 & 0.04 &  0.23 & 0.07 & 0.64 & 0.05 \\
 3 &  67 &  3.04 & 45.0 & 1.53 & 30.0 &  0.03 & 0.12 & 0.90 & 0.07 &  0.06 & 0.08 & 0.62 & 0.05 \\
 4 &  74 &  3.04 & 75.0 & 1.53 & 30.0 & -0.15 & 0.08 & 0.64 & 0.07 & -0.08 & 0.09 & 0.71 & 0.06 \\
 5 &  85 &  4.59 & 11.2 & 1.57 & 22.5 & -0.27 & 0.06 & 0.57 & 0.03 &  0.19 & 0.06 & 0.57 & 0.05 \\
 6 &  77 &  4.59 & 33.7 & 1.57 & 22.5 & -0.08 & 0.07 & 0.63 & 0.05 &  0.13 & 0.06 & 0.57 & 0.08 \\
 7 &  76 &  4.59 & 56.2 & 1.57 & 22.5 & -0.20 & 0.07 & 0.55 & 0.05 &  0.13 & 0.08 & 0.69 & 0.06 \\
 8 &  82 &  4.59 & 78.7 & 1.57 & 22.5 & -0.19 & 0.05 & 0.55 & 0.04 &  0.07 & 0.07 & 0.66 & 0.06 \\
 9 & 105 &  6.31 &  9.0 & 1.86 & 18.0 &  0.00 & 0.06 & 0.60 & 0.04 &  0.26 & 0.05 & 0.50 & 0.04 \\
10 &  88 &  6.31 & 27.0 & 1.86 & 18.0 & -0.13 & 0.07 & 0.61 & 0.04 &  0.13 & 0.05 & 0.48 & 0.05 \\
11 &  70 &  6.31 & 45.0 & 1.86 & 18.0 & -0.28 & 0.07 & 0.58 & 0.07 &  0.23 & 0.06 & 0.50 & 0.06 \\
12 &  72 &  6.31 & 63.0 & 1.86 & 18.0 & -0.25 & 0.05 & 0.45 & 0.04 & -0.01 & 0.06 & 0.53 & 0.05 \\
13 &  65 &  6.31 & 81.0 & 1.86 & 18.0 & -0.25 & 0.07 & 0.58 & 0.05 &  0.05 & 0.06 & 0.45 & 0.03 \\
14 &  95 &  8.49 &  7.5 & 2.52 & 15.0 & -0.04 & 0.05 & 0.56 & 0.04 &  0.22 & 0.04 & 0.38 & 0.02 \\
15 &  88 &  8.49 & 22.5 & 2.52 & 15.0 & -0.09 & 0.05 & 0.46 & 0.04 &  0.10 & 0.07 & 0.53 & 0.07 \\
16 &  91 &  8.49 & 37.5 & 2.52 & 15.0 & -0.15 & 0.05 & 0.49 & 0.04 &  0.14 & 0.04 & 0.41 & 0.03 \\
17 &  73 &  8.49 & 52.5 & 2.52 & 15.0 & -0.31 & 0.06 & 0.51 & 0.06 &  0.19 & 0.05 & 0.44 & 0.03 \\
18 &  72 &  8.49 & 67.5 & 2.52 & 15.0 & -0.35 & 0.05 & 0.44 & 0.04 &  0.14 & 0.06 & 0.54 & 0.05 \\
19 &  61 &  8.49 & 82.5 & 2.52 & 15.0 & -0.40 & 0.07 & 0.58 & 0.05 & -0.03 & 0.07 & 0.48 & 0.04 \\
20 &  88 & 11.54 &  9.0 & 3.56 & 18.0 &  0.02 & 0.05 & 0.44 & 0.04 &  0.20 & 0.05 & 0.46 & 0.04 \\
21 &  95 & 11.54 & 27.0 & 3.56 & 18.0 & -0.17 & 0.04 & 0.42 & 0.04 &  0.17 & 0.05 & 0.49 & 0.04 \\
22 &  64 & 11.54 & 45.0 & 3.56 & 18.0 & -0.24 & 0.05 & 0.44 & 0.04 &  0.18 & 0.05 & 0.41 & 0.03 \\
23 &  85 & 11.54 & 63.0 & 3.56 & 18.0 & -0.41 & 0.05 & 0.44 & 0.03 &  0.05 & 0.04 & 0.43 & 0.03 \\
24 &  68 & 11.54 & 81.0 & 3.56 & 18.0 & -0.36 & 0.05 & 0.43 & 0.03 &  0.05 & 0.05 & 0.46 & 0.03 \\
25 &  58 & 16.64 & 11.2 & 6.64 & 22.5 & -0.02 & 0.06 & 0.40 & 0.04 &  0.19 & 0.06 & 0.41 & 0.05 \\
26 &  74 & 16.64 & 33.7 & 6.64 & 22.5 & -0.14 & 0.06 & 0.48 & 0.05 & -0.01 & 0.06 & 0.45 & 0.04 \\
27 &  79 & 16.64 & 56.2 & 6.64 & 22.5 & -0.17 & 0.05 & 0.46 & 0.03 &  0.04 & 0.04 & 0.41 & 0.04 \\
28 &  92 & 16.64 & 78.7 & 6.64 & 22.5 & -0.21 & 0.05 & 0.43 & 0.03 & -0.05 & 0.04 & 0.35 & 0.03 \\
\hline
\end{tabular}
\end{center}
\end{table*}

\subsection{MGE mass model}
\label{sec:mgemodel}

An MGE-fit is best obtained from a two-dimensional image, which gives
direct information about the flattening and any radial variations in
the two-dimensional structure of the object. Unfortunately, no such
image is available to us, and the only published surface brightness
observations of \oc\ consist of radial surface brightness profiles,
and an ellipticity profile by Geyer et al.\ 
(1983\nocite{1983A&A...125..359G}). We therefore perform a
one-dimensional MGE-fit to the radial surface brightness profile, and
after that use the ellipticity profile to include flattening in the
mass model.

We use the $V$-band surface brightness data from Meylan
(1987\nocite{1987A&A...184..144M}), who combined various published
measurements (Gascoigne \& Burr 1956\nocite{1956MNRAS.116..570G}; Da
Costa 1979\nocite{1979AJ.....84..505D}; King et al.\
1968\nocite{1968AJ.....73..456K}). Their data consists of individual
measurements along concentric rings, while the MGE-algorithm developed
by Cappellari (2002\nocite{2002MNRAS.333..400C}) requires a regular
(logarithmic) spacing of the surface brightness measurements. We
therefore first describe the profile in terms of a fourth-order
polynomial and then fit a set of one-dimensional Gaussians to this
polynomial. Eight Gaussians with different central surface brightness
$\Sigma_{0V,j}$ and dispersion $\sigma'_j$ are required by the MGE-fit
(second and third column of Table~\ref{tab:mge}).
Figure~\ref{fig:mge} shows that this MGE-model provides an excellent
fit, not only to the surface brightness $\Sigma$, but also to
$R'\Sigma$ (cf. Kalnajs 1999\nocite{1999gaha.conf..325K}).

The MGE-parameterisation is converted into a two-dimensional
luminosity distribution by assigning an observed flattening $q'_j$ to
each Gaussian in the superposition. We take into account that the
observed flattening of \oc\ varies as a function of radius (cf. Geyer
et al.\ 1983\nocite{1983A&A...125..359G}).  This is done by assuming
that the flattening of the $j$th Gaussian $q'_j$ is equal to the
observed flattening at a projected radius $R'=\sigma'_j$. This is
justified by the fact that a given Gaussian contributes most at radii
close to its dispersion $\sigma'_j$. Although small deviations from
the true two-dimensional light distribution in \oc\ may still occur,
we showed in \S~\ref{sec:flattening} that this approximation does not
significantly influence the derived intrinsic parameters for \oc.
Moreover, a two-dimensional MGE-fit to the combination of the surface
brightness profile from Meylan (1987\nocite{1987A&A...184..144M}) and
the ellipticity profile from Geyer et al.\ 
(1983\nocite{1983A&A...125..359G}), yields nearly equivalent MGE
parameters as those in Table~\ref{tab:mge}, although the fit to the
observed surface brightness profile is less good.

To conserve the total luminosity, we increase the central surface
brightness of each Gaussian by $1/q'_j$. Taking into account a
reddening of $E(B-V)=0.11$ for \oc\ (Lub
2000\nocite{2002ocuw.conf...95L}), the total $V$-band luminosity of our mass
model, at the canonical distance of $5.0\pm0.2$ kpc, is
$L_V=1.0\pm0.1\times10^6$ \Lsun. This compares well with other estimates
of the total luminosity of \oc\ of $0.8\times10^6$ \Lsun\ (Carraro \&
Lia 2000\nocite{2000A&A...357..977C}), $1.1\times10^6$ \Lsun\ (Seitzer
1983\nocite{1983PhDT........11S}) and $1.3\times 10^6$ \Lsun\ (Meylan
1987\nocite{1987A&A...184..144M}). The most flattened Gaussian in the
superposition ($j=7$) places a mathematical lower limit on the
inclination of 33\dgr. This is safely below the constraint 41--57 degrees
found in \S~\ref{sec:constraintincl}.

\subsection{Mean velocity and velocity dispersion}
\label{sec:meanVsig}

\begin{figure}
\includegraphics{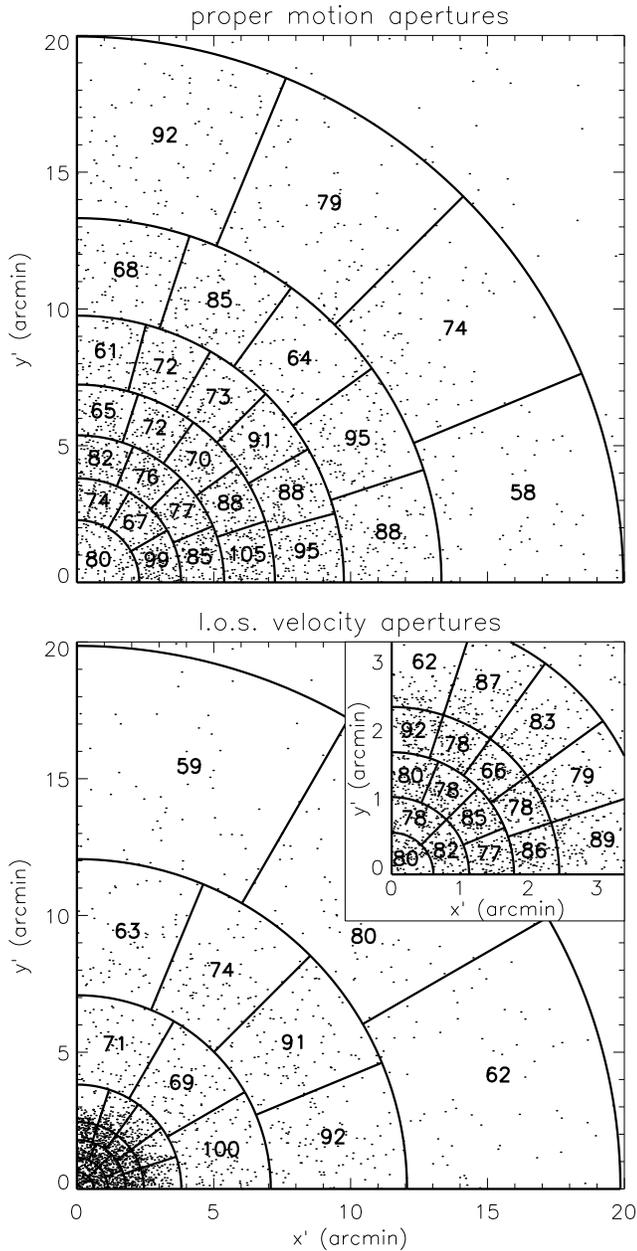}
\caption{
  The polar aperture grid for the proper motions (top panel) and for
  the line-of-sight velocities (bottom panel). The dots represent the
  individual stars, with positions folded to the first quadrant, while
  the solid lines indicate the locations of the apertures. The number
  of stars included are indicated in each aperture. An enlargement of
  the inner part of the line-of-sight polar grid is shown in the
  top-right corner of the bottom panel.}
\label{fig:apgrids}
\end{figure}

We construct a polar aperture grid for the proper motions and
line-of-sight velocities, as shown in Figure~\ref{fig:apgrids}.  The
dots in the top panel represent the positions, folded to the first
quadrant, of the 2295 selected stars with ground-based proper motions.
The overlayed polar grid, extending to about 20 arcmin, consists of 28
apertures. Per aperture, the number of stars is indicated, adding up
to a total of 2223 stars. Similarly, the bottom panel shows the 2163
selected stars with line-of-sight velocities. The different number of
stars and spatial distribution results in a polar grid of 27
apertures, which includes in total 2121 stars. 

For each aperture, we use the maximum likelihood method
(Appendix~\ref{sec:mlvelocitymoments}) to compute the mean velocity
$V$ and velocity dispersion $\sigma$ for both proper motion components
on along the line-of-sight. We calculate corresponding errors by means
of the Monte Carlo bootstrap method.

Each aperture contains around 50 to 100 stars. In
Appendix~\ref{sec:apgrid}, we find that this is a good compromise
between precision in the observables and spatial resolution.
Including more stars per aperture by increasing its size decreases the
uncertainties in the observables (and hence makes the resulting
kinematic fields smoother). At the same time, since the apertures
should not overlap to assure uncorrelated observables, this means less
apertures in the polar grid and hence a loss in spatial resolution.

\begin{figure*}
\includegraphics[width=\textwidth]{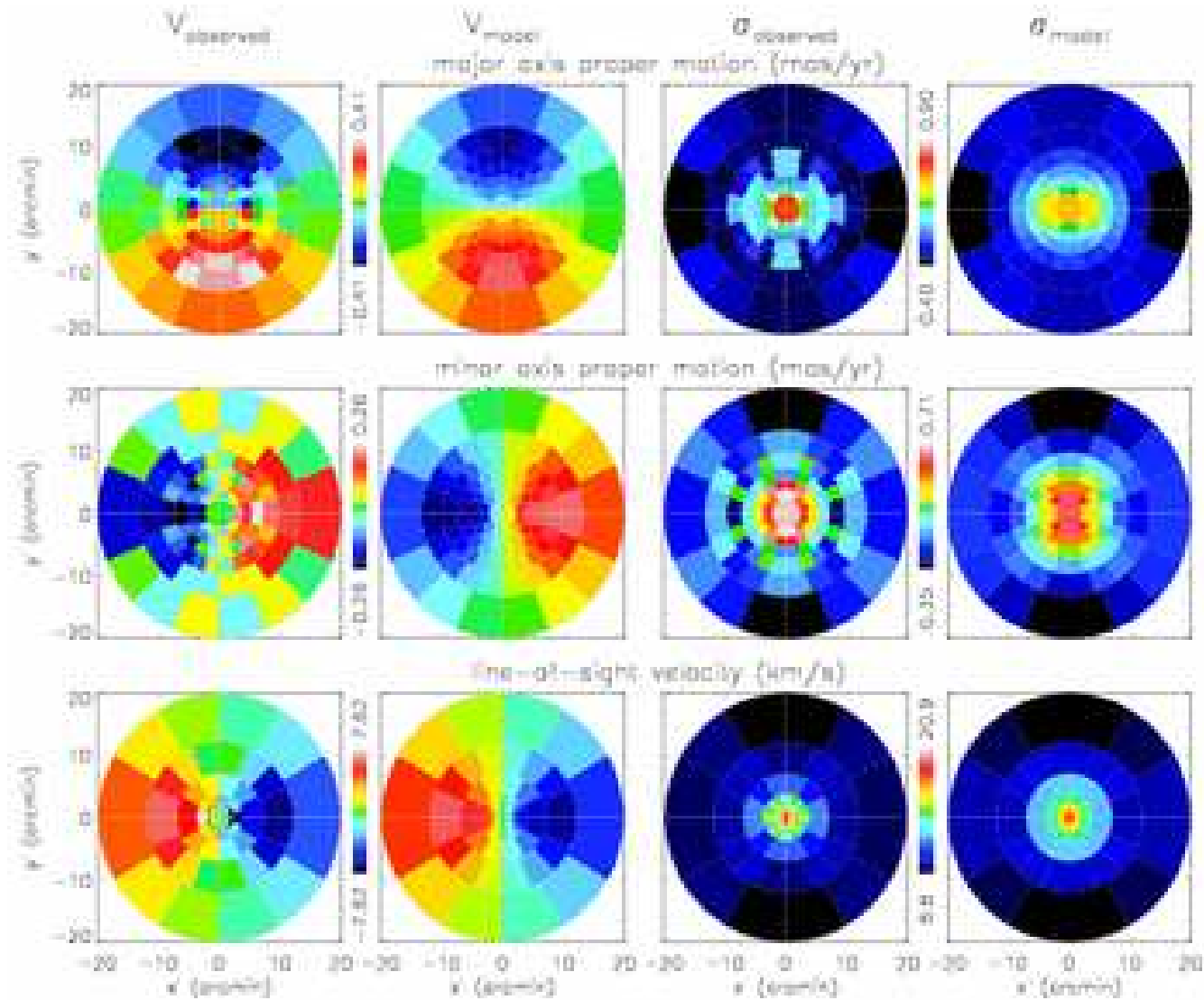}
\caption{
  Mean velocity and velocity dispersion calculated from the
  observations of \oc\ (first and third column) and from the best-fit
  dynamical model with $D=4.8$ kpc, $i=50$\dgr\ and $M/L_V=2.5$
  \MLsun\ (second and fourth column). The mean proper motion
  kinematics in the $x'$-direction (top row) and $y'$-direction
  (middle row), and the mean line-of-sight kinematics (bottom row),
  calculated in polar apertures in the first quadrant, are unfolded to
  the other three quadrants to facilitate the visualisation.}
\label{fig:bin_oc}
\end{figure*}

The properties of the apertures and corresponding mean kinematics are
given in Table~\ref{tab:apVsig_pmxy} for the proper motions and in
Table~\ref{tab:apVsig_vlos} for the line-of-sight velocities. The mean
velocity $V_\mathrm{observed}$ and velocity dispersion
$\sigma_\mathrm{observed}$ fields are shown in the first and third
column of Figure~\ref{fig:bin_oc} respectively. Although the average
kinematics are only calculated in the first quadrant, we use the
assumed axisymmetric geometry to unfold them to the other three
quadrants to facilitate the visualisation.

\begin{table}
\caption{The mean velocity and velocity dispersion calculated in polar
  apertures on the plane of sky from the line-of-sight
  velocity observations. Columns 1--6 are as in
  Table~\ref{tab:apVsig_pmxy} and the remaining columns present the
  average line-of-sight kinematics in \kms.
}
\label{tab:apVsig_vlos}
\begin{center}
\begin{tabular}{*{10}{r@{\hspace{8.0pt}}}}
\hline
& $n_\star$ & $r_0$ & $\theta_0$ & $\Delta r$ & $\Delta\theta$ &
$V_{z'}$ & $\Delta V_{z'}\!\!\!$ & $\sigma_{z'}$ & $\Delta\sigma_{z'}\!\!\!$ \\
\hline
 1 &  80 &  0.31 & 45.0 & 0.61 & 90.0 &  2.4 & 2.2 & 19.0 & 1.5 \\
 2 &  82 &  0.87 & 22.5 & 0.52 & 45.0 & -3.1 & 2.1 & 20.9 & 1.4 \\
 3 &  78 &  0.87 & 67.5 & 0.52 & 45.0 &  0.2 & 1.9 & 19.5 & 1.4 \\
 4 &  77 &  1.46 & 11.2 & 0.66 & 22.5 &  0.0 & 1.9 & 16.7 & 1.3 \\
 5 &  85 &  1.46 & 33.7 & 0.66 & 22.5 & -1.8 & 1.7 & 14.4 & 0.8 \\
 6 &  78 &  1.46 & 56.2 & 0.66 & 22.5 &  1.0 & 1.8 & 15.6 & 1.5 \\
 7 &  80 &  1.46 & 78.7 & 0.66 & 22.5 & -0.7 & 1.7 & 16.2 & 1.2 \\
 8 &  86 &  2.12 &  9.0 & 0.66 & 18.0 & -7.6 & 1.5 & 12.8 & 1.1 \\
 9 &  78 &  2.12 & 27.0 & 0.66 & 18.0 & -6.4 & 1.6 & 14.3 & 0.8 \\
10 &  66 &  2.12 & 45.0 & 0.66 & 18.0 & -3.8 & 1.9 & 16.8 & 1.2 \\
11 &  78 &  2.12 & 63.0 & 0.66 & 18.0 & -3.0 & 1.7 & 15.9 & 1.0 \\
12 &  92 &  2.12 & 81.0 & 0.66 & 18.0 & -0.3 & 1.7 & 14.5 & 1.0 \\
13 &  89 &  3.13 &  9.0 & 1.37 & 18.0 & -7.6 & 1.6 & 15.3 & 1.0 \\
14 &  79 &  3.13 & 27.0 & 1.37 & 18.0 & -2.2 & 1.5 & 14.6 & 1.0 \\
15 &  83 &  3.13 & 45.0 & 1.37 & 18.0 & -1.0 & 1.4 & 14.1 & 0.8 \\
16 &  87 &  3.13 & 63.0 & 1.37 & 18.0 & -2.6 & 1.4 & 15.0 & 0.8 \\
17 &  62 &  3.13 & 81.0 & 1.37 & 18.0 & -2.9 & 1.9 & 13.4 & 1.3 \\
18 & 100 &  5.45 & 15.0 & 3.27 & 30.0 & -5.0 & 1.2 & 12.0 & 1.0 \\
19 &  69 &  5.45 & 45.0 & 3.27 & 30.0 & -3.1 & 1.3 & 10.9 & 1.1 \\
20 &  71 &  5.45 & 75.0 & 3.27 & 30.0 & -1.4 & 1.2 & 11.8 & 1.0 \\
21 &  92 &  9.57 & 11.2 & 4.98 & 22.5 & -6.2 & 1.0 & 10.0 & 0.9 \\
22 &  91 &  9.57 & 33.7 & 4.98 & 22.5 & -5.5 & 1.1 & 10.3 & 1.0 \\
23 &  74 &  9.57 & 56.2 & 4.98 & 22.5 & -2.4 & 1.2 & 10.3 & 0.9 \\
24 &  63 &  9.57 & 78.7 & 4.98 & 22.5 &  0.2 & 1.3 &  9.8 & 0.9 \\
25 &  62 & 15.96 & 15.0 & 7.80 & 30.0 & -4.1 & 1.2 &  9.6 & 1.1 \\
26 &  80 & 15.96 & 45.0 & 7.80 & 30.0 & -1.9 & 1.2 &  9.8 & 0.7 \\
27 &  59 & 15.96 & 75.0 & 7.80 & 30.0 & -0.6 & 1.2 &  8.8 & 0.9 \\
\hline
\end{tabular}
\end{center}
\end{table}

\subsection{Constructing dynamical models}
\label{sec:constructdynmod}

First, we calculate models for a range of values in distance $D$,
inclination $i$ and constant $V$-band mass-to-light ratio $M/L_V$.
Next, fixing $D$ and $i$ at their measured best-fit values, we
also calculate a large set of models in which we allow $M/L_V$ to
vary with radius.

We sample the orbits on a grid of $21\times14\times7$ values in
$(E,L_z,I_3)$ on a radial range from 0.01 to 63 arcmin. This grid
extends beyond the tidal radius of 45 arcmin Trager et al.\
1995\nocite{1995AJ....109..218T}), so that all mass is included.
No PSF-convolution is used and the observables are stored directly
onto the apertures.

We (linearly) sample $D$ between 3.5 and 6.5 kpc in steps of 0.5
kpc, and additionally we refine the grid between 4.0 and 5.5 kpc
to steps of 0.1 kpc. We vary $i$ between 35 (close to the lower
limit of 33 degrees imposed by the flattening, see
\S~\ref{sec:mgemodel}) and 90 degrees in steps of five degrees, and
we refine between 40 and 50 degrees to steps of one degree. We
choose the constant $M/L_V$ values between 2.0 and 4.0 \MLsun\
with steps 0.5 \MLsun, and we refine between 2.0 and 3.0 \MLsun\
to steps of 0.1 \MLsun.

To investigate possible variation in $M/L_V$ with radius, we make use
of the eight Gaussian components of the MGE mass model
(\S~\ref{sec:mgemodel}). In case of constant $M/L_V$, we obtain the
intrinsic density by multiplying all the (deprojected) components with
the same constant $M/L_V$ value. To construct a mass model with a
radial $M/L_V$ profile, we multiply each component with its own
$M/L_V$ value, as in this way the calculation of the potential is
still efficient.

However, to reduce the number of free parameters (to make a search
through parameter space feasible) and to enforce a continuous profile,
we only vary the $M/L_V$ values for the first, second, fourth and
sixth component. For the third and fifth component, we
  interpolate between the $M/L$ values of the neighbouring
  components. To the outer two components we assign the same $M/L_V$
value as the sixth component, because their individual $M/L_V$ values
are not well constrained due to the small number of kinematic
measurements at these radii. With the distance and inclination fixed
at their best-fit values from the case of constant mass-to-light
ratio, we are left with a four-dimensional space to search through,
requiring again a few days on (a cluster of) about 30 computers.

All dynamical models are fitted simultaneously to the two-dimensional
light distribution of \oc\ (\S~\ref{sec:mgemodel}), and to the mean
velocity and velocity dispersion of both proper motions components and
along the line-of-sight, calculated in polar apertures on the plane of
the sky (first and third column of Figure~\ref{fig:bin_oc}).
Comparing the predicted values with the observations, results for each
fitted model in a goodness-of-fit parameter $\chi^2$. We use this
value to find the best-fit model and to determine the accuracy of the
corresponding best-fit parameters.

\begin{figure*}
\includegraphics{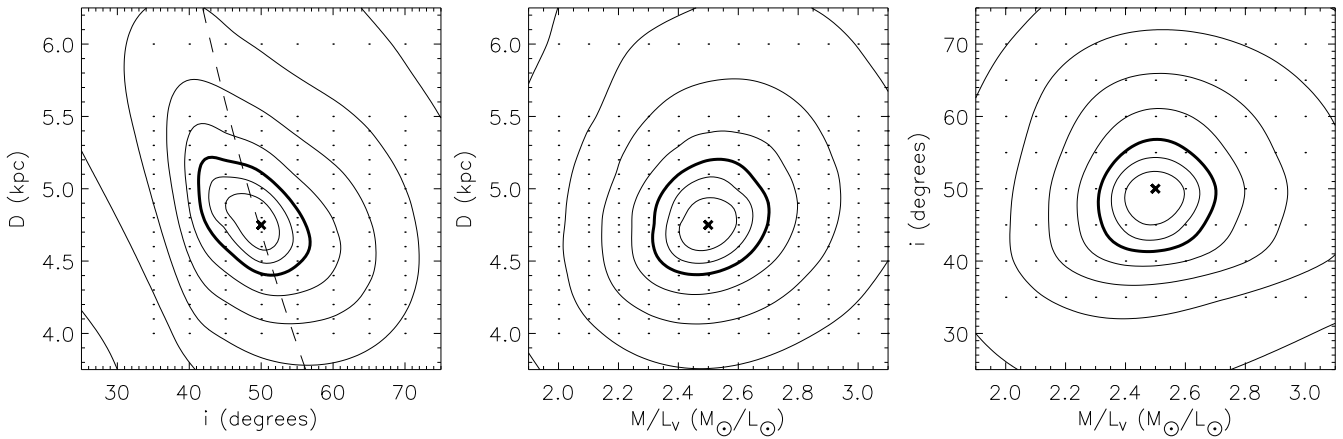}
\caption{
  The (marginalised) goodness-of-fit parameter $\Delta\chi^2$ as a
  function of distance $D$, inclination $i$ and mass-to-light ratio
  $M/L_V$, for different dynamical model fits (indicated by the dots)
  to the kinematics of \oc.  The contours are as in
  Figure~\ref{fig:chi2mod_pl}. The best-fit dynamical model is at
  $D=4.8$ kpc, $i=50$\dgr\ and $M/L_V=2.5$ \MLsun, indicated by the
  cross The dashed curve shows the $D\tan i=5.6$ kpc constraint from
  the mean velocities (\S~\ref{sec:constraintincl}).}
\label{fig:chi2mod_oc}
\end{figure*}


\section{Best-fit parameters}
\label{sec:bestfitpar}

In Figure~\ref{fig:chi2mod_oc}, we show $\Delta\chi^2$ as a
(marginalised) function of the distance $D$, inclination $i$ and
constant mass-to-light ratio $M/L_V$. The dots represent the values at
which dynamical models have been constructed and fitted to the
two-dimensional (photometric and kinematic) observations of \oc.  The
cross indicates the over-all best-fit model. The contours show that
all three parameters are tightly constrained, with at the 68.3\%-level
(99.7\%-level): $D=4.8\pm0.3$ ($\pm0.5$) kpc, $i=50\pm3$ ($\pm5$)
degrees and $M/L_V=2.5\pm0.1$ ($\pm0.2$) \MLsun.  As an illustration
that our best-fit model indeed reproduces the observations, the mean
velocity and velocity dispersion in polar apertures on the plane of
the sky as they follow from this model are shown in respectively the
second and fourth column of Figure~\ref{fig:bin_oc}. The model fits
the observations within the uncertainties given in
Table~\ref{tab:apVsig_pmxy} and \ref{tab:apVsig_vlos}.

After the discussion on the set of models where we allow the
mass-to-light ratio $M/L_V$ to vary with radius, we compare our
best-fit values for the (constant) mass-to-light ratio,
inclination and distance with results from previous studies.

\subsection{Mass-to-light ratio variation}
\label{sec:mlvariation}

\begin{figure}
\includegraphics{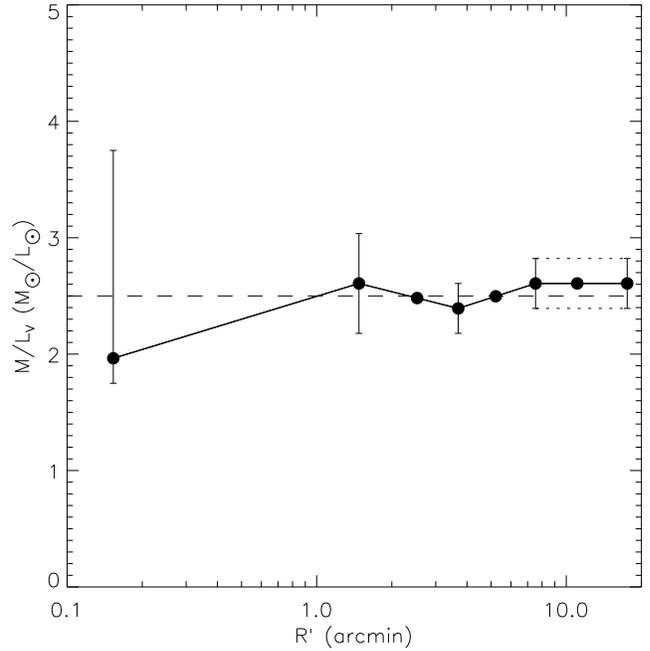}
\caption{
  Variation in mass-to-light ratio $M/L_V$ with projected radius $R'$.
  The filled circles represent the eight Gaussian components of the
  MGE mass model, with the best-fit $M/L_V$ value of each component
  plotted against its dispersion along the major axis. With the
  distance and inclination fixed at $D=4.8$ kpc and $i=50$ degrees, we
  allowed variation in the $M/L_V$ values for the four inner points
  with error bar, while the two outer points were shifted vertically
  similar to the fourth point, and the remaining two points were
  interpolated between the two neighbouring points. Each of the models
  was simultaneously fitted to the photometric and kinematic
  observations of \oc. The error bars represent the 68.3\% confidence
  level for the corresponding $\Delta\chi^2$-distribution with four
  degrees of freedom. The variation in the resulting $M/L_V$ profile
  is small with no significant deviation from the best-fit constant
  $M/L_V$ of 2.5 \Msun\ (horizontal dashed line).}
\label{fig:mlprofile}
\end{figure}

Figure~\ref{fig:mlprofile} summarises the results from fitting models
in which we allowed the mass-to-light ratio $M/L_V$ to vary with
radius in the way described in \S~\ref{sec:constructdynmod}.  The
filled circles represent the eight Gaussian components, with the
best-fit $M/L_V$ value of each component plotted against their
dispersions along the major axis (see column three of
Table~\ref{tab:mge}). The error bars represent the 68.3\% confidence
level.

The uncertainty on the innermost point around 0.15 arcmin is
relatively large since at that small radius there are only a few
observations (see Figure~\ref{fig:apgrids}) to constrain the
$M/L_V$ value. Nevertheless, the resulting $M/L_V$ profile only
shows a small variation, which is not significantly different from
the best-fit constant $M/L_V$ of 2.5 \Msun.

In the above experiment, we fixed the distance and inclination at
the best-fit values of $D=4.8$ kpc and $i=50$ degrees from the
case of constant $M/L_V$. Although an important constraint is that
all eight Gaussian components have to be at the same distance, its
precise value, as well as that of the inclination, is not crucial.
We tested that a reasonable variation in these fixed values
(within the 99.7\% confidence level in
Figure~\ref{fig:chi2mod_oc}) does not significantly change the
best-fit $M/L_V$ profile. We conclude that a constant
mass-to-light ratio for \oc\ is a valid assumption.

\subsection{Mass-to-light ratio}
\label{sec:bestfitml}

Our best-fit mass-to-light ratio of $M/L_V=2.5\pm0.1$ \MLsun\ lies
in between the estimates by Seitzer
(1983\nocite{1983PhDT........11S}) of 2.3 \MLsun\ and by Meylan
(1987\nocite{1987A&A...184..144M}) of 2.9 \MLsun. Meylan et al.\
(1995\nocite{1995A&A...303..761M}) derived a value of 4.1 \MLsun,
based on a spherical, radial anisotropic King-Michie dynamical
model, while we find that \oc\ is flattened and outwards
tangentially anisotropic (see \S~\ref{sec:anisotropy}). Moreover,
their adopted central value of the line-of-sight velocity
dispersion is significantly higher than ours, even if we use the
same data-set by M97.

Meylan et al.\ (1995\nocite{1995A&A...303..761M}) estimated
the total mass of \oc\ to be $5.1\times10^6$ \Msun, which is also
significantly higher than what we derive. After multiplication with
the total luminosity of our mass model of $L=1.0\times10^6$ \Lsun\ (at
the best-fit distance of $D=4.8\pm0.3$ kpc), we find a total mass of
$M=(2.5\pm0.3)\times10^6$ \Msun. This is consistent with the value by
Mandushev et al.\ (1991\nocite{1991A&A...252...94M}) of
$2.4\times10^6$ \Msun\ and Seitzer (1983\nocite{1983PhDT........11S})
of $2.8\times10^6$ \Msun. The estimate by Meylan
(1987\nocite{1983PhDT........11S}) of $3.9\times10^6$ \Msun\ is
higher, but again based on a spherical King-Michie model.

\subsection{Inclination} 
\label{sec:bestfitincl}

The dashed curve in the left panel of Figure~\ref{fig:chi2mod_oc}
shows the $D\tan i=5.6$ kpc constraint from the mean velocities
derived in \S~\ref{sec:constraintincl}. This constraint can be used to
eliminate either the distance or the inclination and hence reduce the
parameter space. Although we do not use this constraint in the
dynamical models, it is clear that the above best-fit $D$ and $i$
yield $D\tan i=5.6\pm 0.2$ kpc, which is consistent with the value
derived from the mean velocities.

The best-fit inclination of $i=50\pm3$ degrees falls within the range
of 30--60 degrees that was derived in Paper~I from the amplitude of
the proper motions, but is slightly higher than the estimate by
van Leeuwen \& Le Poole (2002\nocite{2002ocuw.conf...41V})
between 40 and 60 degrees. However, as discussed in
\S~\ref{sec:constraintincl}, they used models of modest complexity
and freedom which require strong assumptions, whereas our method
is much more general and robust.

Our best-fit inclination implies that \oc\ is intrinsically even more
non-spherical than the average observed flattening of
$q'=0.879\pm0.007$ (Geyer et al.\ 1983\nocite{1983A&A...125..359G})
already indicates. Using the relation $q^2\sin^2i=q'^2-\cos^2i$ for
axisymmetric objects, we find an average intrinsic axial ratio
$q=0.78\pm0.03$.

\subsection{Distance}
\label{sec:bestfitdist}

Adopting a reddening of $E(B-V)=0.11$ for \oc\ (Lub
2000\nocite{2002ocuw.conf...95L}), the best-fit dynamical distance
corresponds to a distance modulus of $(m-M)_V=13.75\pm0.13$ ($\pm0.22$
at the 99.7\%-level). This is consistent with the (canonical)
distance modulus of $(m-M)_V=13.84$ by photometric methods, as given
in the globular cluster catalogue of Harris et al.
(1996\nocite{1996AJ....112.1487H}), together with the uncertainty
estimate of about 0.1 magnitude by Benedict et al.
(2002\nocite{2002AJ....123..473B}), using the absolute magnitude of RR
Lyrae stars. Using the infrared colour versus surface brightness
relation for the eclipsing binary OGLEC 17, Thompson et al.
(2001\nocite{2001AJ....121.3089T}) find a larger distance modulus of
$(m-M)_V = 14.05\pm0.11$. However, their distance modulus estimates
based on the measured bolometric luminosity of the binary components,
are on average lower, ranging from 13.66 to 14.06.

Although our dynamical distance estimate is consistent with that by
other methods, it is at the lower end. A lower value for the distance
is expected if the proper motion dispersion is over-estimated and/or
the line-of-sight velocity dispersion under-estimated (see also
Appendix~\ref{sec:simpledistest}, eq.~\ref{eq:isodist}). As we saw in
\S~\ref{sec:selection}, both are likely in the case of \oc\ if the
kinematic data is not properly selected. The correction in
\S~\ref{sec:kinematics} for perspective rotation and especially for
the residual solid-body rotation is crucial for the construction of a
realistic dynamical model and also to find a reliable distance
estimate.

An impression of the effect of the selection and correction of the
kinematic data on the distance estimate follows from the range of
dynamical models we constructed for \oc. Before any selection and
correction, the kinematics of the cluster stars give rise to a
best-fit dynamical model at a distance as low as $\sim$3.5 kpc.
After removing from the proper motion data-set the stars disturbed
by their neighbours, i.e., only selecting class 0 stars, the
best-fit distance becomes $\sim$4.0 kpc. The correction for
perspective and solid-body rotation increase the best-fit distance
to $\sim$4.5 kpc. Finally, after the additional selection on
velocity errors, we find our best-fit dynamical distance of
$4.8\pm0.3$ kpc.

An even tighter selection does not significantly change the
best-fit dynamical model and corresponding distance. The same is
true if we use a different polar grid, with fewer or more stars
per aperture, and if we restrict to only fitting the average
kinematics in the inner or outer parts. Still, e.g. remaining
interlopers in the proper motion data-set can cause a (small)
under-estimation of the distance. Moreover, Platais et al.\
(2003\nocite{2003ApJ...591L.127P}) argue that possibly a
(non-physical) residual proper motion colour/magnitude dependence
in the data-set of Paper~I causes the systematic offset between
the proper motions of the metal-rich RGB-a stars and those of the
dominant HB and metal-poor RGB stars, noticed by Ferraro,
Bellazzini \& Pancino (2002\nocite{2002ApJ...573L..95F}). Since we
do not correct for this possible systematic offset, the proper
motion dispersion might be over-estimated and hence our distance
estimate can be systematically too low. However, the effect is
expected to be small since the number of RGB-a stars in the
data-set is small. A deeper proper motion catalogue, like that of
King \& Anderson 2002\nocite{2002ocuw.conf...21K}) obtained with
the HST, is needed to better quantify (non-physical and
physical) differences in the proper motions among the multiple
stellar populations observed in \oc.

Although the distance and inclination are tightly linked through
the mean velocities (\S~\ref{sec:constraintincl}), a small
under-estimation of the distance only results in a slight
over-estimation of the inclination (see also the solid curve in
the right panel of Figure~\ref{fig:sbr}). Similarly, the
mass-to-light ratio is nearly insensitive to small changes in the
distance.


\section{Intrinsic structure}
\label{sec:intstructure}

We use the intrinsic velocity moments of our best-fit dynamical
model to investigate the importance of rotation and the degree of
anisotropy in \oc. Additionally, the distribution of the orbital
weights allows us to study the phase-space distribution function
of \oc.

\begin{figure}[t]
\includegraphics{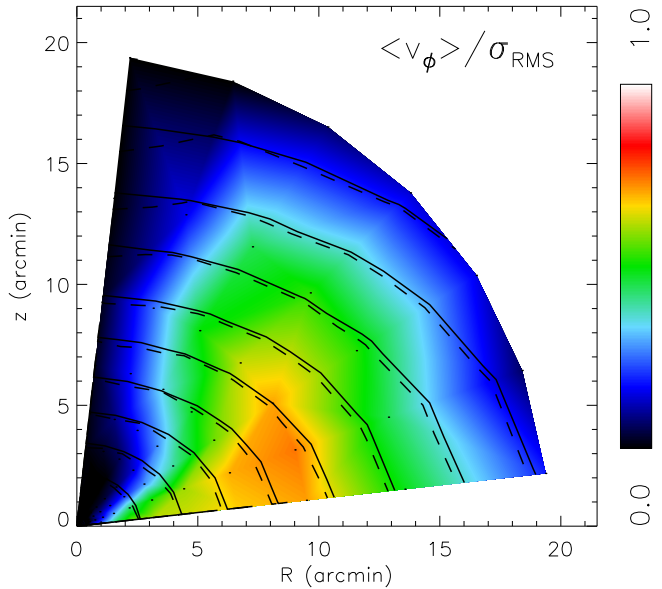}
\caption{
  The colours represent the mean azimuthal rotation
  $\langle v_\phi\rangle$ in the meridional plane as a function of
  equatorial plane radius $R$ and height $z$, and normalised by
  $\sigma_\mathrm{RMS}$ (excluding the axes to avoid numerical
  problems). The black curves are contours of constant mass density in
  steps of one magnitude, from the mass model (solid) and from the
  best-fit model (dashed), showing that the mass is well fitted.}
\label{fig:introtation}
\end{figure}

\subsection{Rotation}
\label{sec:rotation}

We calculate the intrinsic velocity moments of our best-fit model by
combining the appropriate moments of the orbits that receive weight in
the superposition. We consider the first and second order velocity
moments, for which $\langle v_R \rangle = \langle v_\theta \rangle =
\langle v_R v_\phi \rangle = \langle v_\theta v_\phi \rangle = 0$
because of axisymmetry. We define the radial, angular and azimuthal
velocity dispersion respectively as $\sigma^2_R=\langle v_R^2
\rangle$, $\sigma^2_\theta=\langle v_\theta^2 \rangle$,
$\sigma^2_\phi=\langle v_\phi^2 \rangle - \langle v_\phi \rangle^2$.
The only non-vanishing cross-term is $\sigma^2_{R\theta} = \langle v_R
v_\phi \rangle$. The total root-mean-square velocity dispersion
$\sigma_\mathrm{RMS}$ is given by $\sigma^2_\mathrm{RMS}= (\sigma^2_R
+\sigma^2_\theta + \sigma^2_\phi)/3$.

A common way to establish the importance of rotation in elliptical
galaxies and bulges of disk galaxies, is to determine their
position in the $(V/\sigma,\epsilon)$-diagram (e.g. Davies et al.\
1983\nocite{1983ApJ...266...41D}). The \textit{observational}
quantities that are used for $V$, $\sigma$ and $\epsilon$ are
respectively the maximum (line-of-sight) velocity along the major
axis, the average velocity dispersion within half the effective
radius and the ellipticity at the effective radius. We obtain for
\oc\ the observational quantities $V\sim8$ \kms\ (at a radius of
$\sim8$ arcmin), $\sigma\sim16$ \kms\ and $\epsilon\sim0.15$
(Geyer et al.\ 1983\nocite{1983A&A...125..359G}). These values
result in $(V/\sigma,\epsilon)\sim(0.5,0.15)$, placing \oc\ just
above the curve for isotropic oblate rotators.

On the other hand, the intrinsic velocity moments from our best-fit
dynamical model for \oc, allow us to investigate
\textit{intrinsically} the importance of rotation. The colours in
Figure~\ref{fig:introtation} show the ratio of the mean (azimuthal)
rotation $\langle v_\phi\rangle$ over the total root-mean-square
velocity dispersion $\sigma_\mathrm{RMS}$, as function of the position
in the meridional plane. Near the equatorial plane and between radii
of about 5 to 15 arcmin, this ratio is $>0.5$.  The maximum of
$\sim0.7$ around 8 arcmin coincides with the peak in the mean
line-of-sight velocity field. Within this region in the meridional
plane rotational support is important. However, more inwards and
further outwards this ratio rapidly drops below 0.5 and \oc\ is at
least partly pressure supported. We conclude that rotation is
important in \oc, but it is not a simple isotropic oblate rotator.

\subsection{Anisotropy}
\label{sec:anisotropy}

\begin{figure}[t]
\includegraphics{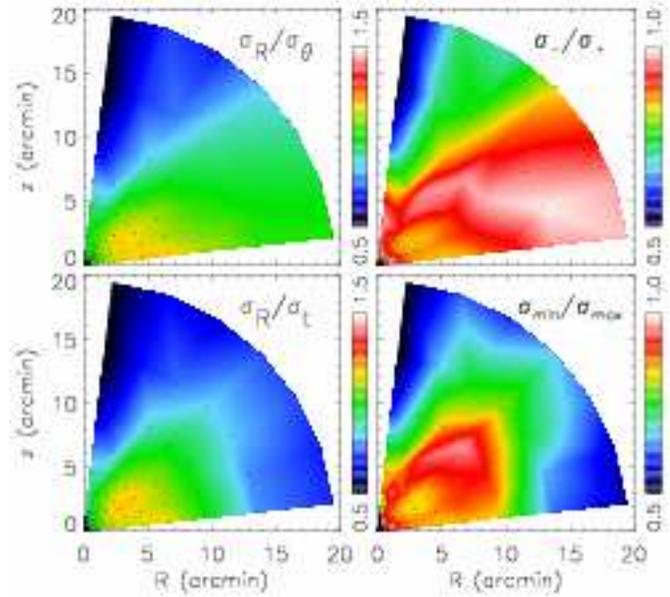}
\caption{
  Degree of anisotropy as function of the equatorial plane
  radius $R$ and height $z$ (excluding the axes to avoid numerical
  problems). The upper panels show the degree of anisotropy in the
  meridional plane: left the radial over the angular velocity
  dispersion and right the minor $\sigma_-$ over the major $\sigma_+$
  semi-axis length of the velocity ellipsoid, taking into account the
  cross-term $\sigma_{R\theta}$. The bottom panels include the
  azimuthal velocity dispersion: left the radial over the tangential
  velocity dispersion, with $\sigma^2_t = (\sigma^2_\theta +
  \sigma^2_\phi)/2$, and right the minimum over the maximum of the
  three semi-axis lengths $\sigma_+$, $\sigma_-$ and $\sigma_\phi$ of
  the velocity ellipsoid. See text for further details.}
\label{fig:intanisotropy}
\end{figure}

For the velocity distribution in \oc\ to be isotropic all three
velocity dispersion components $\sigma_R$, $\sigma_\theta$ and
$\sigma_\phi$ have to be equal and the cross-term
$\sigma_{R\theta}$ has to vanish. Figure~\ref{fig:intanisotropy}
shows that this is not the case.

In the upper panels, we show the degree of anisotropy in the
meridional plane. The top-left panel shows the radial over the angular
velocity dispersion $\sigma_R/\sigma_\theta$. This ratio does however
not include the non-zero cross-term $\sigma_{R\theta}$. The latter
causes the velocity ellipsoid to be rotated with respect to the $R$
and $\theta$ coordinates. Taking this into account the semi-axis
lengths of the velocity ellipsoid in the meridional plane are given by
$\sigma^2_\pm = (\sigma_R^2 +\sigma^2_\theta)/2 \pm \sqrt{(\sigma_R^2
  -\sigma^2_\theta)^2/4 + \sigma_{R\theta}^4}$. In the top-right
panel, we show the ratio of this minor $\sigma_-$ and major $\sigma_+$
semi-axis length of the velocity ellipsoid (which is by definition in
the range from zero to unity). This demonstrates that the velocity
distribution of \oc\ is nearly isotropic near the equatorial plane,
but becomes increasingly tangential anisotropic towards the symmetry
axis.

In the bottom panels we also include the azimuthal velocity
dispersion $\sigma_\phi$. The bottom-left panel shows the radial
over the tangential velocity dispersion, where the latter is
defined as $\sigma^2_t = (\sigma^2_\theta + \sigma^2_\phi)/2$.
Again this ratio does not take into account the cross-term
$\sigma_{R\theta}$. The actual degree of anisotropy is given by
the three semi-axis lengths $\sigma_+$, $\sigma_-$ and
$\sigma_\phi$ of the velocity ellipsoid. In the bottom-right
panel, we show, as a function of the position in the meridional
plane, the minimum over the maximum of these three semi-axis
lengths. Except for the region near the equatorial plane and
within 10 arcmin, the best-fit model for \oc\ is clearly not
isotropic. Even within this region, between about 3 and 5 arcmin,
it is (slightly) radially anisotropic. Outside this region \oc\
becomes increasingly tangentially anisotropic.

Clearly, isotropic models are not suitable to model \oc. Also
dynamical models with a two-integral distribution function of the form
$F(E,L_z)$, with $L_z=R\langle v_\phi \rangle$ the angular momentum
component along the symmetry $z$-axis, are not able to describe the
complex dynamical structure of \oc. For these models the solution of
the Jeans equations can be used to construct dynamical models in a
straightforward way (e.g. Satoh 1980\nocite{1980PASJ...32...41S};
Binney, Davies \& Illingworth 1990\nocite{1990ApJ...361...78B}) and
they allow for azimuthal anisotropy. However, for these models
$\sigma_R=\sigma_\theta$ and $\sigma_{R\theta}=0$, i.e. isotropy in
the full meridional plane, which is not the case for \oc\ (upper
panels of Figure~\ref{fig:intanisotropy}). Our axisymmetric dynamical
models do not have these restrictions as they are based on a general
three-integral distribution function $F(E,L_z,I_3)$, which we
investigate next for our best-fit model.

\begin{figure*}[t]
\includegraphics{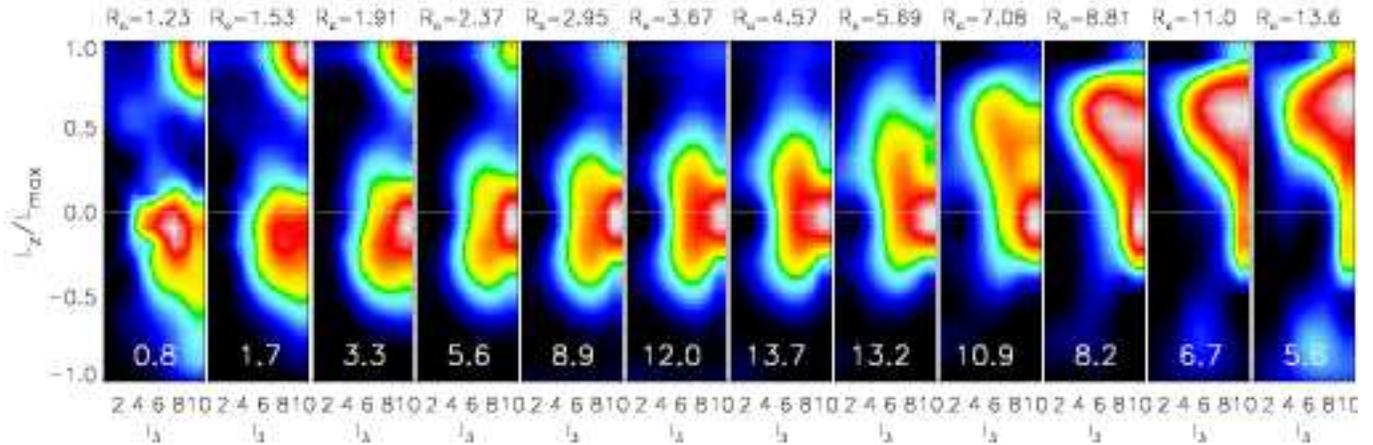}
\caption{
  The orbital weight distribution for our best-fit model of \oc.
  From left to right, the panels show the orbital weight distribution
  at increasing distance from the centre, which corresponds to
  increasing energy. The radius $R_c$ (in arcmin) of the circular
  orbit at the corresponding energy is given above each panel. The
  radial range that is shown is constrained by the observations and
  contains more than 90\% of the total cluster mass. The vertical axis
  represents the angular momentum $L_z$ in units of $L_\mathrm{max}$,
  the angular momentum of the circular orbit. The horizontal axis
  represents the third integral $I_3$, parameterised by the number of
  the (linearly sampled) starting angle of the orbit. Black shading
  corresponds to zero orbital weights, and white corresponds to the
  maximum orbital weight in each panel. At the bottom of each panel
  the fraction of the included mass with respect to the total mass is
  indicated (in \%).}
\label{fig:intspace}
\end{figure*}

\subsection{Distribution function}
\label{sec:distrfunc}

Each orbit in our models is characterised by the three integrals of
motion $E$, $L_z$ and $I_3$. As function of these three integrals, we
show in Figure~\ref{fig:intspace} for our best-fit model of \oc\ the
distribution of the (mass) weights that were assigned to the different
orbits in the NNLS-fit. The energy $E$ is sampled through the radius
$R_c$ (in arcmin) of the circular orbit (different panels), of which
we show the range that is constrained by the observations and that
contains more than 90\% of the total cluster mass.  The angular
momentum $L_z$ (vertical) is in units of $L_\mathrm{max}$, the angular
momentum of the circular orbit. The third integral $I_3$ (horizontal)
is parameterised by the linearly sampled starting angle of the orbit,
from the equatorial plane towards the symmetry axis, and of which the
number is given. \looseness=-1

In each panel, the orbital weights are scaled with respect to the
maximum orbital weight in that panel, indicated by the white colour,
whereas black corresponds to zero orbital weight. The fraction of the
sum of the mass weights in each panel with respect to total mass in
all panels is given at the bottom of each panel (in \%). To avoid an
unrealistic orbital weight distribution that fluctuates rapidly for
adjacent orbits, we regularise our models (\S~\ref{sec:fit2obs}).  For
values of the smoothening parameter below $\Delta=4$ and even without
regularisation, we find the same best-fit parameters and although the
distribution function becomes spiky, the main features of
Figure~\ref{fig:intspace} remain.

Most of the mass in the orbital weight distribution is in the
component that is prominent in all panels. With increasing radius, the
average angular momentum $L_z$ of this component increases from nearly
zero to a significant (positive) value in the outer parts.  This
reflects the outwards increasing tangential anisotropy already seen in
the bottom-left panel of Figure~\ref{fig:intanisotropy}. An almost
non-rotating part is still present beyond 5 arcmin, attached to the
rotating component, which becomes the dominant component (in mass).
There is also a separate component at $L_z/L_\mathrm{max}\sim1$ that
is clearly visible between about 1 and 3 arcmin.
Within this radial range, this maximum rotating component contributes
almost 20\% of the mass, and it includes about 4\% of the total mass,
i.e., its mass is of the order of $10^5$ \Msun.

In the right-most panels of Figure~\ref{fig:intspace} there is a
(weak) signature of a component with $L_z/L_\mathrm{max}\sim-1$, which
we expect to be a spurious feature due to insufficient observational
constraints. Whereas (nearly) circular orbits
($|L_z|/L_\mathrm{max}\sim1$) are confined in radius to $R_c$, orbits
with lower $|L_z|$ can go further inwards, so that they have most of
their contribution (their cusps) at a smaller radius than $R_c$ (e.g.
Cappellari et al.  2004\nocite{2004cbhg.sympE...5C}). Hence, the
apparent feature at $L_z/L_\mathrm{max}\sim-1$ in the most-right panel
is only constrained by data around and beyond the radius $R_c=13.6$
arcmin, where the coverage of the data is sparse with only a few polar
bins (see Figure~\ref{fig:apgrids}). The main component in this panel
at $L_z/L_\mathrm{max}\sim0.5$ is (mostly) constrained by data at
smaller radii, where there is good data coverage. The separate maximum
rotating component between 1 and 3 arcmin is constrained by only a few
proper motion apertures, but is strongly constrained by the
line-of-sight velocity data.

Due to the difference in spatial coverage between the proper
motion and line-of-sight velocity data, the two data-sets (better)
constrain different parts of the orbital weight distribution. By
fitting besides the light distribution of \oc\ the mean velocity
and velocity dispersion of only the proper motion components, we
find a less prominent separate component between 1 and 3 arcmin,
but it is still present. In the case of only fitting the mean
line-of-sight velocity and velocity dispersion, this separate
component is clearly visible and even extends into the outer
rotating main component.
The transition between the main non-rotating and rotating
component is in the case of only line-of-sight data more abrupt
than in Figure~\ref{fig:intspace}. However, the proper motion
data, which has a better coverage in the outer parts, shows a
similar smooth transition. We conclude that, although the spatial
coverage is different, both data-sets give rise to the same main
features in the orbital weight distribution.

\begin{figure*}[th!]
\includegraphics{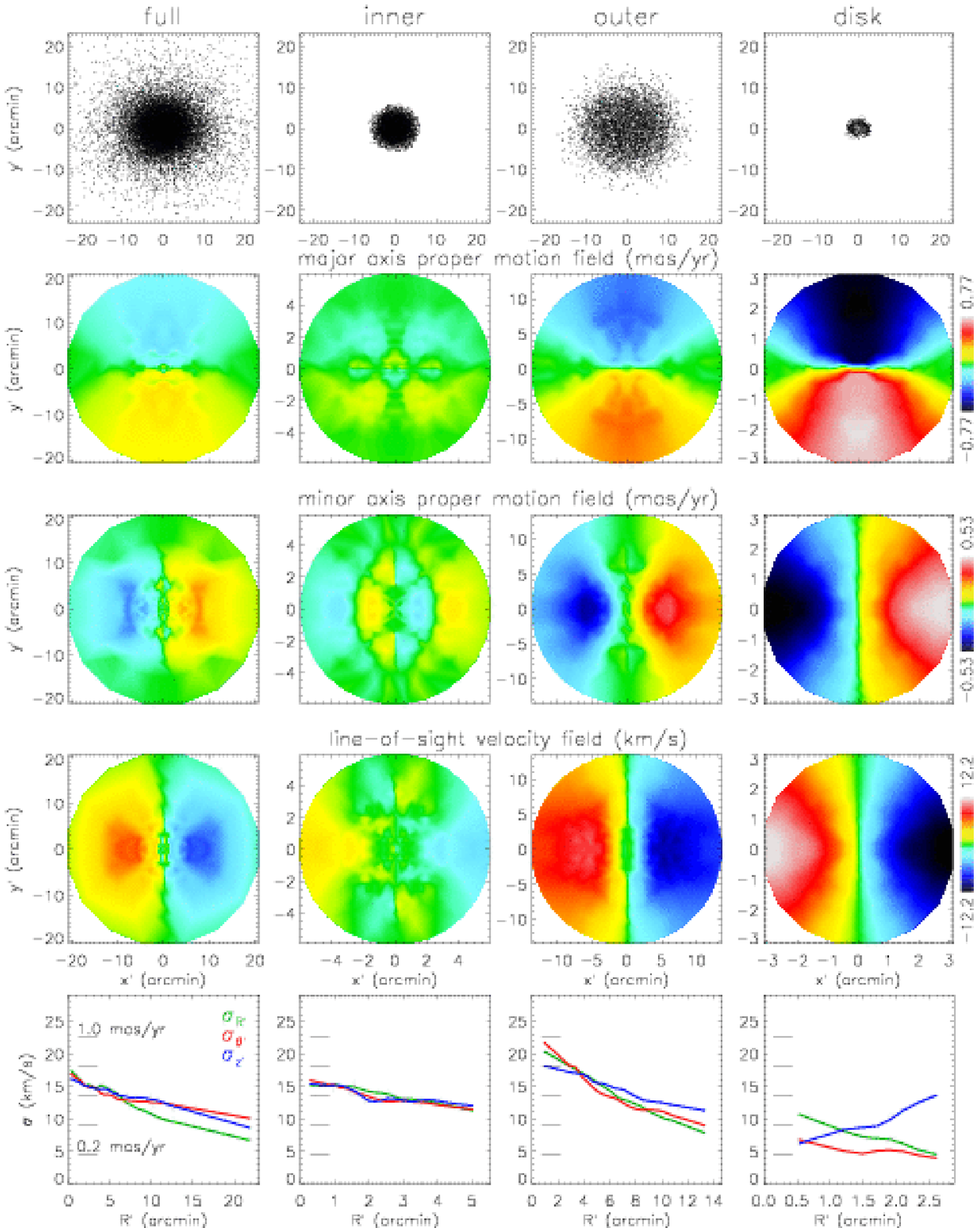}
\caption{
  Kinematics of different components in the distribution function of
  our best-fit model for \oc. From left to right: full distribution
  function, main inner component, main outer component and separate
  disk component between 1 and 3 arcmin (see text for details). From
  top to bottom: spatial distribution, mean velocity fields in the
  direction of the major $x'$-axis, the minor $y'$-axis and the
  line-of-sight $z'$-axis, and mean velocity dispersion profiles. The
  radial dispersion $\sigma_{R'}$ (green) and tangential dispersion
  $\sigma_{\theta'}$ (red) are on the plane of the sky and
  $\sigma_{z'}$ (blue) is the line-of-sight dispersion.  }
\label{fig:df_comp}
\end{figure*}

\subsection{Dynamical substructures}
\label{sec:DFsubstructures}

Within 5 arcmin the main component has on average a high value of
$I_3$. In combination with the low value of $L_z$, we interpret this
as a non-rotating spheroidal structure. Beyond 5 arcmin, $L_z$
increases and $I_3$ decreases, and the main component flattens and
rotates faster. The smaller component attached to it may well be the
signature of the fading non-rotating spheroidal component. 

For the separate component between 1 and 3 arcmin, $L_z$ approaches
its maximum value. As a result, the zero-velocity curve shrinks
towards the circular orbit in the equatorial plane, and the
corresponding orbits are all flat, irrespective of the (high) value of
$I_3$ (see also Figure~3 of Cretton et al.\ 
1999\nocite{1999ApJS..124..383C}). Hence, this fast-rotating component
is likely to be an inner disk, which fades away into the more massive
main rotating component at larger radius.

We compute the spatial distribution and average kinematics of these
possible substructures in the phase-space of \oc. To this end we
select the orbits from our best-fit model that contribute non-zero
weight to three different parts of the distribution function in
Figure~\ref{fig:intspace}. We select the \textit{inner} main component
in the 7 left-most panels, excluding the separate \textit{disk}
component in the 5 left-most panels, and the \textit{outer} main
component in the 3 right-most panels (excluding the weak feature in
the bottom). For each orbit with non-zero weight, we then randomly
draw points along its numerically integrated orbit, with the number of
drawings proportional to its relative weight. In this way, we make an
(N-body) realisation of our best-fit model consisting of a couple of
tens of thousands of particles, representing the stars in \oc.  For
each of these stars, we determine the position on the plane of the sky
and the three velocity components; the two proper motion components in
the plane of the sky and the line-of-sight velocity.  For the stars
that belong to a certain part or substructure of phase-space, we then
calculate the spatial distribution and mean kinematics.

Figure~\ref{fig:df_comp} shows the results for all stars, those in the
inner and outer main component and those in the separate disk
component, respectively, per column from left to right.  The first row
shows the spatial distribution. The flattening of the spatial
distribution of all stars and of the outer main component are both
about 0.88, similar to the average observed flattening for \oc.  The
inner main component, going out to a radius of about 6 arcmin, is
rounder with a flattening of about 0.94. The spatial distribution of
the disk component only extends to a radius of about 3 arcmin, has an
average flattening as lows as 0.60 and is less dense in the centre as
this maximum rotating disk consists of stars on (nearly) circular
orbits which avoid the centre.  The second to fourth row show the mean
velocity fields in respectively the direction of the major $x'$-axis
and the minor $y'$-axis on the plane of the sky and the line-of-sight
$z'$-axis. In each panel the axes are scaled with respect to the
spatial extent of each component.  Whereas the inner main component
indeed hardly show any rotation, the outer main component clearly
rotates and the separate disk component rotates even faster.  In the
last row, the velocity dispersion profiles are presented, radial
(green) and tangential (red) on the plane of the sky and along the
line-of-sight (blue). Even though the outer main component is flatter
and rotates faster than the inner main component, it is not
kinematically colder due to the mixture of orbits with different $L_z$
values. On the other hand, the maximum rotating disk is the
kinematically coldest component. Whereas the inner main component is
nearly isotropic, the outer main component is anisotropic and the disk
component is even stronger anisotropic.

The presence of dynamical substructures implies that the formation
history of \oc\ is more complicated than expected for a typical
globular cluster.  However, the interpretation of these different
components in the distribution function is very difficult. In what
follows we investigate the possible effects due to the tidal
interaction between \oc\ and the Milky Way
(\S~\ref{sec:tidalinteraction}), and the possible link to the observed
multiple stellar populations in \oc\ (\S~\ref{sec:multstellarpop}).

\subsection{Tidal interaction}
\label{sec:tidalinteraction}

Based on its current position and motion in the Milky Way (MW),
Dinescu et al.\ (1999\nocite{1999AJ....117.1792D}) simulated the orbit
of \oc\ around the Galactic Centre (GC). They found that the average
orbit is inclined by only 17\dgr\ with respect to the Galactic plane,
has a period of $P\sim122$ Myr and an angular momentum of about 406
kpc\,\kms. Assuming that the average orbit of \oc\ is circular, we
thus find a radius $R_\mathrm{OC}\sim2.8$ kpc and a velocity of about
143 \kms, of which the component perpendicular to Galactic plane
$v_\perp\sim42$ \kms. Since the scale height of the MW disk is
typically 250 pc, it takes about $t_\mathrm{enc}\sim12$ Myr for \oc\ 
to cross the MW disk.  This means that for nearly 10\% of its time
\oc\ is immersed in the disk and feels the additional gravitational
field.

Based on its current position and motion in the Milky Way (MW),
Dinescu et al.\ (1999\nocite{1999AJ....117.1792D}) simulated the orbit
of \oc\ around the Galactic Centre (GC). They found that the average
orbit is eccentric ($e\sim0.68$), is inclined by only 17\dgr\ with
respect to the Galactic plane and has a period of $P\sim122$ Myr. In
combination with the orbital angular momentum of about 406 kpc\,\kms,
this implies a mean circular radius of $R_\mathrm{OC}\sim2.8$ kpc and
corresponding circular velocity of 143 \kms. The velocity
perpendicular to Galactic plane is on average thus $v_\perp\sim42$
\kms. Since the scale height of the MW disk is typically 250 pc, it
takes about $t_\mathrm{enc}\sim12$ Myr for \oc\ to cross the MW disk.
This means that for nearly 10\% of its time \oc\ is immersed in the
disk and feels the additional gravitational field.

To investigate what effect the MW tidal field has on the stars in \oc,
we use the impulse approximation as described by Binney \& Tremaine
(1987\nocite{1987gady.book.....B}, p.~446), with the typical
properties of the MW from their Tables~1-1~and~1-2. We assume a
Cartesian coordinate system with its origin at the centre of \oc\ and
the $z$-axis perpendicular to the MW disk. If \oc\ goes through the MW
disk, the effect on the velocity component perpendicular to the disk
is the largest. Hence, the velocity of a cluster star changes on
average by $|\Delta v|\sim z |g_z(R)|/v_\perp$, where $g_z$ is the
$z$-component of the gravitational field of the MW disk. The
cumulative effect of successive passages through the MW disk becomes
of the order of the (local) velocity dispersion $\sigma$ of
  the cluster on a timescale of $t_\mathrm{shock} \sim
P\sigma^2v_\perp^2/(8z^2g_z^2)$.

An infinite disk with surface density $\Sigma$ generates a
gravitational field $g_z=2\pi G\Sigma$. In the solar neighbourhood the
MW disk has a surface density of $\Sigma_\odot\sim75$ \Msun\,pc$^{-2}$.
Assuming that the MW disk falls off as $\exp(-R/R_d)$ in the radial
coordinate, with $R_d=3.5$ kpc , we find that at the mean circular
radius $R=R_\mathrm{OC}$ of \oc's orbit around the GC,
$g_z\sim2.9\times10^{-13}$ km\,s$^{-2}$. For a spherical shell of
stars of radius $r$, we have that on average $z^2=r^2/3$. We thus find
that the timescale on which disk shocking becomes important is
\begin{equation}
  \label{eq:t_shock}
   t_\mathrm{shock} \sim 21
   \left(\frac{\sigma}{\mathrm{km\,s}^{-1}}\right)^2
   \left(\frac{r}{\mathrm{arcmin}}\right)^{-2}
   \quad \mathrm{Myr}.
\end{equation}

\begin{figure}
\includegraphics{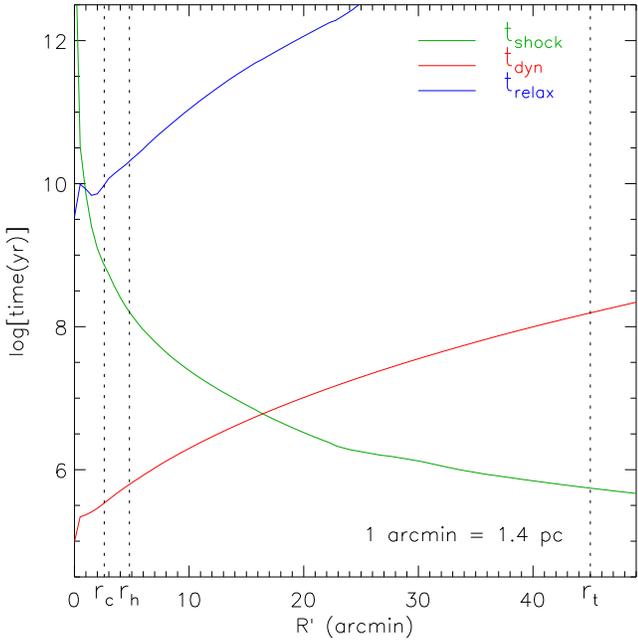}
\caption{
  Timescales as function of the projected radius $R'$.
  The green curve represents the timescale on which shocks, caused by
  successive passages of \oc\ through the MW disk, change the
  mean-squared velocity of a cluster star by the order of the
  (local) velocity dispersion of the cluster. The red and
  blue curves show respectively the dynamical time $t_\mathrm{dyn}$
  and relaxation time $t_\mathrm{relax}$. The vertical dashed lines
  indicate with increasing distance the core radius $r_t$, the
  half-light radius $r_h$ and the tidal radius $r_t$ of \oc. }
\label{fig:timescales}
\end{figure}

Figure~\ref{fig:timescales} shows $t_\mathrm{shock}$ (green curve) as
function of the projected radius $R'$ (in arcmin). We used the
line-of-sight velocity dispersion as given in
Figure~\ref{fig:dispprofile}, smoothed and extrapolated to larger
radii with the help of measurements by Scarpa, Marconi \& Gilmozzi
(2003\nocite{2003A&A...405L..15S}) between about 20 and 30
arcmin\footnote{Taking into account the measurement error of about 1
  \kms\ and the perspective rotation that can be as large as 1.5 \kms\ 
  at those radii (eq.~\ref{eq:velpr}).}. In the same figure we have
also plotted the dynamical time $t_\mathrm{dyn}$ (red curve; Binney \&
Tremaine 1987\nocite{1987gady.book.....B}, eq.~2-30) and the
relaxation time $t_\mathrm{relax}$ (blue curve; Spitzer \& Hart
1971\nocite{1971ApJ...164..399S}; Binney \& Tremaine
1987\nocite{1987gady.book.....B}, eq.~8-71). The three vertical dashed
lines indicate respectively the core radius $r_c=2.6$ arcmin, the
half-light radius $r_h=4.8$ arcmin and the the tidal radius $r_t=45$
arcmin (e.g.  Trager et al.\ 1995\nocite{1995AJ....109..218T}).

Clearly, the impulse approximation is not valid near the centre of
\oc, where the period of the stellar orbits $T\equiv4\,t_\mathrm{dyn}$
is much smaller than the duration of the passage through the disk
$t_\mathrm{enc}\sim12$ Myr. Disk shocking is thus unimportant at the
centre of \oc: the orbits evolve adiabatically and emerge unharmed
from the encounter. Around a radius of 16 arcmin, where $T$ is about
twice $t_\mathrm{enc}$, disk shocks begin to play an important role
since the disk shocking time becomes of the order of the dynamical
time $t_\mathrm{shock} \sim t_\mathrm{dyn} \sim 6$ Myr. At the tidal
radius of 45 arcmin, the MW disk gravitational field becomes dominant.

The effect that the MW tidal field has on the internal dynamics of
\oc\ also strongly depends on the relative orientation and spinning
direction of the angular momentum vector of the stars in \oc\
(internal) and the angular momentum vector of its orbit around the GC
(external). We found that the rotation axis is about 50\dgr\ inclined
with respect the line-of-sight (the $z'$-axis) in the direction
South\footnote{This means that in the common definition of the
  inclination, as in eq.~\eqref{eq:poscartint2obs}, the best-fit
  inclination is -50\dgr.  This also explains the sign difference of
  $\langle v_{z'} \rangle$ in eq.~\eqref{eq:Dtani} and along the
  vertical axis of the plot in the middle panel of
  Figure~\ref{fig:sbr}. However, we decided to adopt the usual
  convention to take the value for the inclination in the range from
  0\dgr\ (face-on) to 90\dgr\ (edge-on).}.  On the plane of sky, the
rotation axis projects onto the minor $y'$-axis, which makes an angle
of about 10\dgr\ away from North in the direction East. The equatorial
coordinates of \oc\ are $\alpha_0=13^h26^m46^s$ and
$\delta_0=-47^\circ28^\prime43^{\prime\prime}$ (J2000), which
correspond to a Galactic longitude and latitude of $l=309$\dgr\ and
$b=15$\dgr. Hence, the rotation axis is nearly parallel (angle
$<3$\dgr) to the equatorial plane, and makes an angle of about 65\dgr\
with respect to the Galactic plane. Seen from the North Galactic pole,
\oc\ is moving in anti-clockwise direction around the GC. The rotation
inside \oc\ is dominated by orbits with positive $L_z$ values in
Figure~\ref{fig:intspace}, which correspond to clockwise rotation.

We thus find that the internal and external angular momentum vector
are for more than 90\% parallel with respect to each other with
opposite spinning direction. From mergers of spinning galaxies it is
well known that if the spins are anti-parallel as in this case, the
orbital disruption is much less than in the case of parallel spins
(e.g. Toomre \& Toomre 1972\nocite{1972ApJ...178..623T}). Hence, in
the past \oc\ might have contained a significant number of stars on
orbits with negative $L_z$ (parallel spin), which then were removed
from the cluster during its successive passages through the MW disk.
On the other hand, stars on orbits with positive $L_z$ (anti-parallel
spin) had a bigger chance to survive.

Furthermore, the stars on more radial orbits (those with smaller
values of $L_z$) cover a broader range in radius, with the influence
of the MW tidal field becoming stronger at increasing radius. In the
course of time, these radial orbits thus have a bigger chance of being
disrupted than the more tangential orbits with similar mean radius.

Both effects (together) might explain the prominent rotating main
component in the distribution function in Figure~\ref{fig:intspace}
beyond a radius of 10 arcmin, while the non-rotating main component
that dominates inwards, fades away. The removal of the more radial
orbits also naturally explains the outwards increasing tangential
anisotropy in our best-fit model of \oc\ (\S~\ref{sec:anisotropy}).

The above analysis shows that the frequent passages of \oc\ through
the MW disk most likely have played a crucial role in the evolution of
this cluster. At least part of the phase-space structure of \oc\ may
well be caused by the tidal field of the MW.  Detailed (N-body)
simulations are needed to further quantify this.

\subsection{Multiple stellar populations}
\label{sec:multstellarpop}

Among the Galactic globular clusters, \oc\ especially stands out
because of its chemical inhomogeneity, first revealed in photometric
investigations by Dickens \& Woolley
(1967\nocite{1967RGOB..128..255D}) and spectroscopically confirmed by
Freeman \& Rodgers (1975\nocite{1975ApJ...201L..71F}). Besides the
main population of metal-poor stars ($\sim$65\% of all stars with
[Ca/H]$\sim-1.4$) and an intermediate population ($\sim$30\%,
[Ca/H]$\sim-1.0$), recently also a separate metal-rich population
($\sim$5\%, [Ca/H]$\sim-0.5$) has been identified (Lee et al.\
1999\nocite{1999Natur.402...55L}; Pancino et al.\
2000\nocite{2000ApJ...534L..83P}), and
even the main sequence of \oc\ is bifurcated (Bedin et al.\
2004\nocite{2004ApJ...605L.125B}).

Theses different stellar populations also appear to have a different
spatial distribution. Whereas the metal-poor stars seems to follow the
observed flattening of \oc\ in the East-West direction, the more
metal-rich stars are elongated in the North-South direction and also
more centrally concentrated (e.g.  Pancino et al.\
2003\nocite{2003MNRAS.345..683P}). There are also indications of
differences in the kinematics of the stellar populations. Norris et
al.\ (1997\nocite{1997ApJ...487L.187N}) find that the metal-poor
populations have on average a higher line-of-sight velocity dispersion
and exhibit a well-defined line-of-sight rotation, while the
metal-rich populations show no significant rotation. Ferraro,
Bellazzini \& Pancino (2002\nocite{2002ApJ...573L..95F}) claim that
the separate metal-rich population has a coherent bulk proper motion
significantly different from the other cluster stars.

We use the empirical relation in eq.~(15) of Paper I to estimate the
[Ca/H] abundances of stars in our analysis with $V$-band magnitude and
$B-V$ colour measurements consistent with the top of the red giant
branch ($V<13.5$ and $B-V>0.7$). The resulting [Ca/H] histograms for
the proper motion and line-of-sight velocity stars both show a
distribution with a broad peak around [Ca/H]$\sim-1.2$ and a long tail
extending beyond [Ca/H]$\sim-0.5$. In both cases the peak shows a
small dip, so that we might divide the stars into a metal-poor
population with [Ca/H]$\le-1.2$ and a metal-rich population with
[Ca/H]$>-1.2$, similar to Norris et al.\ 
(1997\nocite{1997ApJ...487L.187N}).

Comparing the mean line-of-sight kinematics of the metal-poor and
metal-rich stars, we confirm the result of Norris et al.\
(1997\nocite{1997ApJ...487L.187N}) that the more centrally
concentrated metal-rich stars are on \textit{average} kinematically
cooler and nearly non-rotating. The line-of-sight velocity dispersion
\textit{profile} is steeper for the metal-richer stars than for the
metal-poor stars, such that that in the centre the metal-richer stars
are even (slightly) kinematically warmer. The proper motions seems to
imply a similar difference in the slope of the velocity dispersion
profiles. However, with the proper motion errors on average four times
larger than those of the line-of-sight velocities (see also
Figure~\ref{fig:dispprofile}), there are no significant differences
between the kinematics of the metal-poor and metal-rich stellar
populations.

The above correlations between the kinematics and chemical properties
of stars in \oc, are expected to show up in the distribution function
(see also Freeman 2002\nocite{2002ASPC..265..423F}). The centrally
concentrated non-rotating metal-rich stars would lie near the bottom
of the potential well at the lower values of $E$ found in the cluster,
symmetrically distributed over positive and negative values of $L_z$,
and towards higher values of $I_3$. The rotating metal-poor stars
would span the entire range of $E$, with an asymmetric distribution in
$L_z$ and towards lower $I_3$.

These expectations are consistent with the orbital weight distribution
of our best-fit dynamical model of \oc\ 
(Figure~\ref{fig:intspace}~and~\ref{fig:df_comp}). Whereas the
metal-richer stars might well be associated with the inner
non-rotating part of the main component, we might see the kinematical
signatures of the metal-poorer stars becoming dominant when the main
component flattens and rotates faster in the outer parts. Still, we
have to be careful as these are (indirect) indications of a link
between substructures in the distribution function and the different
stellar populations.

To investigate directly the distribution function of the different
stellar populations, once can try to construct separate dynamical
(Schwarzschild) models. However, since the separation into different
stellar populations is not evident, separate mass models are needed
and the separate kinematic constraints are based on much fewer stars,
this is very difficult with the current data-set. A more feasible
approach would be to model together, in a consistent way, the observed
kinematics and physical properties of the stars. For example, by
labelling the orbits in the model with different colours, the observed
colour (averaged per aperture) can be used to constrain the model in
addition to the photometry and kinematics. On the other hand, now that
we have constrained the global parameters (distance, inclination and
mass-to-light ratio) considerably, it has become feasible to use
non-linear maximum likelihood techniques to directly incorporate
discrete stellar measurements. In this way, for the model that best
fits (simultaneously) the measured kinematics and age and metallicity
indicators of individual stars, the different stellar populations can
be cleanly separated in phase-space. This extension, which we leave
for a future paper, will provide an important contribution to solving
the stellar population puzzle in \oc, and clarify its formation
history.


\section{Conclusions}
\label{sec:conclusions}

We used an extension of Schwarzschild's
(1979\nocite{1979ApJ...232..236S}) orbit superposition method to
construct realistic axisymmetric dynamical models for \oc\ with an
arbitrary anisotropic velocity distribution. By fitting these models
simultaneously to proper motion and line-of-sight velocity
measurements, we measured the radial mass-to-light profile, the
inclination and the distance to \oc, which is needed to convert the
proper motions to physical units. This dynamical distance estimate can
provide a useful calibration for the photometric distance ladder.

We used the ground-based proper motions from Paper~I and the
line-of-sight velocities from four independent data-sets. We
brought the kinematic measurements onto a common coordinate system
and carefully selected on cluster membership and on measurement
error. This provided a homogeneous data-set of 2295 stars
with proper motions accurate to 0.20 \masyr\ and 2163 stars with
line-of-sight velocities accurate to 2 \kms, covering a radial
range out to about half the tidal radius of the cluster. We
corrected the kinematic measurements for perspective rotation and
removed a residual solid-body rotation component in the proper
motions. We showed that the latter can be measured without any
modelling other than assuming axisymmetry and at the same time we
obtained a tight constraint on $D\tan i$ of 5.6 (+1.9/-1.0) kpc,
providing a unique way to estimate the inclination $i$ of a nearly
spherical object once the distance $D$ is known. The corrected
mean velocity fields are consistent with regular rotation, and the
mean velocity dispersions display significant deviations from
isotropy.

We binned the individual measurements on the plane of the sky to
search efficiently through the parameter space of the models.
Tests on an analytic model demonstrated that our approach is
capable of measuring the cluster distance to an accuracy of about
6 per cent. Application to \oc\ revealed no dynamical evidence for
a significant radial dependence of the ($V$-band) stellar
mass-to-light ratio $M/L_V$, in harmony with the relatively long
relaxation time of the cluster. We found that our best-fit
dynamical model has $M/L_V=2.5\pm0.1$ \MLsun\ and
$i=50^\circ\pm4^\circ$, which corresponds to an average intrinsic
axial ratio of $0.78\pm0.03$. The best-fit dynamical distance
$D=4.8\pm0.3$ kpc (distance modulus $13.75\pm0.13$ mag) is
significantly larger than obtained by means of simple spherical or
constant-anisotropy axisymmetric dynamical models, and is
consistent with the canonical value $5.0\pm0.2$ kpc obtained by
photometric methods. The total mass of the cluster is
$(2.5\pm0.3)\times10^6$ \Msun.

Schwarzschild's approach also provides an insight into the intrinsic
orbital structure of the cluster. Our best-fit model implies that \oc\ 
is close to isotropic inside a radius of about 10 arcmin and becomes
increasingly tangentially anisotropic in the outer region, which
displays significant mean rotation. We found that this may well be
caused by the effects of the tidal field of the Milky Way.
Furthermore, the best-fit model contains a separate disk-like
component between 1 and 3 arcmin, contributing about 4\% to the total
mass. This phase-space structure, which might be linked to the
multiple stellar populations observed in \oc, is expected to provide
important constraints on its formation history.

We might improve our best-fit dynamical model of \oc\ and better
constrain the distance and the other parameters, by extending the
data-set with for example proper motions derived from HST images.
Whereas with the ground-based proper motions we were unable to
probe the centre of \oc\ due to crowding, the high spatial
resolution and high sensitivity of HST, results in many proper
motion measurements in the very centre, which makes it possible to
investigate a possible central mass concentration in \oc.

We may also increase the kinematic constraints on our dynamical
models by including mean correlated and higher-order velocity
moments. With the parameter range considerably constrained, it now
becomes also feasible to use non-linear maximum likelihood
techniques to directly incorporate the discrete kinematic
measurements. These techniques not only allow correlated and
higher-order velocity moments to be included in a straightforward
way, but also provide a natural way to incorporate measurements of
age and metallicity indicators of individual stars in addition to
their photometry and kinematics. By fitting an orbit-based model
simultaneously to all these observations, different stellar
populations can be separated in phase-space, after which their
structure and dynamics can be studied separately.


We have shown that with the method described in this paper, we were
able to measure the global parameters of \oc, including its distance,
and investigate its intrinsic orbital structure. This method can also
be applied to study other globular clusters and stellar clusters in
the Milky Way, provided that accurate velocity measurements are
available. With the amount of (photometric and kinematic) data quickly
increasing, we expect this method to become an important tool to model
these stellar systems and gain insight in their formation and
evolution.


\section*{Acknowledgements}
\label{sec:acknowledgments}

This project would not have been possible without the extensive plate
collection of \oc\ obtained in the nineteen thirties by
Willem Christiaan Martin, which, together with the plates taken in the
eighties by Ken Freeman and Pat Seitzer and the massive subsequent
effort by Rudolph Le Poole and Floor van Leeuwen, produced the proper
motions reported in Paper I. The line-of-sight velocities used here
include those obtained at CTIO in the early nineties by Renate Reijns
and Pat Seitzer, as reported in Paper II. We are also grateful to Karl
Gebhardt for allowing us to use his unpublished Fabry--Perot
measurements, and to Jay Anderson \& Ivan King for providing their
preliminary HST proper motions. It is a pleasure to thank Thijs
Kouwenhoven for assistance in the initial phase of the work reported
here, to thank Jes\'us Falc\'on-Barroso, Ken Freeman, Floor van
Leeuwen, and Scott Tremaine for useful discussions and suggestions
during the course of this work, and to thank Michele Cappellari,
Richard McDermid and Anne-Marie Weijmans for a critical reading of the
0manuscript. We also thank the referee for constructive comments and
suggestions that improved the presentation of the paper.
This research was supported in part by NWO through grant
614.000.301, and by the Netherlands Research School for Astronomy
NOVA.



\begin{thebibliography}{}

\bibitem{2004ApJ...605L.125B}
{Bedin}, L.~R., {Piotto}, G., {Anderson}, J., {et~al.} 2004, \apjl, 605, L125

\bibitem{2002AJ....123..473B}
{Benedict}, G.~F., {McArthur}, B.~E., {Fredrick}, L.~W.,
{et~al.} 2002, \aj, 123, 473 
 
\bibitem{1987gady.book.....B}
{Binney}, J. \& {Tremaine}, S. 1987, {Galactic Dynamics} (Princeton, NJ,
  Princeton University Press)

\bibitem{1990ApJ...361...78B}
{Binney}, J.~J., {Davies}, R.~L., \& {Illingworth}, G.~D. 1990, \apj, 361, 78

\bibitem{2002MNRAS.333..400C}
{Cappellari}, M. 2002, \mnras, 333, 400

\bibitem{2004cbhg.sympE...5C}
{Cappellari}, M., {van den Bosch}, R.~C.~E., {Verolme}, E.~K., {et~al.} 2004,
  in Carnegie Observatories Astrophysics Series, Vol. 1, Coevolution of Black
  Holes and Galaxies, ed. L.~C. Ho
  (http://www.ociw.edu/\-ociw/\-symposia/\-series/\-symposium1/\-proceedings.html)

\bibitem{2002ApJ...578..787C}
{Cappellari}, M., {Verolme}, E.~K., {van der Marel}, R.~P., {et~al.} 2002,
  \apj, 578, 787

\bibitem{2000A&A...357..977C}
{Carraro}, G. \& {Lia}, C. 2000, \aap, 357, 977

\bibitem{1999ApJS..124..383C}
{Cretton}, N., {de Zeeuw}, P.~T., {van der Marel}, R.~P., \& {Rix}, H. 1999,
  \apjs, 124, 383

\bibitem{1979AJ.....84.1312C}
{Cudworth}, K.~M., 1979, \aj, 84, 1312

\bibitem{1979AJ.....84..505D}
{Da Costa}, G.~S. 1979, \aj, 84, 505

\bibitem{1983ApJ...266...41D}
{Davies}, R.~L., {Efstathiou}, G., {Fall}, S.~M., {Illingworth}, G., \&
  {Schechter}, P.~L. 1983, \apj, 266, 41

\bibitem{1989ApJ...343..113D}
{Dejonghe}, H. 1989, \apj, 343, 113

\bibitem{1967RGOB..128..255D}
{Dickens}, R.~J. \& {Woolley}, R.~v.~d.~R. 1967, Royal Greenwich Observatory
  Bulletin, 128, 255

\bibitem{1999AJ....117.1792D}
{Dinescu}, D.~I., {Girard}, T.~M., \& {van Altena}, W.~F. 1999, \aj, 117, 1792

\bibitem{1997ApJ...481..267D}
{Dull}, J.~D., {Cohn}, H.~N., {Lugger}, P.~M., {et~al.} 1997, \apj, 481, 267

\bibitem{1994A&A...285..723E}
{Emsellem}, E., {Monnet}, G., \& {Bacon}, R. 1994, \aap, 285, 723

\bibitem{1994A&A...285..739E}
{Emsellem}, E., {Monnet}, G., {Bacon}, R., \& {Nieto}, J.-L.
  1994, \aap, 285, 739

\bibitem{1994MNRAS.267..333E}
{Evans}, N.~W. 1994, \mnras, 267, 333

\bibitem{1994MNRAS.271..202E}
{Evans}, N.~W. \& {de Zeeuw}, P.~T. 1994, \mnras, 271, 202

\bibitem{1961MNRAS.122..433F}
{Feast}, M.~W., {Thackeray}, A.~D., \& {Wesselink}, A.~J. 1961, \mnras, 122,
  433

\bibitem{2002ApJ...573L..95F}
{Ferraro}, F.~R., {Bellazzini}, M., \& {Pancino}, E. 2002, \apjl, 573, L95

\bibitem{1993ASPC...48..608F}
{Freeman}, K.~C. 1993, in Astronomical Society of the Pacific Conference
  Series, 608

\bibitem{2002ASPC..265..423F}
{Freeman}, K.~C. 2002, in ASP Conf. Ser. 265: Omega Centauri, A Unique Window
  into Astrophysics, eds. F. van Leeuwen, J.~D. Hughes, G. Piotto, 423

\bibitem{1975ApJ...201L..71F}
{Freeman}, K.~C. \& {Rodgers}, A.~W. 1975, \apjl, 201, L71

\bibitem{1956MNRAS.116..570G}
{Gascoigne}, S.~C.~B. \& {Burr}, E.~J. 1956, \mnras, 116, 570

\bibitem{2003ApJ...583...92G}
{Gebhardt}, K., {Richstone}, D., {Tremaine}, S., {et al.} 2003,
\apj, 583, 92 

\bibitem{1993MNRAS.265..213G}
{Gerhard}, O.~E. 1993, \mnras, 265, 213

\bibitem{1983A&A...125..359G}
{Geyer}, E.~H., {Nelles}, B., \& {Hopp}, U. 1983, \aap, 125, 359

\bibitem{1996AJ....112.1487H}
{Harris}, W.~E. 1996, \aj, 112, 1487

\bibitem{2000yCat.1259....0H}
{Hog}, E., {Fabricius}, C., {Makarov}, V.~V., {et~al.} 2000, VizieR Online Data
  Catalog, 1259, 0

\bibitem{1988A&A...204..115I}
{Icke}, V. \& {Alcaino}, G. 1988, \aap, 204, 115

\bibitem{1999gaha.conf..325K}
{Kalnajs}, A.~J. 1999, in ASP Conf. Ser. 165: The Third Stromlo Symposium: The
  Galactic Halo, 325

\bibitem{1951KK}
{Kenney}, J.~F. \& {Keeping}, E.~S. 1951, {The Distribution of the Standard
  Deviation} (\S~7.8 in \textit{Mathematics of Statistics}, Pt. 2, 2nd ed.
  Princeton, NJ: Van Nostrand)

\bibitem{2002ocuw.conf...21K}
{King}, I.~R. \& {Anderson}, J. 2002, in ASP Conf. Ser. 265: Omega Centauri, A
  Unique Window into Astrophysics, eds. F. van Leeuwen, J.~D. Hughes, G.
  Piotto, 21

\bibitem{1968AJ.....73..456K}
{King}, I.~R., {Hedemann}, E.~J., {Hodge}, S.~M., \& {White}, R.~E. 1968, \aj,
  73, 456

\bibitem{2002MNRAS.330..792K}
{Kleyna}, J., {Wilkinson}, M.~I., {Evans}, N.~W., {Gilmore}, G., \& {Frayn}, C.
  2002, \mnras, 330, 792

\bibitem{2005MNRAS...krajnovic}
{Krajnovi{\' c}}, D., {Cappellari}, M., {Emsellem}, E., {McDermid}, R.~M., \&
  {de Zeeuw}, P.~T. 2005, \mnras, in press

\bibitem{1974slsp.book.....L}
{Lawson}, C.~L. \& {Hanson}, R.~J. 1974, {Solving least squares problems}
  (Prentice-Hall Series in Automatic Computation, Englewood Cliffs:
  Prentice-Hall, 1974)

\bibitem{1999Natur.402...55L}
{Lee}, Y.-W., {Joo}, J.-M., {Sohn}, Y.-J., {et~al.} 1999, \nat, 402, 55

\bibitem{2002ASPC..265..305L}
{Lee}, Y.-W., {Rey}, S.-C., {Ree}, C.~H., {et~al.} 2002, in ASP Conf. Ser. 265:
  Omega Centauri, A Unique Window into Astrophysics, eds. F. van Leeuwen, J.~D.
  Hughes, G. Piotto, 305

\bibitem{2002ocuw.conf...95L}
{Lub}, J. 2002, in ASP Conf. Ser. 265: Omega Centauri, A Unique Window into
  Astrophysics, eds. F. van Leeuwen, J.~D. Hughes, G. Piotto, 95

\bibitem{1974AJ.....79..745L}
{Lucy}, L.~B. 1974, \aj, 79, 745

\bibitem{1991A&A...252...94M}
{Mandushev}, G., {Staneva}, A., \& {Spasova}, N. 1991, \aap, 252, 94

\bibitem{1938AnLei..17b...1M}
{Martin}, W.~C. 1938, Annalen van de Sterrewacht te Leiden, Vol. XVII

\bibitem{1996oedb.conf..190M} 
{Mayor}, M., {Duquennoy}, A., {Udry}, S.  {et al.} 1996, in ASP
  Conf. Ser. 90: The Origins, Evolution, and Destinies of Binary Stars
  in Clusters, eds. E.~F. Milone, J.-C.  Mermilliod, 190

\bibitem{1997AJ....114.1087M}
{Mayor}, M., {Meylan}, G. , {Udry}, S., {et
  al.} 1997, \aj, 114, 1087 [M97] 

\bibitem{2004ApJ...602..264M}
{McNamara}, B.~J., {Harrison}, T.~E., {Baumgardt}, H., 2004, \apj, 602, 264

\bibitem{1990AJ.....99.1548M}
{Merrifield}, M.~R. \& {Kent}, S.~M. 1990, \aj, 99, 1548

\bibitem{1993ApJ...413...79M}
{Merritt}, D. 1993, \apj, 413, 79

\bibitem{1997AJ....114..228M}
---. 1997, \aj, 114, 228

\bibitem{1997AJ....114.1074M}
{Merritt}, D., {Meylan}, G., \& {Mayor}, M. 1997, \aj, 114, 1074

\bibitem{1993ApJ...409...75M}
{Merritt}, D. \& {Saha}, P. 1993, \apj, 409, 75

\bibitem{1987A&A...184..144M}
{Meylan}, G. 1987, \aap, 184, 144

\bibitem{1995A&A...303..761M}
{Meylan}, G., {Mayor}, M., {Duquennoy}, A., \& {Dubath}, P. 1995, \aap, 303,
  761

\bibitem{1992A&A...253..366M}
{Monnet}, G., {Bacon}, R., \& {Emsellem}, E. 1992, \aap, 253, 366

\bibitem{1997ApJ...487L.187N}
{Norris}, J.~E., {Freeman}, K.~C., {Mayor}, M., \& {Seitzer}, P. 1997, \apjl,
  487, L187

\bibitem{2000ApJ...534L..83P}
{Pancino}, E., {Ferraro}, F.~R., {Bellazzini}, M., {Piotto}, G., \& {Zoccali},
  M. 2000, \apjl, 534, L83

\bibitem{2003MNRAS.345..683P}
{Pancino}, E., {Seleznev}, A., {Ferraro}, F.~R., {Bellazzini}, M., \& {Piotto},
  G. 2003, \mnras, 345, 683

\bibitem{1997hity.book.....P}
{Perryman}, M.~A.~C. \& {ESA}. 1997, {The HIPPARCOS and TYCHO catalogues} (ESA
  SP 1200)

\bibitem{1994ApJ...420..612P}
{Peterson}, R.~C.,{Cudworth}, K.~M., 1994, \apj, 420, 612

\bibitem{1995ApJ...443..124P}
{Peterson}, R.~C., {Rees}, R.~F.,{Cudworth}, K.~M., 1995, \apj, 443, 124

\bibitem{2003ApJ...591L.127P}
{Platais}, I., {Wyse}, R.~F.~G., {Hebb}, L., {Lee}, Y., \& {Rey}, S. 2003,
  \apjl, 591, L127

\bibitem{Press92..numrecipies}
{Press}, W.~H., {Teukolsky}, S.~A., {Vettering}, W.~T., \& {Flannery}, B.~P.
  1992, {Numerical Recipes} (Cambridge Univ. Press, Cambridge)

\bibitem{1997ASPC..127..109R}
{Rees}, R.~F., 1997, in ASP Conf.\ Ser.\ 127: Proper Motions and
Galactic Astronomy, ed.\ R.M.\ Humphreys, 109

\bibitem{2005A&A...paperII}
{Reijns}, R.~A., {Seitzer}, P., {Arnold}, R., {et~al.} 2005, \aap, submitted
  [Paper II]

\bibitem{1988ApJ...327...82R}
{Richstone}, D.~O. \& {Tremaine}, S. 1988, \apj, 327, 82

\bibitem{1997ApJ...488..702R}
{Rix}, H., {de Zeeuw}, P.~T., {Cretton}, N., {van der Marel}, R.~P., \&
  {Carollo}, C.~M. 1997, \apj, 488, 702

\bibitem{2001ApJ...553..722R}
{Romanowsky}, A.~J. \& {Kochanek}, C.~S. 2001, \apj, 553, 722

\bibitem{1980PASJ...32...41S}
{Satoh}, C. 1980, \pasj, 32, 41

\bibitem{2003A&A...405L..15S}
{Scarpa}, R., {Marconi}, G., \& {Gilmozzi}, R. 2003, \aap, 405, L15

\bibitem{1979ApJ...232..236S}
{Schwarzschild}, M. 1979, \apj, 232, 236

\bibitem{1982ApJ...263..599S}
---. 1982, \apj, 263, 599

\bibitem{1993ApJ...409..563S}
---. 1993, \apj, 409, 563

\bibitem{1983PhDT........11S}
{Seitzer}, P.~O. 1983, Ph.D.~Thesis, Australian National Univ.

\bibitem{1971ApJ...164..399S}
{Spitzer}, L.~J. \& {Hart}, M.~H. 1971, \apj, 164, 399

\bibitem{1996AJ....111.1913S}
{Suntzeff}, N.~B. \& {Kraft}, R.~P. 1996, \aj, 111, 1913 [SK96]

\bibitem{2001AJ....121.3089T}
{Thompson}, I.~B., {Kaluzny}, J., {Pych}, W., {et~al.} 2001, \aj, 121, 3089

\bibitem{1972ApJ...178..623T}
{Toomre}, A. \& {Toomre}, J. 1972, \apj, 178, 623

\bibitem{1995AJ....109..218T}
{Trager}, S.~C., {King}, I.~R., \& {Djorgovski}, S. 1995, \aj, 109, 218

\bibitem{2002ApJ...574..740T}
{Tremaine}, S., {Gebhardt}, K., {Bender}, R., {et al.} 2002, \apj, 574, 740

\bibitem{2004MNRAS.350.1141T}
{Tsuchiya}, T., {Korchagin}, V.~I., \& {Dinescu}, D.~I. 2004, \mnras, 350, 1141

\bibitem{2004cbhg.symp...37V}
{van der Marel}, R.~P. 2004,
  in Carnegie Observatories Astrophysics Series, Vol. 1, Coevolution of Black
  Holes and Galaxies, ed. L.~C. Ho
  (http://www.ociw.edu/\-ociw/\-symposia/\-series/\-symposium1/\-proceedings.html)

\bibitem{1998ApJ...493..613V}
{van der Marel}, R.~P., {Cretton}, N., {de Zeeuw}, P.~T., \& {Rix}, H. 1998,
  \apj, 493, 613

\bibitem{1993ApJ...407..525V}
{van der Marel}, R.~P. \& {Franx}, M. 1993, \apj, 407, 525

\bibitem{2002ocuw.conf.....V}
{van Leeuwen}, F., {Hughes}, J.~D., \& {Piotto}, G., eds. 2002, {Omega
  Centauri, A Unique Window into Astrophysics, ASP Conf. Ser. 265}

\bibitem{2002ocuw.conf...41V}
{van Leeuwen}, F. \& {Le Poole}, R.~S. 2002, in ASP Conf. Ser. 265: Omega
  Centauri, A Unique Window into Astrophysics, eds. F. van Leeuwen, J.~D.
  Hughes, G. Piotto, 41

\bibitem{2000A&A...360..472V}
{van Leeuwen}, F., {Le Poole}, R.~S., {Reijns}, R.~A., {Freeman}, K.~C., \& {de
  Zeeuw}, P.~T. 2000, \aap, 360, 472 [Paper I]

\bibitem{1984ApJ...287..475V}
{Vandervoort}, P.~O. 1984, \apj, 287, 475

\bibitem{1979A&AS...37..333V}
{Vasilevskis}, S., {van Leeuwen}, F., {Nicholson}, W., \& {Murray}, C.~A. 1979,
  \aaps, 37, 333

\bibitem{2002MNRAS.331..959V}
{Verolme}, E.~K. \& {de Zeeuw}, P.~T. 2002, \mnras, 331, 959

\bibitem{2002MNRAS.335..517V}
{Verolme}, E.~K., {Cappellari}, M., {Copin}, Y., {et al.} 2002, \mnras, 335, 517

\bibitem{1987ApJ...317..246W}
{White}, R.~E. \& {Shawl}, S.~J. 1987, \apj, 317, 246

\bibitem{1966ROAn....2....1W}
{Woolley}, R.~v.~d.~R. 1966, Royal Observatory Annals, 2, 1

\end{thebibliography}


\appendix


\section{Maximum likelihood estimation velocity moments}
\label{sec:mlvelocitymoments}

We use the average kinematics of stars that fall within
apertures on the plane of the sky. This is comparable
to the kinematics from the integrated spectra of galaxies in an
aperture. A very important difference is, however, that we have to
take into account the errors on the individual velocity measurements.

A possible way to measure the mean velocity and velocity dispersion,
is to fit a Gaussian distribution to the velocity histogram of the
stars that fall within an aperture. Whereas the mean velocity $V$ is
well estimated, the best-fit mean velocity dispersion
$\sigma_\mathrm{fit}$ is too large, as the Gaussian distribution is
broadened due to the velocity errors. This additional 'instrumental'
dispersion $\sigma_\mathrm{ins}$ can be estimated by the mean of the
velocity errors. The corrected mean velocity dispersion $\sigma$ then
follows from $\sigma^2=\sigma_\mathrm{fit}^2-\sigma_\mathrm{ins}^2$.
Since this is only an approximate correction, we use a maximum
likelihood estimate of the velocity moments that at the same time
corrects for each individual velocity error.

Suppose $\mathcal{L}(v)$ is the (intrinsic) velocity distribution of
the stars in an aperture, in one of the three principal directions.
We can consider each stellar velocity measurement $v_i$ in that
aperture as drawn from this distribution, or alternatively, the
product of $\mathcal{L}(v)$ with a delta function around $v_i$,
integrated over all velocities. Due to (instrumental) uncertainties
this delta-function is broadened, and we assume that it can be
described by a Gaussian around $v_i$, with the corresponding velocity
error $\sigma_i$ as the standard deviation. For a sufficient number of
draws $N$, i.e. velocity measurements in the aperture, we can then
recover the (unknown) velocity distribution $\mathcal{L}(v)$ by
maximising the likelihood
\begin{equation}
  \label{eq:deflikelihood}
  L(V,\sigma,\dots)
  = \prod_{i=1}^N \int_{-\infty}^{\infty}
  \mathcal{L}(v)
  \frac{e^{-\frac12\left(\frac{v_i-v}{\sigma_i}\right)^2}}
  {\sqrt{2\pi}\sigma_i} \,\d v,
\end{equation}
or, equivalently, minimising $\Lambda\equiv-2\ln(L)$, with respect
to the mean velocity $V$, mean velocity dispersion $\sigma$ and
possible higher-order velocity moments.

It is possible to recover $\mathcal{L}(v)$ in a non-parametric way
using (extensions of) Lucy's (1974\nocite{1974AJ.....79..745L})
method, but exploiting the fact that Gaussians are good low-order
approximations, the velocity distribution is often parameterised by a
Gauss-Hermite (GH) series (van der Marel \& Franx,
1993\nocite{1993ApJ...407..525V}; Gerhard,
1993\nocite{1993MNRAS.265..213G}). It has the advantage that it only
requires the storage of the velocity moments ($V$, $\sigma$, $h_3$,
$h_4$, \dots) instead of the full velocity distribution. Furthermore,
it allows a simple velocity scaling of the model, which is useful when
investigating the effect of a change in the stellar mass-to-light ratio.

Another advantage of parameterising $\mathcal{L}(v)$ comes from the
observation that the integral in (\ref{eq:deflikelihood}) is the
\textit{convolution} of the velocity distribution and the Gaussian of
each velocity measurement. For a Gaussian velocity distribution this
convolution is straightforward, but also in the case that
$\mathcal{L}(v)$ is described by a GH series, the convolution can be
carried out analytically. This makes it feasible to apply the method
to a large number of discrete measurements and to estimate the
uncertainties on the extracted velocity moments by means of the Monte
Carlo bootstrap method (\S~15.6 of Press et al.\ 
1992\nocite{Press92..numrecipies}).

In the case of no measurement errors, the maximum likelihood estimator
of the standard deviation $\sigma$, given by
\begin{equation}
  \label{eq:MLest_stddev}
  \hat{\sigma} =
  \sqrt{\frac1n\,\sum_{i=1}^n\left(v_i-\overline{v}\right)^2},
  \qquad \mathrm{with}\;\;\overline{v}=\sum_{i=1}^n v_i,
\end{equation}
is a biased estimator, underestimating the true $\sigma$ by a
factor (see also e.g. Kenney \& Keeping 1951\nocite{1951KK}, p.~171)
\begin{equation}
  \label{eq:functionbn}
  b(n) = \sqrt{\frac2n}\frac{\Gamma\left(\frac{n}{2}\right)}
  {\Gamma\left(\frac{n-1}{2}\right)} = 1 - \frac{3}{4n} -
  \frac{7}{32n^2} - \dots.
\end{equation}
where $\Gamma$ is the gamma function. When we take into account the
measurement errors $\sigma_i$, there is no such simple analytical bias
correction as \eqref{eq:functionbn}. However, we can use the latter
result to derive the following approximate corrected standard
deviation estimator
\begin{equation}
  \label{eq:MLest_stddev_correction}
  \tilde{\sigma} \approx \frac{1}{b(n)}
  \sqrt{\hat{\sigma}^2 + [1-b^2(n)]\,\overline{\sigma^2}},
\end{equation}
where $\hat{\sigma}$ is the maximum likelihood estimated
dispersion and $\overline{\sigma^2}=\frac1n\,\sum_{i=1}^n
\sigma_i^2$ the average measurement error.



\begin{figure}
\includegraphics{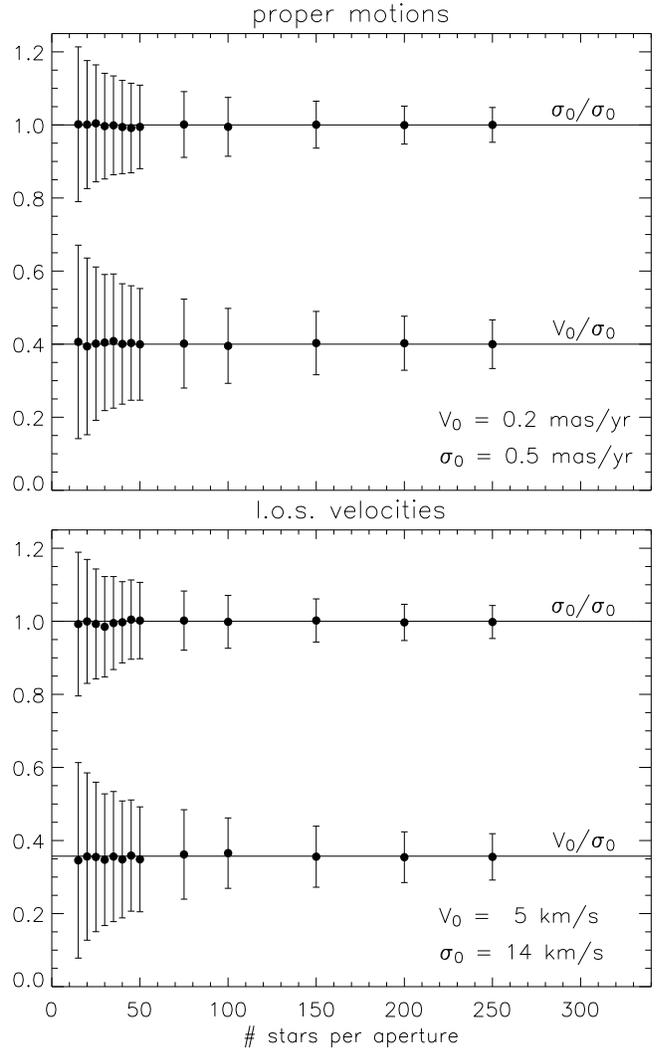}
\caption{ 
  Recovery of maximum-likelihood-estimated kinematics from
  proper motions (top panel) and line-of-sight velocities (bottom
  panel). For a given number of stars per aperture, velocities and
  corresponding errors are simulated by randomly drawing from an
  intrinsic Gaussian distribution with mean velocity $V_0$ and
  velocity dispersion $\sigma_0$, broadened by velocity errors
  randomly drawn from the observed velocity error distributions (left
  panels Figure~\ref{fig:velhist}). Each filled circle with
  error bar shows the mean and standard deviation of the measured
  kinematics from 500 such simulations. As a compromise between
  lower precision (larger error bars) for a small number of stars per
  aperture, and lower spatial resolution (larger bins) for a larger
  number of stars, we choose to have between 50 and 100 stars per bin.
}
\label{fig:kinrecovery}
\end{figure}

\section{Polar grid of apertures}
\label{sec:apgrid}


We use Monte Carlo simulations of the observed stellar velocities and
corresponding errors to investigate the recovery of their average
kinematics. We mimic the stellar velocity observations by randomly
drawing from an assumed intrinsic Gaussian velocity distribution, with
given mean velocity $V_0$ and velocity dispersion $\sigma_0$.  This
set of intrinsic velocities, is then 'instrumentally' broadened by
adding to each velocity a random drawing from a Gaussian with zero
mean and the velocity error as standard deviation. These velocity
errors are simulated by randomly drawing from the observed velocity
error distribution (right panels of Figure~\ref{fig:velhist}). For the
latter we use the \textit{rejection method} (\S~7.3 of Press et al.\ 
1992\nocite{Press92..numrecipies}), with a Lorentzian distribution as
comparison function. In this way, we create, for a given number of
stars, 500 sets of simulated velocities and corresponding errors.

Next, we use the maximum likelihood method of
Appendix~\ref{sec:mlvelocitymoments} to calculate the mean velocity
and velocity dispersion for each simulated set separately.  In
Figure~\ref{fig:kinrecovery}, we compare the (biweight\footnote{The
  biweight mean and biweight standard deviation (e.g. Andrews et al.
  1972; Beers et al. 1999) are robust estimators for a broad range of
  non-Gaussian underlying populations and are less sensitive to
  outliers than other moment estimators.}) mean (filled circles) of
these 500 mean velocity and velocity dispersion measurements with
$V_0$ and $\sigma_0$ (horizontal lines) of the given intrinsic
Gaussian velocity distribution. The error bars are the (biweight)
standard deviation of the kinematic measurements, and indicate the
precision with which the kinematics can be measured, given the
observed velocity error distribution. The precision increases with
increasing number of stars per bin. The precision also increases with
decreasing intrinsic mean velocity dispersion $\sigma_0$. To remove
the latter dependency, we give \textit{relative} kinematic
measurements and corresponding errors, i.e., divided by the
(arbitrarily) chosen value for $\sigma_0$.

Both the mean velocity and velocity dispersion are recovered well.  To
obtain a better precision, we can increase the number of stars per
aperture, but at the same time the spatial resolution decreases, as we
have to increase the size of the apertures. We find that between 50
and 100 stars per aperture is a good compromise. For the proper
motions this implies a (relative) precision for the mean velocity $V$
and velocity dispersion $\sigma$ of respectively $\Delta
V/\sigma\sim0.12$ and $\Delta\sigma/\sigma\sim0.09$. For the
line-of-sight velocities we find similar values, respectively $\Delta
V/\sigma\sim0.12$ and $\Delta\sigma/\sigma\sim0.08$.

Given the average proper motion dispersion of about 0.5 \masyr\ for
\oc\ (\S~\ref{sec:meanVsig}), this means we expect to measure the mean
proper motion and dispersions with an average (absolute) precision of
respectively 0.06 \masyr\ and 0.05 \masyr. Similarly, with an average
line-of-sight velocity dispersion of about 14 \kms\ for \oc, we expect
to measure the mean line-of-sight velocity and dispersion with an
average precision of respectively 1.7 \kms\ and 1.1 \kms.

Indeed, the average of the uncertainties in the kinematics given in
Table~\ref{tab:apVsig_pmxy}~and~\ref{tab:apVsig_vlos} are consistent
with these expectations. Moreover, as predicted, the decrease in the
uncertainties with radius is proportional to the decrease in
dispersion. In other words, if we divide the uncertainties by the
corresponding dispersions, we find nearly constant (relative)
precisions, $\Delta V/\sigma\sim0.11$ and
$\Delta\sigma/\sigma\sim0.08$ for both proper motions and
line-of-sight velocities, consistent with the above simulated
precisions.

To enhance the signal-to-noise of the observations, we first reflect
all measurements back to the first quadrant ($x'\ge0$, $y'\ge0$). We
exploit the fact that for an axisymmetric object, the proper motions
in the $x'$-direction are symmetric in the projected minor axis, while
the proper motions in the $y'$-direction as well as the line-of-sight
velocities are symmetric in the projected major axis. Since our models
are intrinsically axisymmetric, it is equivalent to fit either to the
original or to the reflected data.

We use a polar grid of apertures on (the first quadrant of) the plane
of the sky to better approximate the shape of photometric and
kinematic observations. Every aperture is characterised by its central
radius $r_0>0$ and angle $0^\circ<\theta_0<90^\circ$, together with
its radial and angular width, denoted by $\Delta r$ and
$\Delta\theta$, respectively. We construct the polar grids such that
each aperture has (at least) 50 stars, together with the requirement
that the apertures are as 'round' as possible in the sense that
$\Delta r \approx r_0\Delta\theta$. The latter avoids (very) radial or
angular elongated apertures, which would include stars from (very)
different positions, with probably different (kinematical) properties
than the stars near the centre of the aperture.


\begin{figure*}
\includegraphics{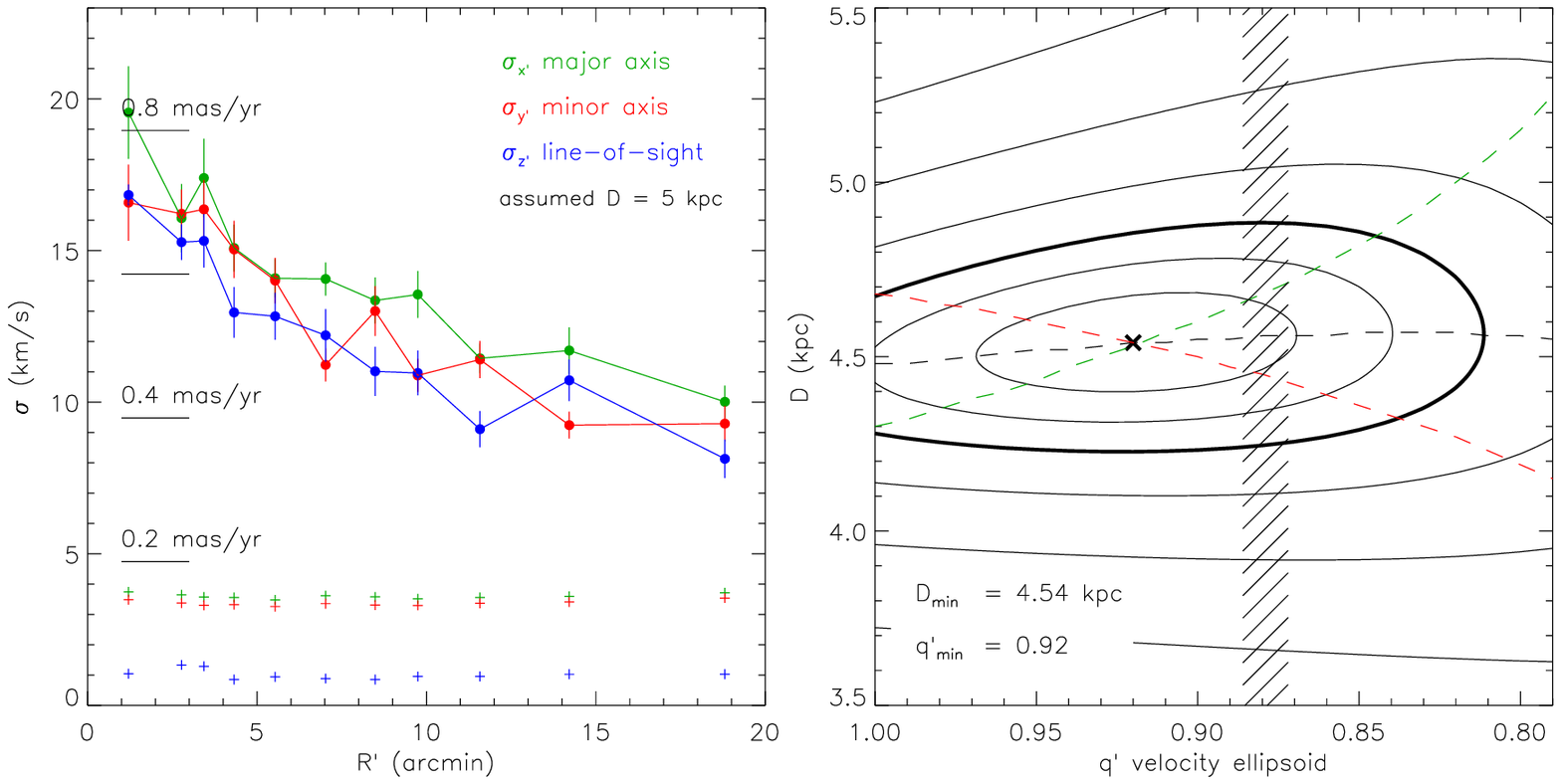}
\caption{
  \emph{Left panel}: velocity dispersion profiles calculated along
  concentric rings. Assuming the canonical distance of 5 kpc, the
  profiles of the proper motion components in the $x'$-direction
  (green) and $y'$-direction (red) are converted into the same units
  of \kms\ as the line-of-sight profile in the $z'$-direction (blue).
  The black horizontal lines indicate the corresponding scale in
  \masyr. Below the profiles, the mean velocity error per ring is
  indicated. \emph{Right panel}: Assuming an oblate velocity ellipsoid
  with constant (projected) flattening, the ratio of the line-of-sight
  over the proper motion velocity dispersion profiles yields an
  estimate for the dynamical distance $D$. The best-fit values
  correspond to the minimum (cross) in the $\Delta\chi^2$ contour
  plot, where the inner three contours are drawn at the 68.3\%, 95.4\% and
  99.7\% (thick contour) levels, and subsequent contours correspond to
  a factor of two increase in $\Delta\chi^2$. For increasing
  flattening of the velocity ellipsoid, starting with the isotropic
  case on the left axis, the green (red) dashed curve shows the
  corresponding best-fit distance if only the profile of the proper
  motion in the $x'$-direction ($y'$-direction) is used, and the black
  dashed curve if both are used.  The observed flattening from the
  stellar photometry (Geyer et al.\ 1983) is indicated by the hashed
  region.}
  \label{fig:constanisomodel}
\end{figure*}

\section{Simple distance estimate}
\label{sec:simpledistest}

The most straightforward way to obtain a dynamical distance
estimate is from the ratio of the line-of-sight velocity
dispersion $\sigma_\mathrm{los}$ and the proper motion velocity
dispersion $\sigma_\mathrm{pm}$ for spherically symmetric objects
(e.g. Binney \& Tremaine 1987\nocite{1987gady.book.....B}, p.~280)
\begin{equation}
  \label{eq:isodist}
  D\;(\mathrm{kpc}) = \frac{\sigma_\mathrm{los}\;(\mathrm{km\,s}^{-1})}
    {4.74\,\sigma_\mathrm{pm}\;(\mathrm{mas\,yr}^{-1})}.
\end{equation}
Using, from the 2295 selected stars with proper motions and 2163
selected stars with line-of-sight velocities, the 718 stars for which
all three velocity components are measured, we find for the two mean
proper motion dispersion components $\sigma_{x'}=0.58\pm0.02$ \masyr\
and $\sigma_{y'}=0.55\pm0.02$ \masyr, and for the mean line-of-sight
velocity dispersion $\sigma_{z'}=12.3\pm0.3$ \kms.
Substituting the latter value together with the average proper motion
dispersion in (\ref{eq:isodist}), we obtain a distance of
$D=4.6\pm0.2$ kpc.\looseness=-1

This value is below the canonical distance $D=5.0\pm0.2$ (Harris et
al.\ 1996\nocite{1996AJ....112.1487H}).  The above simple distance
estimate is not valid for \oc, which is not spherically symmetric.
Moreover, although the above average values for $\sigma_{x'}$ and
$\sigma_{y'}$ are just consistent with each other, from the left panel
of Figure~\ref{fig:constanisomodel} it is clear that the profile of
the mean proper motion dispersion profile of $\sigma_{x'}$ (green)
lies systematically above that of $\sigma_{y'}$ (red). A non-spherical
anisotropic model is needed to explain these observations. Here we
consider a simple model with constant anisotropy.

If we make the (ad-hoc) assumption that the velocity ellipsoid is
oblate with intrinsic semi-axis lengths
$\sigma_x=\sigma_y\equiv\sigma$ and $\sigma_z=q_\mathrm{ve}\sigma$
(all in \kms), where $q_\mathrm{ve}$ is the average intrinsic
flattening, the observed velocity dispersions are given by
\begin{eqnarray}
  \label{eq:velellipsoid}
  \sigma_{x'} & = & \sigma\,/\,4.74\,D \quad \mathrm{mas\,yr}^{-1}, \nonumber \\
  \sigma_{y'} & = & q'_\mathrm{ve}\,\sigma\,/\,4.74\,D \quad \mathrm{mas\,yr}^{-1}, \\
  \sigma_{z'} & = & \left[ 1 - (1-q'^{\,2}_\mathrm{ve})\cot^2i \right]^{1/2}\,\sigma
  \nonumber \quad \mathrm{km\,s}^{-1},
\end{eqnarray}
where we have used eq.~\eqref{eq:poscartint2obs} and the relation
$q^2\sin^2i=q'^2-\cos^2i$. Using the best-fit value for $D\tan i$
of 5.6 kpc (\S~\ref{sec:constraintincl}), we eliminate the
inclination $i$. Next, by fitting the ratios of the line-of-sight
velocity dispersion over the proper motion dispersion components,
$\sigma_{z'}/\sigma_{x'}$ and $\sigma_{z'}/\sigma_{y'}$, to the
observations in the left panel of Figure~\ref{fig:dispprofile}, we
determine the best-fit values for the remaining two free
parameters: the distance $D$ and the (projected) flattening of the
velocity ellipsoid $q'_\mathrm{ve}$.

Since we use the full dispersion profiles and we allow for an
anisotropic velocity distribution, this simple way to obtain a
dynamical distance estimate goes beyond the above spherical
symmetric approach. If $q'_\mathrm{ve}=1$ in
eq.~\eqref{eq:velellipsoid}, we recover this spherical symmetric
approach in which both ratios are equal and the distance follows
from eq.~\eqref{eq:isodist}.

We show in the right panel of Figure~\ref{fig:constanisomodel} the
$\Delta\chi^2$ contours for a range of $q'_\mathrm{ve}$ and $D$.
The overall minimum, indicated by a cross, corresponds to the
best-fit values $q'_\mathrm{ve}=0.92\pm0.05$ and $D=4.54\pm0.14$
kpc. The isotropic case ($q'_\mathrm{ve}=1$) is excluded at about
the 95.4\%-level.  The best-fit (projected) flattening of the
velocity ellipsoid is less than the average observed flattening
$q'=0.879\pm0.007$ (hashed region) from the stellar photometry of
\oc\ (Geyer et al.\ 1983\nocite{1983A&A...125..359G}), although an
equivalent value is not excluded (at the 68.3\%-level). The
velocity distribution is expected to be less flattened, since it
traces more directly the potential, which in general is rounder
than the light distribution (see e.g. p. 48 of Binney \& Tremaine
1987\nocite{1987gady.book.....B}).

If we only fit the ratio $\sigma_{z'}/\sigma_{x'}$, the green dashed
curve shows the best-fit distance at given flattening.  While in this
case the distance increases with flattening, almost exactly the
opposite happens if we only fit the ratio $\sigma_{z'}/\sigma_{y'}$
(red dashed curve). Simultaneously fitting both ratios does not
provide a good fit (the $\chi^2$ value is significantly larger than
the number of degrees of the freedom) and the resulting best-fit
distance (black dashed curve) of about 4.5 kpc is significantly below
the canonical distance of 5.0 kpc. 

We conclude that both the simple distance estimate \eqref{eq:isodist}
and the above constant-anisotropy axisymmetric model are not valid for
\oc\ and underestimate its distance. To explain the observed
kinematics of \oc\ and obtain a reliable distance estimate, one needs
a non-spherical dynamical model with varying anisotropy, like the
Schwarzschild modelling technique used in this paper.

Finally, it is interesting to note that Peterson \& Cudworth
(1994\nocite{1994ApJ...420..612P}) reported rotation in the
line-of-sight velocities and proper motions of M22, and found that its
dynamical distance increased slightly after an approximate correction
based on a comparison of dispersion profiles. Peterson et al.\ 
(1995\nocite{1995ApJ...443..124P}) saw no evidence for rotation in M4,
but did note that their resulting dynamical distance was smaller than
the canonical value. Both studies used the simple distance estimates
described here. It will be interesting to reanalyse these clusters
with the comprehensive method we have presented in this paper.


\end{document}